\documentclass[a4,epsf,12pt]{article}
\usepackage{axodraw}
\newcommand{\ep}{\varepsilon}
\newcommand{\msb}{\overline{\rm MS}}
\newcommand{\MSb}{$\overline{\rm MS}$ }
\newcommand{\Li}[2]{{\mbox{Li}}_{#1\!}\left(#2\right)}
\newcommand{\Cl}[2]{{\mbox{Cl}}_{#1}\left(#2\right)}

\newcommand{\tfrac}[2]{{\textstyle{\frac{#1}{#2}}}}

\newcommand{\lapprox}{\raisebox{-.2ex}{$\stackrel{\textstyle<}{\raisebox{-.6ex}[0ex][0ex]{$\sim$}}$}}
\newcommand{\gto}{\raisebox{-.2ex}{$\stackrel{\textstyle>}{\raisebox{-.6ex}[0ex][0ex]{$\to$}}$}}

\begin{document}
%=====================================================================
\begin{flushleft}
{\normalsize \rm
DESY 01-067  \hspace{9cm} hep-ph/0105304 \\[3mm]
% {May 2001}
%DESY 01-067 [hep-ph/0105304]\\
May 2001 (rev. February 2002) \\
\vspace*{0.5cm}}
\end{flushleft}

 \begin{center}
 {\large \bf
 $\overline{{\rm MS}}$ vs. Pole Masses of Gauge Bosons: \\
 Electroweak Bosonic Two-Loop  Corrections}
 \end{center}
 \vspace{1cm}
 \begin{center}
{\large
F.~Jegerlehner%
\footnote{~E-mail: fjeger@ifh.de},
M.~Yu.~Kalmykov%
\footnote{~E-mail: kalmykov@ifh.de}
\footnote{
~On leave of absence from BLTP, JINR, 141980, Dubna (Moscow Region),
Russia},~
O.~Veretin%
\footnote{~E-mail: veretin@ifh.de}
}

\vspace{5mm}

{\it ~DESY Zeuthen, Platanenallee 6, D-15738, Zeuthen,
  Germany}
\end{center}
\hspace{3in}
\begin{abstract}

The relationship between $\overline{\rm MS}$ and pole masses of the
vector bosons $Z$ and $W$ is calculated at the two-loop level in the
Standard Model. We only consider the purely bosonic contributions
which represent a gauge invariant subclass of diagrams. All
calculations were performed in the linear $R_\xi$ gauge with three
arbitrary gauge parameters utilizing the method of asymptotic
expansions. The results are presented in analytic form as series in
the small parameters $\sin^2\theta_W$ and the mass ratio $m_Z^2/m_H^2$.
We also present the corresponding on-shell mass counter-terms for the
massive gauge bosons, which will be needed for the calculation of
observables at two-loops in the on-shell renormalization scheme. 
%As
%a byproduct we obtain the bosonic two-loop contributions to the
%renormalization of the weak mixing parameter $\sin^2\theta_W$ and of
%the Fermi constant $G_F$. The running of $G_F$ will become important
%at high energy colliders.
\end{abstract}
%%%%%%%%%%%%%%%%%%%%%%%%%%%%%%%%%%%%%%%%%%%%%%%%%%%%%%%%%%%%%%%%%%%%%%
%\thispagestyle{empty}
%\setcounter{page}{0}
\section{Introduction}
Precision Physics of the electroweak gauge bosons $Z$ and $W$ started
about 12 years ago at the LEP storage ring with the ALEPH, DELPHI, L3
and OPAL experiments and ended just recently with the dismantling of
the LEP installation. In particular the very accurate determination of
the masses and the couplings to the fermions revealed unexpectedly
rich information about the quantum correction of the Standard Model
(SM)~\cite{Abbaneo:2001ix}. Calculations of higher order corrections
thus gained increasing importance. At the one--loop level these
calculations for the relevant ``2 fermions into 2 fermions'' processes
were completed before LEP started operating in 1989 \cite{YR89}. These
SM predictions enabled an indirect determination of the top mass which
culminated in the top discovery at the Tevatron. Now after the top
mass has been fixed with rather good accuracy, the indirect bound to
the Higgs mass, the last missing SM parameter, is the main goal.  The
knowledge of the actual value of the Higgs mass is extremely important
because it determines how Higgs physics will look like at future
colliders like the LHC or TESLA~\cite{Aguilar-Saavedra:2001rg}.  Since
the sensitivity of SM predictions on the Higgs mass is weak the
precise meaning of the indirect Higgs mass bounds depend crucially on
the accuracy of the theoretical predictions. Fortunately a lot of
important theoretical progress has been made in the last decade with
the calculation of leading and some sub-leading two--loop effects~
\cite{QED-running}-\cite{2loop-muon:QED}. However, no complete 
two--loop calculation could be achieved so far, because such
calculations are hampered by the dramatic increase in complexity
encountered in such calculations. How important the precise evaluation
of radiative corrections is may be illustrated by the following fact: taking
only the leading corrections, the shift $\Delta \alpha_{\rm em}$ in
the fine structure constant and the quadratic top mass correction
$\Delta \rho_{\rm top}$ in the relationship between neutral and
charged current effective couplings, predictions are about 10$\sigma$
off from the data for most of the precisely known observables like $\sin^2
\theta^\ell_{\rm eff}$ or $M_W$~\cite{FJLL94,ASLP99}. Thus the
sub-leading effects are huge in relation to current experimental
precision. Therefore the issue of sub-leading two--loop
corrections has to be taken very seriously. They easily may obscure
the interpretation of the indirect Higgs mass bound obtained from LEP
experiments by using SM predictions which are incomplete at the
two--loop level.

Although we are a long way from filling the gap, as a first step we
calculate the full bosonic two-loop electroweak corrections to the
on-shell masses of gauge bosons. In other words, we evaluate the
two-loop renormalization constants, the relations between bare, \MSb
and on-shell masses, which are required as an essential part for the
two-loop renormalization program in the on-shell scheme, with the fine
structure constant $\alpha$ and the gauge boson masses $M_W$ and $M_Z$
as independent input parameters.

The programs developed for performing the present calculations can be
applied without modifications for the calculation of the on-shell
wave-function renormalization constants of the gauge bosons. Since these
quantities are gauge dependent we don't reproduce corresponding
results here. As yet we have not calculated any observable quantity.
However, the new set of corrections will be important for future,
complete electroweak two-loop calculations of physical
observables. Also after the shutdown of LEP it is important to
continue such calculations because the question how additional
corrections affect the Higgs mass bound can be answered
retrospectively, once given the precise LEP/SLC results. Full two-loop
calculations would be indispensable in future in any case if a project
like TESLA with the Giga$Z$ option would be realized.

The most recent essential progress here was achieved in the
calculation of the top--quark contributions to the two-loop
electroweak corrections.  The corresponding contributions to the
$\rho$-parameter were considered in~\cite{2loop-rho}, the one's to
$\Delta r$, which determines the $M_W-M_Z$
relationship, in~\cite{2loop-muon:SM} and to the $Z$ boson
partial widths in \cite{2loop:Z}\footnote{ Another important step forward was the
calculation of the two--loop QED correction to the muon decay width
\cite{2loop-muon:QED}.}. 
Two different approaches---the asymptotic expansion method
\cite{asymptotic} and numerical integration\cite{2loop-muon:numeric}
---have been used to perform these calculations. One of the important
steps when performing these calculations is the two-loop renormalization of the
gauge boson masses, which also is contributing to the $\sin^2 \theta_W$
renormalization~\cite{onshell-scheme}-\cite{FJ}.  In the SM so far no
complete analytical calculation of the two-loop renormalized
propagator has been carried out~\cite{ftt}. The first available
results were given for zero external momentum~\cite{LTL}, 
when the original diagrams may be reduced to a set of
bubble-type integrals with different masses~\cite{bubble}. 
In~\cite{2loop-fermion} the two-loop unrenormalized
fermion corrections to the gauge boson propagator have been presented
for off-shell momentum in the general linear $R_\xi$ gauge.  The
results are presented in terms of scalar master integrals with
several different mass scales, the masses of fermions and bosons. For the
evaluation of these master diagrams analytical
results~\cite{2loop-analytic-a,2loop-analytic-b} or one-fold 
integral representations are available~\cite{2loop-numeric}.

The main aim of the present paper is to present a calculation within
the Standard Model of the two-loop bosonic contributions to the
relationship between $\overline{\rm MS}$ and on-shell masses of the gauge
bosons ($W,Z$). This relationship, alternatively, will be represented
in terms of on-shell renormalization constants.  We shall discuss in
some detail the algorithm used for calculating  two-loop electroweak
corrections for on-shell quantities in an arbitrary gauge.

The paper is organized as follows. In Section~2 we briefly reconsider
the definition of the pole mass of the massive gauge bosons within
the Standard Model. The calculations have been performed with the help of
computer programs which will be described in some detail in Section~3.
In Section~4 we discuss the UV renormalization of the pole mass and
the interrelation of our results with the standard renormalization
group approach. In particular we make several cross-checks of the
singular $1/\ep^2$- and $1/\ep$-terms. The numerical results for the
finite parts are presented and discussed in Section~5. For further
technical details and some useful formulae we refer to four
appendices. In Appendix~A we collect results for the one-loop
propagator type diagrams. Special attention is given here to the
$\ep$-parts of the corresponding integrals which are needed for the
two-loop calculation. Appendix~B and C collect one-loop results in
$d=4$ and $d\neq4$, respectively. They are included for completeness.
In Appendix~D we present the analytical coefficients which are the
main results of our investigation.

%%%%%%%%%%%%%%%%%%%%%%%%%%%%%%%%%%%%%%%%%%%%%%%%%%%%%%%%%%%%%%%%%%%%%%%
\section{Pole mass}
\setcounter{equation}{0}
%%%%%%%%%%%%%%%%%%%%%%%%%%%%%%%%%%%%%%%%%%%%%%%%%%%%%%%%%%%%%%%%%%%%%%%

The position of the pole $s_P$ of the propagator of a massive gauge boson 
in a quantum field theory is a solution for $p^2$ at which the inverse
of the connected full propagator equals zero, i.e.,
\begin{equation}
%\left. p^2 - m^2 - \Pi(p^2,m^2,\cdots)\right|_{p^2=\tilde{M}^2} = 0,
s_P - m^2 - \Pi(s_P,m^2,\cdots) = 0,
\label{pole}
\end{equation}
where $\Pi(p^2,\cdots)$ is the transversal part of the one-particle
irreducible self-energy. The latter depends on all SM parameters but,
in order to the keep notation simple, we have indicated explicitly only
the dependence on the external momentum $p$ and in some cases also
$m$, where $m$ is the mass of the particle under consideration. This can be
either the bare mass $m_0$ or the renormalized mass defined in some
particular renormalization scheme.

Generally, the pole $s_P$ is located in the complex plane of $p^2$ 
and has a real and an imaginary part. 
We write 
\begin{equation}
s_P \equiv M^2 - {\rm i } M \Gamma .
\label{def}
\end{equation}
The real part of (\ref{def}) defines $M$ which we call 
the pole mass\footnote{Throughout this paper we identified the terms 
pole mass and on-shell mass.}, 
while the imaginary part is related to the width $\Gamma$ of the particle. 
This is the natural generalization of the physical mass
of a stable particle, which is defined by the mass of its asymptotic
scattering state. 

For the remainder of the paper we will adopt the following notation: 
capital $M$ always denotes the pole mass; 
lower case $m$ stands for the renormalized mass in the 
$\overline{\rm MS}$ scheme, 
while $m_0$ denotes the bare mass. In addition we use $e$
and $g$ to denote the $U(1)_{\rm em}$ and $SU(2)_{\rm L}$ couplings of
the SM in the $\overline{\rm MS}$ scheme.

In perturbation theory (\ref{pole}) is to be solved order by order.
To two loops we have the following solution of (\ref{pole})
\begin{eqnarray}
\hspace{-1cm}
s_P &=& m^2 + \Pi^{(1)}(m^2,m^2,\cdots) + \Pi^{(2)}(m^2,m^2,\cdots) 
+ \Pi^{(1)}(m^2,m^2,\cdots) \Pi^{(1)}{}'(m^2,m^2,\cdots),
\label{polemass}
\end{eqnarray}
which yields the pole mass $M^2$ and the width $\Gamma$ at this order.
$\Pi^{(L)}$ is the bare ($m=m_0$) or {\MSb}-renormalized ($m$ the 
{\MSb}-mass) $L$-loop contribution to $\Pi$, and the prime denotes the
derivative with respect to $p^2$.  In this way we need to evaluate
propagator type diagrams and their derivatives at
$p^2=m^2$. Diagrammatically the self-energy contributions are shown in
Fig.~\ref{Pi}.

The (on-shell) mass counter terms $\delta M^2$ for the on-shell
renormalization scheme are obtained by considering the real part of
(\ref{polemass}) upon identifying the r.h.s with the bare expression, setting
$m^2 \equiv m_0^2=M^2+\delta M^2$ and solving for $\delta M^2=(\delta
M^{2})^{(1)}+(\delta M^{2})^{(2)}$ to two loops.

% NEW: Corrected by 10.02.02 
In this paper we show by explicit calculation at the two-loop level
that the bosonic correction to the pole $s_P$ (and hence to the
on-shell mass counter-term) is a gauge invariant and infrared stable
quantity\footnote{From general considerations we know (see~\cite{FJ},
for example), that after summing all diagrams, including the tadpoles,
at a given order of perturbation theory the location of the pole of
the propagator is a gauge invariant quantity.  The gauge invariance of
the 2-loop massless fermion correction to the pole mass of the gauge
bosons was verified in~\cite{2loop-fermion}.}.  The free propagator of
a massive vector boson in the linear $R_\xi$ gauge reads
\begin{equation}
\label{freepropagator}
D_{\mu\nu}(p) = \frac{i}{p^2-m^2} \left(
  - g_{\mu\nu} + (1-\xi) \frac{p_\mu  p_\nu}{p^2 - \xi m^2} \right),
\end{equation}
where $\xi$ is a gauge parameter.
We decompose the vector boson self-energy  $\Pi_{\mu \nu}(p^2,m^2,\cdots)$ 
into a transverse $\Pi(p^2)$ and a longitudinal $L(p^2)$ part
\begin{equation}
\Pi_{\mu \nu}(p^2,m^2,\cdots) =
     \left( g_{\mu \nu} - \frac{p_\mu  p_\nu}{p^2} \right) \Pi (p^2)
    + \frac{p_\mu  p_\nu}{p^2} L(p^2).
\end{equation}
In (\ref{polemass}) only the transverse part contributes
and the dressed propagator reads 
\begin{equation}
{D}_{\mu \nu}(p) = \frac{-i }{p^2-m^2 - \Pi(p^2)}\left( g_{\mu \nu} -
\frac{p_\mu p_\nu}{p^2} \right) + \frac{p_\mu p_\nu}{p^2}\; \cdots
\end{equation}
The simple relation between the full propagator and the irreducible
self-energy only holds if there is no mixing, like for the $W$-boson.
In the neutral sector, because of $\gamma-Z$ mixing, we cannot
consider the $Z$ and $\gamma$ propagators separately.  They form a
$2\times2$ matrix propagator, so that (\ref{pole}) is modified into (see
details in \cite{FJ,gammaZ}) 
\begin{equation}
\label{Zgamma}
s_P - m_Z^2 - \Pi_{ZZ}(s_P) 
- \frac{\Pi^2_{\gamma Z}(s_P)}{s_P-\Pi_{\gamma \gamma}(s_P)} = 0.
\end{equation}
We note that the $\Pi^2_{\gamma Z}$ mixing term starts to contribute
at the two-loop level. Obviously, we do not need to compute
$\Pi_{\gamma\gamma}$ here since it starts to play a role only beyond 
the two-loop approximation. In the sequel we will denote the
self-energies by $\Pi_V$ ($V=W,Z$) with
$$\Pi_W(p^2,\cdots)=\Pi_{WW}(p^2,\cdots)$$ and
$$\Pi_{Z}(p^2,\cdots)= \Pi_{ZZ}(p^2,\cdots) 
+ \frac{\Pi^2_{\gamma Z}(p^2,\cdots)}{p^2-\Pi_{\gamma
\gamma}(p^2,\cdots)}\;.$$ Thus, formally, the form (\ref{pole})
applies for both the $W$ and the $Z$.

The non-zero imaginary part (width) (\ref{def}) of the on-shell gauge
boson self-energy appears as soon as the fermions are included.  For
the bosonic contributions alone the imaginary part of
$\Pi(p^2)$ on the mass-shell is zero at the two-loop level (see below).

%%%%%%%%%%%%%%%%%%%%%%%%%%%%%%%%%%%%%%%%%%%%%%%%%%%%%%%%%%%%%%%%%%%%%
\section{Program part}

\begin{figure}[th]
\begin{center}
\centerline{\vbox{\epsfysize=100mm \epsfbox{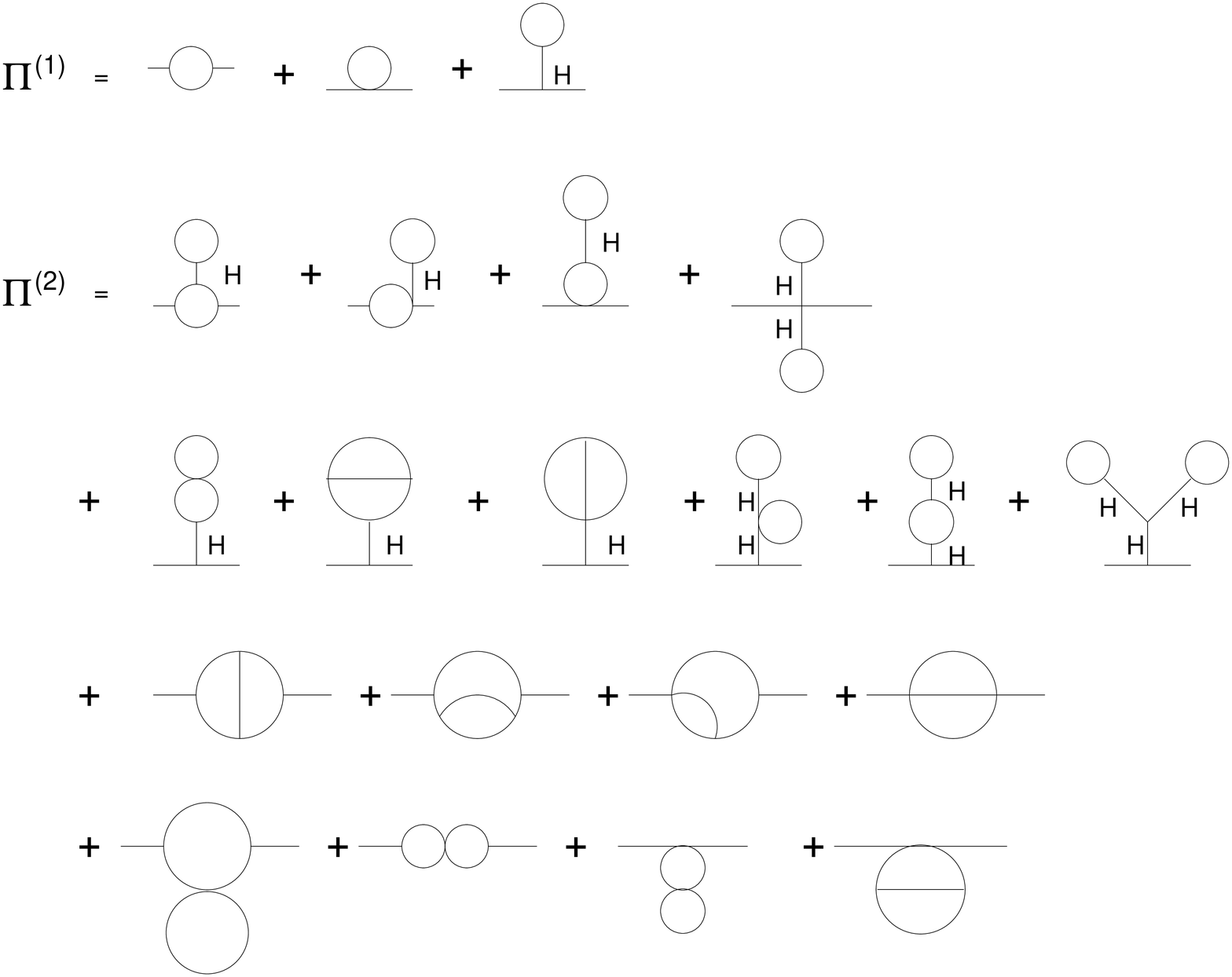}}}
\caption{\label{Pi} 
One- and two-loop contributions to the massive boson self-energies.}
\end{center}
\end{figure}

In order to find the relations between the pole masses $M_Z^2,\,M_W^2$
and the $\overline{\rm MS}$ masses $m_Z^2,\,m_W^2$ we have to compute
the one- and two-loop self-energies for $Z$- and $W$-bosons at
$p^2=m_Z^2$ and $p^2=m_W^2$, respectively.  The complete set of
topologies that occurs in this calculation is shown in
Fig.~\ref{Pi}. In order to be able to work with manifestly gauge 
parameter independent 
renormalization constants we have to include the Higgs tadpole diagrams.
%
%  FIGURE: TOPOLOGIES

While at one-loop order we have about 50 diagrams, in the two-loop
approximation the number of diagrams is about 1000, which requires an
automatized generation and evaluation of diagrams.  We use ${\bf
QGRAF}$ \cite{qgraf} to generate the diagrams and then the C-program
{\bf DIANA} \cite{diana} to produce for each diagram an input suitable
for our {\bf FORM} \cite{FORM} packages%
\footnote{{\bf DIANA} generates 
additional information, e.g. identifying symbols for the particles of the 
diagram  and their masses, distribution of integration momenta, number of 
fermion loops etc.}.  

For two-loop propagator type diagrams with several masses a complete
set of recurrence relations is given in \cite{tarasov-propagator}.  It
allows us to reduce all tensor integrals to a small set of so-called
master-integrals. However, the master-integrals which show up in the
SM are not expressible in terms of known functions but may be written
e.g. as one-fold integrals \cite{2loop-numeric}.  Instead of using
these explicit formulae we resort to some approximations here, namely,
we perform an appropriate series expansion in mass ratios\footnote{For
diagrams with several different masses, there may exist several small
parameters.  In this case we apply different asymptotic expansions
(see \cite{2region}) one after another.}.  Each coefficient of this
series can be calculated analytically by means of the asymptotic expansion
algorithm described in \cite{asymptotic}.

To keep control of gauge invariance we work in a $R_\xi$ gauge with
three different gauge parameters $\xi_W,\,\xi_Z$ and $\xi_\gamma$.
The corresponding free vector boson propagators (\ref{freepropagator})
thus exhibit the new masses $\sqrt{\xi_W}m_W$ and $\sqrt{\xi_Z}m_Z$ in
the propagators.  This complicates the calculation enormously both in
Tarasov's algorithm as well as in the asymptotic expansion approach.
With the first method the presence of new masses in both the reduction
formulae and the Gram determinants leads to cumbersome expressions
which are difficult to simplify.  In the asymptotic expansion approach
the unphysical parameters $\xi_W m_W^2$ and $\xi_Z m_Z^2$ define two
new scales.  Of course, in order to keep things manageable, we have to
keep the number of scales as small as possible.  This can be done by
expanding diagrams about some fixed values of the gauge parameters.
Three different regimes of expansion are feasible: a)~$\xi\to\infty$,
b)~$\xi\to0$ and c)~$\xi\to1$.  We choose the last possibility
expanding the original propagators at $\xi_i=1$ in a Taylor
series. For the purpose of checking the gauge invariance of our
results it is sufficient to keep the first three terms of the
expansion, so that the propagators of the vector bosons and associated
Higgs scalar ghosts look like
\begin{eqnarray}
D_{\mu \nu}^V(p) & = & \frac{i}{p^2-m_V^2} \left(
  - g_{\mu \nu} + (1-\xi_V)\frac{p_\mu  p_\nu}{p^2 - m_V^2} 
  - (1-\xi_V)^2 \frac{m_V^2 p_\mu p_\nu}{(p^2 - m_V^2)^2} 
  + \dots \right), 
\nonumber \\
\Delta_V(p) & = & \frac{i}{p^2-m_V^2} \left( 1 - (1-\xi_V) \frac{m_V^2}{p^2 - m_V^2} 
         + (1-\xi_V)^2 \left( \frac{m_V^2}{p^2 - m_V^2} \right)^2 
         + \dots \right), 
\label{xiexpand}
\end{eqnarray}
where $V=W,Z$ and  the dots stand for the terms of higher order in
$(1-\xi_V)$, which we don't take into account. At the same time we do
not have any problems with the photon propagator and use it in its
usual form.

One drawback of our choice of the expansion about $\xi_i=1$ is that
possible unphysical thresholds do not show up in our expansion and
hence will not produce the correct imaginary part for a given diagram.
Examples are the thresholds $p^2=4\xi_Z M_Z^2$ ($Z \to \phi^0\phi^0$
production, $\phi^0$ the neutral Higgs ghost) which is below the $Z$
mass-shell $p^2=M_Z^2$ when $\xi_Z < \frac14$ or $p^2=\xi_W
M_W^2$($W^\pm \to \phi^\pm \gamma$ production, $\phi^\pm$ the charged
Higgs ghosts) which is below the $W$ mass-shell $p^2=M_W^2$ when
$\xi_W < 1$. In such cases, the above expansion does not preserve the
analytical properties of diagrams exhibiting ghost particles. However,
in any case unphysical degrees of freedom should not contribute to the
physical width. While individual diagrams exhibit an imaginary part,
in the sum of all diagrams it has to cancel. This cancellation is a
consequence of the Slavnov-Taylor identities, which tell us how Higgs
ghost, Faddeev-Popov ghosts and scalar components contained in the
gauge boson fields decouple from observables like the physical
width. Therefore, the above expansion indeed reproduces the correct
gauge invariant result. At the one-loop level one can do an exact
analytic calculation for the $R_\xi$ gauge with an arbitrary value of
the gauge parameter. This has been done long time ago~\cite{FJ} and,
not surprisingly, the on-shell self-energies of the $W$- and
$Z$-bosons are gauge invariant and do not exhibit any unphysical
threshold (possible problems related to unphysical thresholds in the
non-gauge invariant wave-function renormalization factor are also
discussed in~\cite{FJ}). For the two-loop contribution, as we should,
we get a gauge invariant on-shell limit which is real. Gauge
cancellations are highly non-trivial and only happen if one is doing a
perfectly consistent calculation. We have verified that for gauge
parameters $\xi > 1$ the imaginary part of the $W$ and $Z$
self-energies in the bosonic sector up to two-loops is zero, by
applying the Cutkowsky rules and inspecting all possible two and three
particle intermediate states allowed by the SM Lagrangian. While for
$\xi > 1$ the imaginary part is zero for each individual diagram, for
small enough values of the gauge parameters a non-trivial cancellation
must take place.  An independent direct check of this is possible by
considering the problem in the limit $\xi \to 0$, for example.

After we have made the expansion~(\ref{xiexpand}), the bosonic
contributions we are considering only depend on the three different
masses $m_W^2, m_Z^2$ and $m_H^2$. One natural small parameter is the
weak mixing parameter $\sin^2\theta_W = 1 - m_W^2/m_Z^2 <
0.25$.  We expand in this parameter and get rid in this way of $m_W$
(or $m_Z$).  For diagrams which contain Higgs boson lines\footnote{We
have 280 and 357 two-loop one-particle irreducible diagrams for the $Z$-
and $W$-boson, respectively.}  we apply an asymptotic expansion with
respect to a heavy Higgs mass. Taking into account the most recent
lower bound on the value of the Higgs boson mass we are dealing with an
expansion parameter $m_Z^2/m_H^2 \lapprox 0.64$. This implies that
we have to calculate quite a number of coefficient in the expansion in
order to get a convincing result. We should mention that these
expansions are well behaved because $p^2=M_V^2$ in any case is below the
possible thresholds.  The convergence of such expansions can be easily
checked for the one-loop case where the exact analytic result is
available.

\begin{figure}[th]
\begin{center}
\centerline{\vbox{\epsfysize=80mm \epsfbox{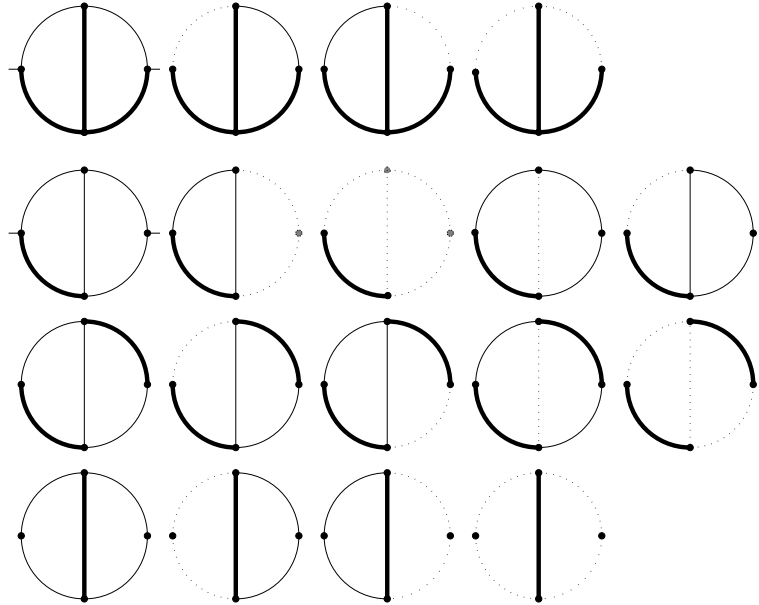}}}
\caption{\label{prototypes} The prototype diagrams and their subgraphs
contributing to the large mass expansion for two-loop diagrams with
heavy propagators. Thick and thin lines correspond to the heavy- and
light-mass (massless) particle propagators, respectively.  Dotted
lines indicate the lines omitted in the subgraph.}
\end{center}
\end{figure}

In our case we find four different prototype structures with Higgs
lines inside of the loops.  The large mass expansion has been
performed with the help of the packages {\bf TLAMM} \cite{tlamm} and
\cite{top} once for Euclidean \cite{diagramatic} and once for
Minkowski space-time, respectively. The corresponding topologies and
set of subgraphs are given in Fig.~\ref{prototypes}.
%
%  FIGURE: HIGGS PROTOTYPES
The diagrams without Higgs\footnote{At the two-loop level we find 
336 one-particle irreducible diagrams without a Higgs line for 
the $Z$-boson and 435 for the $W$-boson.}  
are nothing but single scale massive diagrams,
when all internal masses are equal to the external momentum or zero. Such
diagrams can be calculated analytically. For this purpose we use the
packages {\bf ONSHELL2} \cite{onshell2} and another one written by
O.V. \cite{oleg-onshell} for the calculation of the set of the master
integrals given in \cite{onshell-master}.  We find that only the following
four prototypes are required (in terms of the notation used in
\cite{onshell2}): {\bf F11111}, {\bf F11110}, {\bf F01111} and {\bf
F01101}.

For independent verification of the input, the Feynman rules and the
evaluation we performed calculations independently in Euclidean
(M.Yu.K) and Minkowski (O.V.) space-time and got full agreement
between them~\footnote{The packages used in the calculations and the
results can be found at the following URL address {\tt
http://www-zeuthen.desy.de/$\;\widetilde{}\;$kalmykov/pole/pole.html}.}.

%%%%%%%%%%%%%%%%%%%%%%%%%%%%%%%%%%%%%%%%%%%%%%%%%%%%%%%%%%%%%%%%%%%%%%
\section{UV renormalization in the $\overline{\rm MS}$ scheme}
\setcounter{equation}{0}
%%%%%%%%%%%%%%%%%%%%%%%%%%%%%%%%%%%%%%%%%%%%%%%%%%%%%%%%%%%%%%%%%%%%%%

Here we describe in more detail the renormalization procedure. It is
well known that the pole mass in QED/QCD is a gauge independent and
infrared stable quantity to all orders of the loop expansion
\cite{pole:QCD}. To renormalize the pole mass at the two-loop level 
requires to calculate the one-loop renormalization constants for all
physical parameters (charge and masses), and the two-loop
renormalization constant only for the mass itself. Not needed are the
wave-function renormalizations or ghost (unphysical) sector
renormalizations. The above mentioned basic properties of the pole
mass are valid also in the SM. In order to obtain a gauge invariant
result in the SM, however, we have to add in a proper way the tadpole
contributions~\cite{FJ}. The tadpole terms are due to the vacuum
expectation value (VEV) of the Higgs field, which does not vanish
automatically. By a constant shift we can adjust the Higgs field to
have vanishing VEV, however. Since here the Higgs field is integrated
out in the path integral the result cannot depend on whether we
perform a shift or not. Indeed, after we take into
account all diagrams shown in Fig.~\ref{Pi} we find a gauge
invariant result for the pole mass up to two loops in terms of the
bare parameters. The tadpole contribution can be calculated either
from tadpole diagrams\footnote{These are the two-loop bubble diagrams for
the process $H$ in terms of {\bf QGRAF} notation.} or from Ward
identities, which connect the tadpoles with the one-particle
irreducible self-energies of the pseudo-Goldstone bosons $\Pi_{\phi
\phi}$ at zero momentum
\begin{equation}
\Pi_{\phi \phi}^{\rm 1PI} (0) + T = 0, 
\end{equation}
where $\phi = \phi^+,\phi^-, \phi^0$. 
We performed both types of calculations and obtained full agreement. 
Diagrams contributing to the $Z$-boson pole mass do not contain 
infrared  singularities at the two-loop level. 
Infrared finiteness of the $W$-boson mass was proven in 
\cite{pole:SM}. We also give an alternative proof of this 
statement for our case. 

In our calculation dimensional regularization \cite{dimreg} is used,
which allows one to regularize both UV and IR singularities by the
same parameter $\ep$ related to the dimension of space-time by
$d=4-2\ep$.  We first perform the UV-renormalization within the
$\overline{\rm MS}$ scheme in order to obtain finite results. In a
next step we find the relation between the pole- and $\overline{\rm
MS}$ parameters.  We adopt the convention that the $\overline{\rm
MS}$-parameters are defined by multiplying each $L$-loop integral by
the factor $(\exp({\gamma})/4\pi)^{\varepsilon L}$.  Each loop picks
an additional factor $1/16\pi^2$.

\subsection{One-loop charge renormalization}
The bare charge $e_0$ and the $\overline{\rm MS}$ charge\footnote{
All $\overline{\rm MS}$-parameters, like $e,g,g'$ and all renormalization 
constants, like $Z_{\overline{\rm MS}}$ or $z_{\rm OS}$ are $\mu$-dependent
quantities.} 
$e$ are related
via $e_0= \mu^\ep e\:\left(1+Z_{\overline{\rm MS}}^{(1)}/\varepsilon
+O(e^4) \right)$, 
with the appropriate constant $Z_{\overline{\rm MS}}^{(1)}$.  At the same
time, the relation between the $\overline{\rm MS}$ and the on-shell charge reads
$$
e = e_{\rm OS}\:\left(1+z_{\rm OS}^{(1)}+O(e^4_{\rm OS})\right), 
$$ 
where $e_{\rm OS}$ is defined by the Thompson limit of Compton scattering. 
The electromagnetic Ward--Takahashi identity implies that some of the 
diagrams cancel, such that $z_{\rm OS}^{(1)}$ at the one-loop level 
can be written in terms of self-energies only \cite{Sirlin}
\begin{equation}
z_{\rm OS}^{(1)} = - 
  \frac{1}{2} \Pi^{(1)}_{\gamma \gamma}{}'(0) 
      + \frac{\sin\theta_W}{\cos\theta_W} 
          \frac{\Pi^{(1)}_{\gamma Z}(0)}{M_Z^2} 
         - \frac{1}{\varepsilon} Z_{\overline{\rm MS}}^{(1)}\;.
\label{def_e}
\end{equation}
UV-finiteness of $z_{\rm OS}^{(1)}$ implies
\begin{equation}
e_0 = \mu^\ep e \left( 1 - \frac{1}{\ep}\;\frac{7}{2}\; \frac{e^2}{16\pi^2} \right)\;. 
\label{e0to0}
\end{equation}
This may be confirmed also by using a renormalization group analysis
of the SM keeping the Yang--Mills and the Higgs sector only. From
the relation
\begin{equation}
\frac{1}{e^2} = \frac{1}{g^2} + \frac{1}{g'^2} \;, 
\label{coupling}
\end{equation}
where $g'$ and $g$ are the $U(1)$ and the $SU(2)$ gauge coupling
constants, respectively, it is easy to deduce that
\begin{equation}
\beta_e = e^3 \left( \frac{\beta_g}{g^3} 
   + \frac{\beta_{g'}}{g'^3} \right)
   = - \frac{7}{2} \frac{e^3}{16\pi^2} + {\it O} (e^5)\;. 
\label{beta_e}
\end{equation}
The $\beta$-functions $\beta^{(1)}_{g,g'}$  may be calculated in the unbroken theory.  
They have been calculated in  \cite{RG_1loop}. 
We see that the above result is in agreement with (\ref{e0to0}) if we take
into account that 
$(\mu^2 d/d\mu^2)\: e  = - \frac{\ep}{2} e  + \beta_e \;.$

%%%%%%%%%%%%%%%%%%%%%%%%%%%%%%%%%%%%%%%%%%%%%%%%%%%%%%%%%%%%%%%%%%%%%%%%%%
\subsection{Mass renormalization}
We introduce the following notation for the mass renormalization
constants 
\begin{equation}
m_{V,0}^2 =  m_V^2(\mu)\:
\left( 1 + \frac{g^2(\mu)}{16\pi^2\ep} Z_V^{(1,1)} 
+ \frac{g^4(\mu)}{(16\pi^2)^2\ep} Z_V^{(2,1)} +
\frac{g^4(\mu)}{(16\pi^2)^2\ep^2} Z_V^{(2,2)}
\right), 
\label{baremsb}
\end{equation}
where $V$ stands for any of the bosons $Z$, $W$ or $H$.  In addition
to the masses we have one coupling constant as a free parameter of the
SM which we have chosen above to be $e=g\sin\theta_W$.  The one-loop
mass counter-terms are well known \cite{FJ}. For the purely bosonic
contributions we have
\begin{eqnarray}
Z_H^{(1,1)} & = & 
  - \frac{3}{2}
  - \frac{3}{4} \frac{m_Z^2}{m_W^2} 
  + \frac{3}{4} \frac{m_H^2}{m_W^2} , 
\label{Z_11_h}
\\
Z_W^{(1,1)} & = & 
  - \frac{3}{4} \frac{m_H^2}{m_W^2} 
  - 3 \frac{m_W^2}{m_H^2} 
  - \frac{3}{2} \frac{m_Z^4}{m_H^2 m_W^2}
  + \frac{3}{4} \frac{m_Z^2}{m_W^2} 
  - \frac{17}{3} ,
\\
Z_Z^{(1,1)} & = &  
  - \frac{3}{4} \frac{m_H^2}{m_W^2} 
  - 3 \frac{m_W^2}{m_H^2}
  - \frac{3}{2} \frac{m_Z^4}{m_W^2 m_H^2} 
  + \frac{11}{12} \frac{m_Z^2}{m_W^2}
  - 7 \frac{m_W^2}{m_Z^2} 
  +  \frac{7}{6}.
\label{one-loop}
\end{eqnarray}
All masses here are \MSb masses and depend on the renormalization
scale $\mu$: $m^2_V=m^2_V(\mu)$. It should be noted that unlike in the
case of the couplings the mass renormalization constants cannot be
calculated from the unbroken gauge theory. Here in any case the
calculation of the Feynman diagrams in the Standard Model is
required. 

Let us comment about the somewhat unusual looking dependence of our
\MSb renormalization constants on the particle masses. Since masses
are induced by the Higgs mechanism we have the mass coupling relations
$$m_W=g\: \frac{v}{2}\;,\;\;m_Z=\sqrt{g^2+g^{'2}}\:\frac{v}{2}\;,\;\;
m_H=\sqrt{2\lambda}\: v$$ where $v$ is the Higgs VEV and $\lambda$ the
Higgs self-coupling.  One peculiarity of the ``spontaneous symmetry
breaking'' and the related mixing of states, like the $\gamma-Z$
mixing, leads to the non-polynomial nature of the perturbation
expansion in the SM. Due to the mixing, the actual $Z$ coupling reads
$\sqrt{g^2+g^{'2}}=g/\cos \theta_W$ etc. In the dimensionless mass
ratios the factors $v^2$ drop out and we have in fact just ratios of
couplings. To a large extent this is a trivial consequence of
factorizing out powers of $g^2$ which cancels against such factors
which appear in the denominators of the $Z_V^{(1,1)}$'s.  However,
there are also true inverse powers of the Higgs self-coupling present.
They originate in the tadpoles which we need taking into account for
the sake of gauge invariance.

Our results for the two-loop mass
renormalization constants are as follows:
\begin{eqnarray}
Z_W^{(2,1)} & = & \frac{63}{64} \frac{m_H^4}{m_W^4}
- \frac{3}{4} \frac{m_H^2}{m_W^2}  
- \frac{3}{8} \frac{m_H^2 m_Z^2}{m_W^4} 
- \frac{301}{192} \frac{m_Z^4}{m_W^4} 
+ \frac{17}{12}\frac{m_Z^2}{m_W^2} 
- \frac{53}{3}
\nonumber \\ &&
+ \frac{31}{12} \frac{m_Z^4}{m_H^2 m_W^2}
- \frac{17}{2}  \frac{m_Z^2}{m_H^2} 
- \frac{176}{3} \frac{m_W^2}{m_H^2} 
+ \frac{59}{24} \frac{m_Z^6}{m_H^2 m_W^4} , 
\label{Z_21_w}
\\
%%%%%%%%%%%%%%%%%%%%%%%%%%%%%%%%%%%%%%%%%%%%%%%%%%%%%%%%%
Z_W^{(2,2)} & = & -\frac{9}{32} \frac{m_H^4}{m_W^4} + \frac{43}{8} \frac{m_H^2}{m_W^2}
+ \frac{65}{32} \frac{m_Z^4}{m_W^4} 
- \frac{35}{8} \frac{m_Z^2}{m_W^2}
+ \frac{334}{9} 
+ \frac{27}{4} \frac{m_Z^4}{m_H^2 m_W^2} 
\nonumber \\ && 
+ 6 \frac{m_Z^2}{m_H^2} 
+ 34 \frac{m_W^2}{m_H^2}
- \frac{5}{2} \frac{m_Z^6}{m_H^2 m_W^4}
+ 9 \frac{m_W^4}{m_H^4} 
+ 9 \frac{m_Z^4}{m_H^4} 
+ \frac{9}{4} \frac{m_Z^8}{m_H^4 m_W^4} , 
\end{eqnarray}

\begin{eqnarray}
Z_Z^{(2,1)} & = & \frac{63}{64} \frac{m_H^4}{m_W^4}
- \frac{3}{4} \frac{m_H^2}{m_W^2}  
- \frac{3}{8} \frac{m_H^2 m_Z^2}{m_W^4} 
+ \frac{11}{3}
+ \frac{7}{6} \frac{m_Z^2}{m_W^2} 
- \frac{64}{3} \frac{m_W^2}{m_Z^2}
- \frac{253}{192} \frac{m_Z^4}{m_W^4}
\nonumber \\ && 
- \frac{176}{3} \frac{m_W^2}{m_H^2}
- \frac{17}{2} \frac{m_Z^2}{m_H^2}
+ \frac{31}{12} \frac{m_Z^4}{m_H^2 m_W^2}
+ \frac{59}{24} \frac{m_Z^6}{m_H^2 m_W^4} , 
\\
%%%%%%%%%%%%%%%%%%%%%%%%%%%%%%%%%%%%%%%%%%%%%%%%%%%%%%%%%%%%%%%%%%%%%%%%%%
Z_Z^{(2,2)} & = & -\frac{9}{32} \frac{m_H^4}{m_W^4} 
+ \frac{1}{4} \frac{m_H^2}{m_W^2}
- \frac{1}{8} \frac{m_H^2 m_Z^2}{m_W^4} 
+ \frac{21}{4} \frac{m_H^2}{m_Z^2}
- \frac{55}{6} 
+ \frac{11}{12} \frac{m_Z^2}{m_W^2}
+ \frac{629}{288} \frac{m_Z^4}{m_W^4}
\nonumber \\ &&
+ \frac{245}{6} \frac{m_W^2}{m_Z^2}
+ \frac{27}{2} \frac{m_W^2}{m_H^2}
+ 16 \frac{m_Z^2}{m_H^2} 
+ 21 \frac{m_W^4}{m_Z^2 m_H^2}
- \frac{7}{2} \frac{m_Z^4}{m_H^2 m_W^2}
- \frac{11}{4} \frac{m_Z^6}{m_H^2 m_W^4}
\nonumber \\ &&
+ 9 \frac{m_W^4}{m_H^4}
+ 9 \frac{m_Z^4}{m_H^4}
+ \frac{9}{4} \frac{m_Z^8}{m_H^4 m_W^4} .
\label{Z_22_z}
\end{eqnarray}

Again the higher order pole terms $1/\ep^2$ can be checked by means of
the appropriate renormalization group equation. Let us write the
relation $m_{V,0}^2 = m_V^2 \left(1 + \sum_n Z_V^{(n)}/\ep^n \right)$
which connects bare and renormalized masses and introduces the
anomalous dimension of the mass 
$\gamma_V = (\mu^2\,d/d\mu^2)\ln m_V^2$.  
Taking into account that $(\mu^2 d/d\mu^2)\:m_{V,0}^2 = 0$
and repeating the calculations given in \cite{tHooft:RG}
we find 
\begin{eqnarray}
\gamma_V & = &  \frac{1}{2} g \frac{\partial}{\partial g } Z_V^{(1)},
\\
\left(
\gamma_V + \beta_g \frac{\partial}{\partial g } 
+ \sum_i \gamma_i m_i^2 \frac{\partial}{\partial m_i^2} \right) Z_V^{(n)}
& = &  \frac{1}{2} g \frac{\partial}{\partial g } Z_V^{(n+1)}.
\label{RG:SM}
\end{eqnarray}
For each function like $\gamma_V$ or $Z_V^{(n)}$ 
we may perform a loop expansion 
$\gamma_V = \sum_{k=1}^\infty (g^2/16\pi^2)^k \gamma_V^{(k)}$
or 
$Z_V^{(n)} = \sum_{k=n}^\infty (g^2/16\pi^2)^k Z_V^{(n,k)}$
and (\ref{RG:SM}) can be written for each loop
correction separately. In particular, for $n=1$, we have
\begin{eqnarray}
\gamma_V^{(1)} = Z_V^{(1,1)},
\nonumber \\
\gamma_V^{(2)} = 2 Z_V^{(2,1)},
\nonumber 
\end{eqnarray}
such that the coefficient of the $n=2$ poles can be checked via
\begin{equation}
\left( Z_V^{(1,1)} \right)^2 + 
\frac{16 \pi^2 }{g^2} 
2 \frac{\beta_g^{(1)} Z_V^{(1,1)}}{g}
+ \sum_i Z_{m_i}^{(1,1)}  m_i^2 \frac{\partial}{\partial m_i^2} 
          Z_V^{(1,1)} = 2  Z_V^{(2,2)}\;.
\end{equation}
The value of $\beta_g^{(1)} = - \frac{43}{12} \frac{g^3}{16\pi^2}$
may be calculated from the relation
\begin{equation}
\beta_g^{(1)} \sin\theta_W = 
   \frac{1}{2} g \frac{ \cos^2 \theta_W}{\sin \theta_W} 
    \left( \frac{g^2}{16 \pi^2 } \right)
    \left(\gamma_W^{(1)}-\gamma_Z^{(1)} \right) + \beta_e^{(1)}, 
\end{equation}
where $\beta_e^{(1)}$ is given in (\ref{beta_e}), 
and $\cos^2 \theta_W = m_W^2/m_Z^2$.

An independent verification of the $1/\ep$ terms
can be obtained from the relationship $e^2=g^2 \sin^2 \theta_W$, which is
valid for bare and $\overline{\rm MS}$ renormalized quantities.
Differentiating the renormalized quantities with respect to $\ln \mu^2 $
we find
\begin{equation}
e^4 \left( \frac{\beta_g}{g^3} + \frac{\beta_{g'}}{g'^3} \right) 
= g \beta_g \sin^2\theta_W
- \frac{1}{2} g^2 \left( \gamma_W - \gamma_Z\right) \cos^2\theta_W , 
\end{equation}
or for the two-loop case 
\begin{equation}
g^2 \frac{\beta_{g'}^{(2)}}{g'^3} \sin^4\theta_W
- \frac{\beta_g^{(2)}}{g} \sin^2\theta_W  \cos^2\theta_W
= - \left( \frac{g^2}{16 \pi^2} \right)^2
     \left( Z_W^{(2,1)} -  Z_Z^{(2,1)}\right) \cos^2\theta_W .
\end{equation}
The two-loop $\beta$-functions for $g$ and $g'$ are given in 
\cite{RG_2loop} and read
\begin{eqnarray}
\beta_{g'} & = & \frac{1}{12} \frac{g'^3}{16\pi^2}
    + \frac{1}{4} \frac{g'^5}{(16\pi^2)^2}
    + \frac{3}{4} \frac{g'^3 g^2}{(16\pi^2)^2}\;\:, 
\nonumber \\
\beta_g & = & - \frac{43}{12} \frac{g^3}{16\pi^2}
   - \frac{259}{12} \frac{g^5}{(16\pi^2)^2}
   + \frac{1}{4} \frac{g^3 g'^2}{(16\pi^2)^2}\;\:. 
\label{beta}
\end{eqnarray}

The fact that after UV renormalization we get a finite result confirms
the infrared finiteness of the bosonic contribution to the pole mass.

Let us write now the RG equation for the effective Fermi constant
$G_F$. $G_F$ is usually defined as a low energy constant in
one-to-one correspondence with the muon lifetime. However, if we
consider physics at higher energies a parameterization in terms of low
energy constants may lead to large radiative corrections. Much in the
same way as the fine structure constant $\alpha$ often is replaced by
the effective running fine structure constant $\alpha(\mu)$ we expect
that $G_F$ should be replaced by an effective version of it at higher
energies. Unlike in the case of $\alpha$, however, because of the
smallness of the light fermion Yukawa couplings, $G_F$ starts to run
effectively only at scales beyond the $W$--pair production threshold
(see below).
The bare relations are
\begin{eqnarray}
m_{W,0}^2 & =  & \frac{g_0^2 v_0^2}{4}, 
\\
\frac{1}{v^2_0} & = & \sqrt{2}\: G_{F,0}.
\label{GF}
\end{eqnarray}
Both these relations are valid also for  $\overline{\rm MS}$ renormalized 
parameters, so that their differentiation with respect to $\ln \mu^2$ 
gives rise to the relation
\begin{eqnarray}
\gamma_W = 2 \frac{\beta_g}{g} - \gamma_{G_F}, 
\end{eqnarray}
where we introduced the anomalous dimension of the Fermi constant
$\gamma_{G_F} = (\mu^2\,d/d\mu^2)\ln G_F$. Its loop expansion looks like 
$\gamma_{G_F} = \gamma_{G_F}^{(1)} + \gamma_{G_F}^{(2)} + \cdots$. 
At the one-loop level we have
\begin{eqnarray}
\gamma_{G_F}^{(1)} & = & 2 \frac{\beta_g^{(1)}}{g} - Z_W^{(1,1)}
\nonumber \\
& = & 
 \frac{g^2}{16\pi^2} \frac{1}{4 m_W^2}\:\Biggl \{
 \frac{2}{m_H^2} \:\left( 3\:(2
m_W^4+m_Z^4)+m_H^4-4\sum_fm_f^4\right) \nonumber  \\
&-& \left( 3\:(2m_W^2+m_Z^2)-m_H^2-2\sum_f m_f^2\right)  \Biggr\}
\label{fermi1}
\end{eqnarray}
and we used  the results of Ref.~\cite{FJ} for the fermion contributions. 
The first term proportional $1/m^2_H$ is the contribution from the
tadpoles. The appearance of the tadpole terms is somewhat mysterious,
since we know that in renormalized observables tadpoles drop out. 
Here they seem to contribute to the renormalization group evolution 
of the Fermi constant. In any case the tadpoles are present in the 
relationship between the bare and the renormalized parameters.
At the two-loop level, our results allows us to write the bosonic
corrections only. They are given by
\begin{eqnarray}
\gamma_{G_F}^{(2),{\rm boson}} & = & 2 \frac{\beta_g^{(2)}}{g} - 2 Z_W^{(2,1)}
\nonumber \\ 
& = & -\frac{2 g^4}{(16 \pi^2)^2} \Biggl\{ 
\frac{63}{64} \frac{m_H^4}{m_W^4}
- \frac{3}{4} \frac{m_H^2}{m_W^2}  
- \frac{3}{8} \frac{m_H^2 m_Z^2}{m_W^4} 
- \frac{301}{192} \frac{m_Z^4}{m_W^4} 
+ \frac{7}{6}\frac{m_Z^2}{m_W^2} 
+ \frac{25}{6}
\nonumber \\ && \hspace{1.5cm}
+ \frac{31}{12} \frac{m_Z^4}{m_H^2 m_W^2}
- \frac{17}{2}  \frac{m_Z^2}{m_H^2} 
- \frac{176}{3} \frac{m_W^2}{m_H^2} 
+ \frac{59}{24} \frac{m_Z^6}{m_H^2 m_W^4} 
\Biggr \}  \;\;.
\label{fermi2}
\end{eqnarray}
The equations~(\ref{fermi1}) and (\ref{fermi2}) are written in
$\overline{\rm MS}$ scheme.  As usual in this scheme, in solving the
renormalization group equation
$$
\mu^2\, \frac{d}{d\mu^2} \: G_F(\mu) = G_F(\mu) \:\gamma_{G_F} 
$$
the decoupling of the heavy particles has to be performed ``by hand''.
This means that for low values of the energy scale $\mu$, when $\mu~<~
m_H,m_W,m_Z$, the bosonic terms on the r.h.s. are equal to zero while
the light fermion contributions proportional to $G_F m_f^2$ are tiny.
Consequently, below the $W$ mass, the effective Fermi constant does
practically not change with scale. Obviously, the running of $G_F$
only starts at about $\mu
\sim m_Z$, when the scale of a process exceeds the masses of the
bosons. Also the top quark will contribute once we have passed its
threshold.
%%%%%%%%%%%%%%%%%%%%%%%%%%%%%%%%%%%%%%%%%%%%%%%%%%%%%%%%%%%
\section{Results, discussion and conclusion}
\setcounter{equation}{0}

After UV renormalization the pole mass (see (\ref{polemass}))
\begin{equation}
M_V^2  = m_V^2 + \hat{\Pi}_V^{(1)} 
+ \hat{\Pi}_V^{(2)} 
+ \hat{\Pi}_V^{(1)} \hat{\Pi}_V^{(1)}{}'
%+ \widehat{\Pi_V^{(1)} \Pi_V^{(1)}{}'}
\label{MMtomm}
\end{equation}
is represented in terms of finite \MSb renormalized amplitudes
$$
\hat{\Pi}_V^{(i)}
= \Pi_V^{(i)}(p^2,m_V^2,\cdots)\Bigr|_{p^2=m_V^2,\:{\rm F.P.}}\;\;.
$$  
F.P. denotes the \MSb finite part prescription.
The calculation of the one-loop \MSb renormalized amplitude is well known.
We get it by rewriting the bare expression in terms of \MSb parameters according to
(\ref{baremsb})  and (\ref{X0:def})
\begin{eqnarray}
\hat{\Pi}_V^{(1)} & \equiv & 
\lim_{\ep \to 0} \left( m_{0,V}^2 -m^2_V +  m_{0,V}^2 \frac{g_0^2}{16\pi^2} X_{0,V}^{(1)}\right)  
\nonumber \\
& = & 
m_V^2(\mu) \frac{e^2}{16\pi^2 \sin^2 \theta_W} \lim_{\ep \to 0} 
 \left( \frac{1}{\ep} Z_V^{(1,1)} + X_{0,V}^{(1)}\right)  
=  m_V^2(\mu) \frac{e^2}{16\pi^2 \sin^2 \theta_W} X_{V}^{(1)} \;.
\end{eqnarray}
We restrict ourselves to a consideration of the bosonic part
$X_{0,V}^{(1),\mbox{boson}}$.  The renormalization of the
off-shell self-energy functions $\Pi_{0,V}^{(1)}{}'$ and
$\Pi_{0,V}^{(2)}$ is more complicated.  Besides the renormalization of
the physical parameters, it would require us to perform
order-by-order, the wave-function renormalization as well as the
renormalization of the ghost sector, in particular of the unphysical
gauge parameters. However, the \MSb renormalization of the combination
$
\Pi_{0,V}^{(2)}+\Pi_{0,V}^{(1)} \Pi_{0,V}^{(1)}{}'
$ it much simpler. The subtraction of sub-divergencies in this case is
reduced to the one-loop renormalization of the charge and the physical masses
only, while the wave-function renormalization or the renormalization
of the unphysical sector is not needed. Accordingly, as a genuine two-loop
counter-term, only the mass renormalization occurs. 
 
The full two-loop \MSb renormalized amplitude can be
written in the form
\begin{eqnarray}
&& 
\hat{\Pi}_V^{(2)} + \hat{\Pi}_V^{(1)} \hat{\Pi}_V^{(1)}{}'
\equiv 
\Biggl \{ \Pi_{0,V}^{(2)} + \Pi_{0,V}^{(1)} \Pi_{0,V}^{(1)}{}' \Biggr\}_{\it MS}
=  \lim_{\ep \to 0}  \Biggl( 
\Pi_{0,V}^{(2)} + \Pi_{0,V}^{(1)} \Pi_{0,V}^{(1)}{}'
\nonumber \\ && 
+ m_V^2(\mu) \frac{1}{\ep} \left( \frac{e^2}{16\pi^2 \sin^2 \theta_W} \right)^2
\Biggl[Z_V^{(1,1)}
+ \Biggl[ \frac{\Delta g^2}{g^2} \Biggr]
+ \sum_j Z_{m^2_j}^{(1,1)} \frac{\partial}{\partial m_j^2} \Biggr] X_{0,V}^{(1)}
\nonumber \\  &&
+ m_V^2(\mu) \left( \frac{e^2}{16\pi^2 \sin^2 \theta_W} \right)^2
\Biggl[ \frac{1}{\ep} Z_V^{(2,1)}  + \frac{1}{\ep^2} Z_V^{(2,2)}
\Biggr]
\Biggr)
\label{MS2:subtracted}
\end{eqnarray}
where the sum runs over all species of particles $j=Z,\,W,\,H$
and 
$$
\Biggl[ \frac{\Delta g^2}{g^2} \Biggr] = 
\frac{\cos^2 \theta_W}{\sin^2 \theta_W} \left(Z_W^{(1,1)} - Z_Z^{(1,1)} 
\right) - 7 \sin^2 \theta_W \;.  $$ 
The functions $Z_V^{(i,j)}, X_{V}^{(1)}$ and $ X_{0,V}^{(1)}$ are
defined in (\ref{Z_11_h})-(\ref{one-loop}),
(\ref{Z_21_w})-(\ref{Z_22_z}), (\ref{X:def}) and (\ref{X0:def}),
respectively.  Again, throughout, we take into account only the
bosonic corrections. The r.h.s. of (\ref{MS2:subtracted}) is given
by the bare two-loop contributions, the two-loop contribution obtained
by expanding the bare parameters in the one-loop amplitude
(subtraction of the sub-divergences) and the genuine two-loop
subtractions. The second type of contribution follows from
writing $\Pi^{(1)}=m_0^2 \left(\frac{g_0^2}{16 \pi^2}\right)\:
X^{(1)}_0$ and by utilizing (\ref{baremsb}).  We would like to stress,
that each of the one-loop bare amplitudes $\Pi_{0,V}^{(1)}$ and
$\Pi_{0,V}^{(1)}{}'$ has to be expanded up to linear terms in
$\ep$. Together with the singular $1/\ep$ terms the contributions
linear in $\ep$ yield additional finite contributions in the limit
$\ep \to 0$. In the product $\Pi_{0,V}^{(1)}
\Pi_{0,V}^{(1)}{}'$ the parameters may be identified by the \MSb
renormalized quantities, but all functions, $A_0$ and $B_0$ have to be
taken in $d=4-2\ep$ dimensions and must be expanded up to terms
linear in $\ep$ (see Appendix~A).

As mentioned 
earlier, for the purely bosonic contributions alone the imaginary part
of $\Pi(p^2)$ on the mass-shell is zero at the two-loop level. This is
due to the fact that in the bosonic sector we have the physical masses
$m_\gamma=0$, $M_Z$, $M_W$ and $M_H$ and by inspection of the possible
two and three particle intermediate states one observes that all
physical thresholds lie above the mass shells of the $W$ and $Z$
bosons, i.e., the self-energies of the massive gauge bosons develop an
imaginary part only at $p^2 > M_V^2$ (to two loops in the SM). On
kinematical grounds imaginary parts could show up from the Higgs or
Faddeev-Popov ghosts, which have square masses $\xi_V M_V^2$, for
small values of the gauge parameter. However, as we have verified, the
two-loop on-shell self-energies are gauge independent. This implies
that ghost contributions have to cancel and hence cannot contribute to
the imaginary part. Thus $s_P=M_V^2$ in our case. In higher orders for
the $Z$--propagator one
gets an imaginary part as soon as $p^2 >0$, from diagrams like\\[1cm]

\centerline{
\begin{picture}(120,60)(60,0)
{\SetScale{2.0}
{%\SetColor{Red} 
\Photon(00,30)(15,30){1}{5}
\Photon(28,37.5)(43,37.5){1}{4}
\Photon(28,22.5)(43,22.5){1}{4}
\Photon(56,30)(71,30){1}{5}
\Photon(15,30)(28,37.5){1}{5}
\Photon(28,37.5)(28,22.5){1}{5}
\Photon(28,22.5)(15,30){1}{5}
\Photon(56,30)(43,37.5){1}{5}
\Photon(43,37.5)(43,22.5){1}{5}
\Photon(43,22.5)(56,30){1}{5}
} % End RedColor
\Vertex(15,30){1}
\Vertex(28,37.5){1}
\Vertex(28,22.5){1}
\Vertex(56,30){1}
\Vertex(43,37.5){1}
\Vertex(43,22.5){1}
} % End SetScale
\Text(14,52)[]{$Z$}
\Text(129,52)[]{$Z$}
\Text(49,60)[]{$W$}
\Text(96,60)[]{$W$}
\Text(73,38)[]{$\gamma$}
\Text(73,66)[]{$\vdots$}
\Text(73,85)[]{$\gamma$}
\Text(149,58)[]{$.$}
\end{picture}
}
\noindent
For the $W$--propagator an imaginary part is only possible for $p^2 >
M_W^2$, because charge conservation requires at least one $W$ in any
physical intermediate state.

For the massive gauge bosons $Z$ and $W$ we write
\begin{equation}
\frac{M_V^2}{m_V^2} =  1 
  + \left( \frac{e^2}{16\pi^2\sin^2 \theta_W} \right)\: X^{(1)}_V 
  + \left( \frac{e^2}{16\pi^2\sin^2\theta_W} \right)^2 \: X^{(2)}_V \,,
%\frac{m_H^4}{m_V^4} \sum\limits_{k=0}^6 A_k^V 
%                   \left( \sin^2 \theta_W \right)^k \,,
\label{result}
\end{equation}
where both $e$ and $\sin\theta_W$ are to be taken in the
$\overline{\rm MS}$ scheme.

The one-loop coefficients $X^{(1)}_V $ for $Z,\,W$ and $H$ are known of
course as exact results. We write them down for completeness in
Appendix B. For the coefficients $X^{(2)}_V$ we make an expansion
and perform them as double series: in $\sin^2\theta_W$ and in the
mass ratio $m_V^2/m_H^2$. We have calculated the first six terms
of the expansion with respect to $\sin^2\theta_W$ and the first six terms 
with respect to mass ratio $m_V^2/m_H^2$. 
The analytical values of these coefficients are presented in Appendix D.  
These represent our main result.

Sometimes in massive multi-loop calculations the so-called modified
$\overline{\rm MS}$ scheme ($\overline{\rm MMS}$) is used
\cite{modify:MS}. The difference between $\overline{\rm MS}$ and
$\overline{\rm MMS}$ is that in the former scheme each loop is multiplied
by $(e^{\gamma}/4\pi)^\varepsilon$ while in the latter the
normalization factor is $1/(4\pi)^\ep/\Gamma(1+\varepsilon)$, which
yields a difference at the two-loop level. It has been shown that in
QCD both schemes reproduce the same formula for the mass relation
analogous to (\ref{MMtomm})~\cite{mass-ratio}.  We have checked that the
same holds true for the pole masses in the Standard Model.

Very often the inverse of (\ref{MMtomm}) is required.  To that
end we have to express all $\overline{\rm MS}$ parameters in terms of
on-shell ones. Thus
%NEW: corrected 10.02.02
\begin{eqnarray}
\label{inverse}
m^2_V &=& M_V^2 
- \hat{\Pi}_V^{(1)} 
- \Biggl \{ \Pi_{0,V}^{(2)} + \Pi_{0,V}^{(1)} \Pi_{0,V}^{(1)}{}' \Biggr\}_{\it MS} 
% - \hat{\Pi}_V^{(2)} - \hat{\Pi}_V^{(1)} \hat{\Pi}_V^{(1)}{}'
\nonumber\\
&& 
- \sum\limits_j (\Delta m^2_j)^{(1)} \frac{\partial}{\partial m_j^2} \hat{\Pi}_V^{(1)}
       - (\Delta e)^{(1)} \frac{\partial}{\partial e} \hat{\Pi}_V^{(1)}
      \Biggr|_{m_j^2=M_j^2,\, e=e_{\rm OS}},
\label{reverse}
\end{eqnarray}
where the sum runs over all species of particles $j=Z,\,W,\,H$
and 
$$
(\Delta m^2_j)^{(1)}=
-{\rm Re} \hat{\Pi}^{(1)}_j
\Biggr|_{m_j^2=M_j^2,\, e=e_{\rm OS}}
\equiv 
- M_V^2 \frac{e^2_{\rm OS}}{16 \pi^2 \sin^2 \theta_W} X_V^{(1)} \Biggr|_{m_j^2=M_j^2}
$$ stands for the
self-energy of the $j$th particle at $p^2=m_j^2$ in the \MSb scheme
and parameters replaced by the on-shell ones. 
Note that in the above relation we had to perform 
a change from the $\overline{\rm MS}$ to the on-shell scheme also for
the electric charge
\begin{equation}
e(\mu^2) = e_{\rm OS} \left[ 1 + \frac{e_{\rm OS}^2}{16\pi^2} \left(
\frac{7}{2}
\ln \left( \frac{M_W^2}{\mu^2} \right)-\frac{1}{3}  \right) \right]
\end{equation}
with $e_{\rm OS}^2/4\pi  = \alpha \sim 1/137$. Accordingly, since 
$\hat{\Pi}^{(1)}$ depends on $e$ by an overall factor $e^2$ only,  
$$
(\Delta e)^{(1)} \frac{\partial}{\partial e}\hat{\Pi}_V^{(1)} 
= \frac{e^2}{16\pi^2} \left[ 7 \ln \left( \frac{M_W^2}{\mu^2} \right)
- \frac{2}{3} \right]\:\hat{\Pi}_V^{(1)}\;.
$$

By identifying $m^2_V=m_{V,0}^2=M_V^2+\delta M_V^2$  (\ref{inverse})
is the relationship appropriate to obtain the on-shell gauge-boson mass
counter-terms $\delta M_V^2$ :
\begin{eqnarray}
\delta M_V^2 &=& - {\rm Re} \Biggl[ 
 \check{\Pi}^{(1)}_{V,0} + \check{\Pi}_{V,0}^{(2)} 
+\check{\Pi}^{(1)}_{V,0} \check{\Pi}_{V,0}^{(1)}{}'
\nonumber\\
&& ~~~~~~~~
+ \sum\limits_j (\delta M^2_j)^{(1)} \frac{\partial}{\partial m_{j,0}^2}
\check{\Pi}_{V,0}^{(1)}
       + (\delta e)^{(1)} \frac{\partial}{\partial e_0}\check{\Pi}_{V,0}^{(1)}
      \Biggr]\:
      \Biggr|_{m_{j,0}^2=M_j^2,\, e_0=e_{\rm OS}} 
\nonumber\\
&=& \Biggl(Z_{\msb}\cdot Z_{\rm OS}-1 \Biggr)\:M_V^2
\label{mct}
\end{eqnarray}
in terms of the original bare on-shell amplitudes
$$\check{\Pi}_{V,0}^{(i)}=\Pi_{V,0}^{(i)}(p^2,m_{V,0}^2,\cdots)
\Bigr|_{p^2=M_V^2,m_{j,0}^2=M_j^2,\, e_0=e_{\rm OS}}$$
and the bare on-shell counter-terms $\delta M^2_j$ and $\delta e$.
The second equality gives $\delta M_V^2$ in terms of the singular
factor $Z_{\msb}=m_{V,0}^2/m_V^2(\mu)$ given in (\ref{baremsb}) and
the finite factor $Z_{\rm OS}=m^2_V/ M_V^2$ given in (\ref{inverse}).
These will be needed in two-loop calculations of observables in the
on-shell scheme.

We now turn to a discussion of our results which we obtained for the
relationship (\ref{reverse}).  For our numerical analysis we used the
following values of the pole masses: $M_W = 80.419$ GeV and
$M_Z=91.188$ GeV and $\alpha=1/137.036.$ We first investigate
numerically the relationship between the $\overline{\rm MS}$--mass
$m_V(\mu)$ and the pole-mass $M_V$.  In
Figs.~\ref{zlight}~and~\ref{zheavy} we plot $\Delta_Z \equiv
m^2_Z(M_Z)/M_Z^2-1$ as function of the Higgs mass $M_H$ for
intermediate and heavier Higgs masses, respectively. For the one-loop
corrections the exact analytical functions are evaluated while for the
two-loop results we utilize all coefficients of our
expansion. Analogously, in Figs.~\ref{wlight} and~\ref{wheavy} the
Higgs mass dependence of $\Delta_W \equiv m^2_W(M_W)/M_W^2-1$ is
depicted for the same ranges of the Higgs mass. As we can see, for a
``light'' Higgs of mass less than about $200$ GeV the two-loop
corrections are small as compared to the one-loop ones. However, at a
Higgs mass of about $220$ GeV the absolute value of the two-loop
correction is of the same size as the one-loop result, such that the
two-loop corrections start to play an essential role.

%NEW: corrected by 02.05.02.
Our analysis shows (see  Figs.~\ref{zheavy} and ~\ref{wheavy}) that
the perturbative expansion looses its meaning for large Higgs masses
(strong coupling regime).
In the relationship between the \MSb-- and the pole--masses corrections
exceed the 50\% level around 880 GeV (for $\mu=M_V$). The size of the
corrections depend on the choice of the renormalization scale $\mu$. 

Since our results have been obtained by an expansion in $m_V^2/m_H^2$
(i.e., for a heavy Higgs), one of the main questions which remains to
be considered is the validity of our results as we approach lighter
Higgs masses. Note that we are dealing with an asymptotic expansion
(which means the radius of convergence is zero) and thus we can answer
the question about where the expansion is reliable only empirically;
at least as long as an exact result is not available for a direct
comparison. Since the structure of the one- and two-loop corrections
for both massive gauge bosons is very similar, we are going to
investigate in the following the problems of convergence of our
results for the $Z$--boson only. For the $W$--boson the convergence is
better because of the smaller value of the $W$--mass. Obviously our
series--expansion breaks down for a ``light'' Higgs, when 
$ m_H  {\tiny \gto} m_Z$. Unexpectedly, we find that the two-loop
corrections remain numerically small for values of the Higgs mass even
down to 100 GeV (below about 0.2\%). However, we observe a steep raise
of the correction which signals that the expansion becomes unreliable
below about 130 GeV\footnote{We have checked the convergence of the
corresponding expansion at the one-loop level, where the exact result
is available. Deviations start to show up below about 145 GeV and grow
to about 30\% at 100 GeV. The best approximation in this case is based on the 
first 3 or 4 coefficients. In this case, beyond the first few terms, the series
starts to diverge in a way typical for an asymptotic expansion}.  We also should keep in mind that, on the
level of the size of the two-loop corrections, which are very small in
the region around 150 GeV for $\mu \sim m_Z$, the latter depend
substantially on the choice of the \MSb renormalization--scale $\mu$,
which causes an essentially constant shift in the quantities shown in the
Figures which follow.

In Fig.~\ref{z-higgs-light} we show the dependences of the two-loop
corrections to $\Delta_Z=m^2_Z(M_Z^2)/M_Z^2-1$ as a function of the
Higgs mass and the number of coefficients of the expansion used for
their evaluation. For a ``light'' Higgs the difference between the
result, including six coefficients of the expansion, and
the results obtained by including only the first three of them
(leading, next-to-leading and next-to-next-to-leading) are numerically
small.  However, the convergence of the series is not
satisfactory. The coefficients grow fast beyond the first four terms
and the correction starts growing fast.  For a heavy Higgs with a mass
of more than $300$ GeV the convergence is much better, so that we omit
the corresponding plot. We only mention, that there is no essential
differences between the full result and the next-to-leading one.
Similarly, the dependence of the two-loop corrections on the number of
coefficients of the expansion with respect to $\sin^2\theta_W$ for a
``light'' Higgs is illustrated in Fig.~\ref{z-sin-light}.  Here unlike
in the previous ``light'' case, the first coefficient is relatively large
already, while the higher terms of the expansion do not alter the
result in a significant manner. Again, we observe that results become
unreliable below about 130 GeV.

Finally we analyze the Higgs mass dependence of $\sin^2\theta_W$. The
relation between the $\overline{\rm MS}$ weak mixing parameter and its
version in terms of the pole masses reads
\begin{eqnarray}
\label{sinus}
\sin^2 \theta_W 
  & = & 1 - \frac{m_W^2}{m_Z^2}
   =  1 - \frac{M_W^2}{M_Z^2} \left(
   \frac{1+\delta^{(1)}_W+\delta^{(2)}_W }{1+\delta^{(1)}_Z+\delta^{(2)}_Z} 
                  \right)
\nonumber \\ & = & 
\left(1 - \frac{M_W^2}{M_Z^2} \right)
-\frac{M_W^2}{M_Z^2}
\left[ 
    ( \delta^{(1)}_W -\delta^{(1)}_Z )
    (  1 - \delta^{(1)}_Z )
    + \delta^{(2)}_W - \delta^{(2)}_Z 
   \right]
\end{eqnarray}
where we adopted the notation, $m_V^2/M_V^2 = 1 + \delta^{(1)}_V +
\delta^{(2)}_V$.  It turns out that in (\ref{sinus}) all $M_H^4$ terms
cancel. The corresponding results are depicted in
Figs.~\ref{sin-light} and~\ref{sin-heavy}. Again for a ``light'' Higgs
the two-loop corrections are small, while for a heavier Higgs particle
the corrections become large. Again, below about 120 GeV  we cannot trust the
series--expansion any longer.

We would like to mention that the presence of $m_H^4$ corrections in
the relation (\ref{result}) does not contradict Veltman's screening
theorem \cite{Veltman:screening} which states, that the L-loop Higgs
dependence of a physical observable is at most of the form
$(m_H^2)^{L-1} \ln^L m_H^2 $ for large Higgs masses.  This theorem
applies to physical observables like cross sections and asymmetries,
whereas our formula is nothing but a relation between parameters of
two different schemes.

%NEW: corrected by 10.02.02
In conclusion, the main results of our paper are the following: (i) we
have presented an independent proof of gauge invariance and infrared
stability of the 2-loop electroweak bosonic corrections to the pole of
the gauge boson propagators; (ii) analytical results for a number of
coefficients of the expansion in $\sin^2\theta_W = 1 - m_W^2/m_Z^2$
and $m_V^2/m_H^2$ for the 2-loop electroweak bosonic corrections are
given for the on-shell mass counter-terms (\ref{mct}) and the
relationship between $\overline{\rm MS}$ and on-shell masses of the
gauge bosons $W$ and $Z$. 
%(iii) our calculation allowed us to write
%the RG equation for the effective Fermi constant (\ref{GF}), which
%could play an interesting role in the analysis of TeV energy collider
%data. 
All calculations have been performed in the electroweak Standard
Model.\\ 

\noindent
Note added: After completion of our paper we received the
preprint~\cite{FHWW02}, which presents a complete SM calculation of the
Higgs mass dependent terms to the observable $\Delta r$ which
determines the $M_W-M_Z$ relationship given $\alpha$, $G_\mu$ and
$M_Z$ as input parameters.

\vspace{1.0cm}
%%%%%%%%%%%%%%%%%%%%%%%%%%%%%%%%%%%%%%%%%%%%%%%%%%%%%%%%%%%%%%%%%%%%%%%%%%%%%%
\noindent
{\bf Acknowledgments.}  We are grateful to D.~Bardin, A.~Davydychev,
J.~Fleischer, O.V.~Tarasov and G.~Weiglein for useful discussions.  We
thank A.~Freitas for pointing out some misprints in the original
version of the preprint and to the referee for valuable comments.
We especially want to thank M.~Tentukov for his help in working
with DIANA. We also thank C.~Ford for carefully reading the
manuscript.  M.~K.'s research was supported in part by INTAS-CERN
grant No.~99-0377.

%%%%%%%%%%%%%%%%%%%%%%%%%%%%%%%%%%%%%%%%%%%%%%%%%%%%%%%%%%%%%%%%%%%%%%%%%%%%%%
\appendix
%%%%%%%%%%%%%%%%%%%%%%%%%%%%%%%%%%%%%%%%%%%%%%%%%%%%%%%%%%%%%%%%%%%%%%%%%%%%%%
\section{The one-loop master integral and its $\ep$-expansion}
\setcounter{equation}{0}

For the two-loop calculation we have to take into account the 
part proportional to $\ep$ of the one-loop propagator type integral%
\footnote{In \cite{one-loop-propagator} it is
          denoted as $J^{(2)}(4-2\varepsilon;1,1)$.
}
\begin{equation}
J = \int
\frac{\mbox{d}^d q}{\bigl( q^2-m_1^2 + \mbox{i} 0 \bigr)
           \bigl( (k-q)^2-m_2^2 + \mbox{i} 0 \bigr)} \; ,
\end{equation}
where $d=4-2\varepsilon$ and  ``+i0'' is the causal 
prescription for the propagator.
Its finite part has been presented in \cite{finite}, 
the $O(\ep)$-part in \cite{one-loop-propagator}, 
the terms up to order $\ep^3$ can be extracted 
by means of Eq.~(A.3) of~\cite{oleg-onshell}
and an all order $\ep$-expansion was obtained in \cite{D-ep}. 
We write the part of $J$ linear in $\ep$ in a form 
suitable for the implementation  in ${\bf FORM}$  
\footnote{The higher order $\ep$ terms can be extracted from 
\cite{D-ep,1loop-analytic}.} 

\begin{eqnarray}
\label{2pt_res2}
J & = & \mbox{i}\pi^{2-\varepsilon}
  \frac{\Gamma(1+\varepsilon)}{2(1-2\varepsilon)}
  \Biggl\{ \frac{m_1^{-2\varepsilon} \!+\! m_2^{-2\varepsilon}}{\varepsilon} 
   + \frac{m_1^2\!-\!m_2^2}{\varepsilon \; k^2}
   \left( m_1^{-2\varepsilon} \!-\! m_2^{-2\varepsilon} \right)
\nonumber \\ && 
- 2 \frac{ \sqrt{-\lambda(m_1^2,m_2^2,k^2)}}{k^2}
   \Biggl[ \arccos \left( \frac{m_1^2+m_2^2-k^2}{2 m_1 m_2} \right)
   \left( 1 - \ep \ln \left( \frac{-\lambda(m_1^2,m_2^2,k^2)}{k^2}\right)
    \right)
\nonumber \\ && 
   + 2 \ep \Bigl( \Cl{2}{ \tau_1}  - \Cl{2}{ \pi- \tau_1}
              + \Cl{2}{ \tau_2}  - \Cl{2}{ \pi- \tau_2} \Bigr)
  \Biggr ] + {\cal O}(\varepsilon^2) 
  \Biggr\} ,
\end{eqnarray}
where 
$\lambda(m_1^2,m_2^2,k^2) =  \left( m_1^4 + m_2^4 + k^4 
- 2 m_1^2 k^2 - 2 m_2^2 k^2 - 2 m_1^2 m_2^2  \right)$
and the  angles $\tau_i$ are defined (see \cite{D-ep}) via
\begin{equation}
\cos\tau_1 = \frac{k^2-m_1^2+m_2^2}{2m_2 \sqrt{k^2}} \; , \quad
\cos\tau_2 = \frac{k^2+m_1^2-m_2^2}{2m_1 \sqrt{k^2}} \; .
\end{equation}
$\Cl{2}{\theta}$ is the Clausen function
$\Cl{2}{\theta} =  \frac{1}{2 {\rm i}} 
\left[\Li{2}{e^{{\rm i}\theta}} - \Li{2}{e^{-{\rm i}\theta}}\right]$.
The expansion (\ref{2pt_res2}) is directly applicable in the region
where $\lambda \leq 0$, i.e.\ when 
$(m_1-m_2)^2 \leq k^2 \leq (m_1+m_2)^2$. 
For the region $\lambda > 0$ we need the proper
analytic continuation which has been given in
\cite{1loop-analytic}. Let us briefly describe it here. 
First of all, we rewrite (\ref{2pt_res2}) 
in the form (see section (2.2) of \cite{1loop-analytic} for details)
\begin{eqnarray}
&& J = \mbox{i}\pi^{2-\varepsilon}\;
  \frac{\Gamma(1+\ep)}{2(1-2\ep)} \;
  \Biggl( \frac{m_1^{-2\ep} \!+\! m_2^{-2\ep}}{\ep}
  + \frac{m_1^2\!-\!m_2^2}{\ep k^2}
  \left( m_1^{-2\varepsilon} \!-\! m_2^{-2\varepsilon} \right)
\nonumber \\ 
&& -  {\rm i} 
\frac{\left[-\lambda(m_1^2,m_2^2,k^2) \right]^{1/2-\varepsilon}}{(k^2)^{1-\varepsilon}}
\Biggl\{
\frac{2 \Gamma^2(1-\ep)}{\ep \Gamma(1-2\ep)} \; (1-\rho_1-\rho_2)
+ \frac{1}{\ep} \sum_{i=1}^2 
   \Biggl ( \rho_i (-z_i)^{-\ep} - (1-\rho_i)(-z_i)^{\ep} \Biggr)
\nonumber \\ 
&& + 2  \ep \sum_{i=1}^2 \Biggl(
   \rho_i (-z_i)^{-\ep}
   \Li{2}{z_i} - (1-\rho_i)(-z_i)^{\ep} \Li{2}{1/z_i}
    + {\cal O}(\varepsilon) 
    \Biggr) \Biggr\} \Biggr) , 
\label{general}
\end{eqnarray}
where 
$\rho_1$ and $\rho_2$ are some numbers, which we will define later
and 
\begin{equation}
z_1 =
\frac{\left[\sqrt{\lambda(m_1^2,m_2^2,k^2)}+m_1^2-m_2^2-k^2\right]^2}
     {4m_2^2 k^2}, \quad
z_2 =
\frac{\left[\sqrt{\lambda(m_1^2,m_2^2,k^2)}-m_1^2+m_2^2-k^2\right]^2}
     {4m_1^2 k^2} .
\end{equation}
Firstly, we note that the causal prescription amounts to 
the following rule for $\lambda$ ($\lambda > 0$)
\begin{eqnarray}
\ln(-\lambda(m_1^2,m_2^2,k^2)) & = & 
\ln(\lambda(m_1^2,m_2^2,k^2)) - {\rm i} \pi ,
\nonumber \\
\sqrt{-\lambda(m_1^2,m_2^2,k^2)} & = & 
- {\rm i} \sqrt{ \lambda(m_1^2,m_2^2,k^2) } .
\nonumber 
\end{eqnarray}
The function $\Li{2}{z}$ is real for real $z$ and $|z| \leq 1$. 
For real $z$ and $|z| > 1$
we change argument $z \to 1/z$ using the relation 
\footnote{For the higher order poly-logarithm the relation is \cite{Lewin}
$$
\Li{n}{z} + (-1)^n \Li{n}{\frac{1}{z}} = - \frac{1}{n!} \ln^n (-z) 
- \sum_{j=1}^{[n/2]} \frac{\ln^{n-2r} (-z) }{(n-2r)!} \zeta_{2r}.
$$
}
$$
\Li{2}{z} + \Li{2}{\frac{1}{z}} = - \frac{1}{2} \ln^2 (-z) - \zeta_2,
$$
by which an imaginary part shows up.
This change of variables can be done from the very beginning in (\ref{general})
by an appropriate choice of the values of the coefficients $\rho_j$:

\begin{eqnarray}
0<z_j<1 & \Rightarrow & 
\rho_j = 1; ~~~
\ln(-z_j) =  \ln(z_j) + {\rm i} \pi , 
\nonumber \\
z_j>1 & \Rightarrow &
\rho_j = 0; ~~~
\ln(-z_j) =  \ln(z_j) - {\rm i} \pi .
\nonumber 
\end{eqnarray}

Assuming $m_1<m_2$ in the following we have
\begin{itemize}
\item
for $k^2 < (m_1-m_2)^2 \Rightarrow  z_1 < 1,  z_2 > 1$
\begin{eqnarray}
J & = & 
\mbox{i}\pi^{2-\varepsilon}\;
  \frac{\Gamma(1+\ep)}{2(1-2\ep)} \;
  \Biggl( \frac{m_1^{-2\ep} \!+\! m_2^{-2\ep}}{\ep}
  + \frac{m_1^2\!-\!m_2^2}{\ep k^2}
  \left( m_1^{-2\varepsilon} \!-\! m_2^{-2\varepsilon} \right)
\nonumber \\  && 
\hspace{-1cm}
+ \frac{\sqrt{\lambda(m_1^2,m_2^2,k^2)}}{k^2} 
\Biggl\{ \ln (z_1 z_2) 
+ \ep \Biggl[ 2 \Li{2}{\frac{1}{z_2}} - 2 \Li{2}{z_1}
         - \frac{1}{2} \ln^2 z_1 + \frac{1}{2} \ln^2 z_2                  
\nonumber \\  && \hspace{2.2cm}
         - \ln (z_1 z_2) \ln \left( \frac{\lambda(m_1^2,m_2^2,k^2)}{k^2} \right)
+ {\cal O}(\varepsilon^2) 
      \Biggr]
\Biggr \}   \Biggr)
\end{eqnarray}

\item
for $k^2 > (m_1+m_2)^2 \Rightarrow  z_1 < 1,   z_2 < 1$
\begin{eqnarray}
J & = & 
\mbox{i}\pi^{2-\varepsilon}\;
  \frac{\Gamma(1+\ep)}{2(1-2\ep)} \;
  \Biggl( \frac{m_1^{-2\ep} \!+\! m_2^{-2\ep}}{\ep}
  + \frac{m_1^2\!-\!m_2^2}{\ep k^2}
  \left( m_1^{-2\varepsilon} \!-\! m_2^{-2\varepsilon} \right)
\nonumber \\  && 
\hspace{-1cm}
+ \frac{\sqrt{\lambda(m_1^2,m_2^2,k^2)}}{k^2} 
\Biggl\{ \ln (z_1 z_2) 
+ \ep \Biggl[ - 8 \zeta_2 
- 2 \Li{2}{z_1} - 2 \Li{2}{z_2} 
\nonumber \\  && 
 - \frac{1}{2} \ln^2 z_1 - \frac{1}{2} \ln^2 z_2                  
 - \ln (z_1 z_2) \ln \left( \frac{\lambda(m_1^2,m_2^2,k^2)}{k^2} \right)
\Biggr]
\nonumber \\  && 
+ \mbox{i} \pi \Biggl[ 2 -  2 \ep \ln \left( \frac{\lambda(m_1^2,m_2^2,k^2)}{k^2} \right)
               \Biggr]
+ {\cal O}(\varepsilon^2) 
\Biggr \}   \Biggr) .
\end{eqnarray}
\end{itemize}
In a similar manner, starting from Eq.~(2.17) of \cite{1loop-analytic}
and performing an analytical continuation,
it is possible to obtain the higher order terms of the $\ep$
expansion \footnote{A collection of useful expressions  for
the one-loop two-point function is given also in Appendix A of \cite{BDS}.}. 
In particular, the imaginary part of $J$ in each order of $\ep$ 
coincides with that obtained from the exact result \cite{2loop-analytic-b}
$$
{\rm Im} J = \mbox{i} \pi \theta \left( k^2-(m_1+m_2)^2 \right) 
\frac{\sqrt{\lambda(m_1^2,m_2^2,k^2)}}{k^2}
\left(\frac{\lambda(m_1^2,m_2^2,k^2)}{k^2} \right)^{-\ep}
\frac{\Gamma(1-\ep)}{\Gamma(2-2\ep)} .
$$
In the limit, when one of the masses vanishes, the result is \cite{1loop-analytic}

\begin{equation}
\label{m2=0}
\left. J \right|_{m_1=0, \; m_2\equiv m} =
\mbox{i}\pi^{2-\varepsilon} m^{-2\ep}
\frac{\Gamma(1+\varepsilon)}{(1-2\varepsilon)}
\Biggl\{ \frac{1}{\ep} - \frac{1-u}{2u\ep}
\left[ (1-u)^{-2\ep} - 1 \right]
- \frac{(1\!-\!u)^{1\!-\!2\ep}}{u} \ep \Li{2}{u}
+ {\cal O}(\varepsilon^2) 
\Biggr\} ,
\end{equation}
with $u = k^2/m^2$.

The transition from the
bare parameters to the renormalized ones requires differentiations of
the one-loop propagators with respect to all parameters, couplings,
masses and external momentum.  The integrals obtained thereby can be
reduces again to integrals of type (\ref{general}) plus simpler
bubble integrals. The expansion of the propagators with respect to
small parameters (ratios of the masses or momenta and masses) can be
extracted from the exact analytical results written in terms of
hyper-geometric functions (see \cite{BD-TMF}).

%%%%%%%%%%%%%%%%%%%%%%%%%%%%%%%%%%%%%%%%%%%%%%%%%%%%%%%%%%%%%%%%%%%%%%%%%%%%%%
\section{$\overline{{\rm MS}}$ vs. pole masses at one-loop}
\setcounter{equation}{0}
In this Appendix we present, for completeness, the well know \cite{FJ} 
one-loop relations between pole and ${\overline{\rm MS}}$ masses of gauge bosons.
Using the following notation
\begin{equation}
\frac{M_V^2}{m_V^2} =  1 + \left( \frac{e^2}{16\pi^2\sin^2 \theta_W} \right) X^{(1)}_V 
\label{X:def}
\end{equation}
we have
\begin{eqnarray}
X^{(1)}_H & = & \frac{1}{2} - \frac{1}{2} \ln\frac{M_W^2}{\mu^2} - B(m_W^2,m_W^2;m_H^2)
\nonumber \\ 
& + & \frac{m_H^2}{m_W^2} 
\left( - \frac{3}{2} + \frac{9}{8} \frac{\pi}{\sqrt{3}} 
       + \frac{3}{8} \ln\frac{M_H^2}{\mu^2}
       + \frac{1}{4} B(m_W^2,m_W^2;m_H^2) + \frac{1}{8} B(m_Z^2,m_Z^2;m_H^2)  \right)
\nonumber \\ 
& + & \frac{m_Z^2}{m_W^2} \left( \frac{1}{4} - \frac{1}{4} \ln\frac{M_Z^2}{\mu^2}
       - \frac{1}{2} B(m_Z^2,m_Z;m_H^2) \right)
\nonumber \\ 
& + &  \frac{m_W^2}{m_H^2} 
\left( 3 - 3 \ln\frac{M_W^2}{\mu^2} + 3 B(m_W^2,m_W;m_H^2) \right)
\nonumber \\ 
& + & \frac{m_Z^4}{m_W^2 m_H^2} \left( \frac{3}{2} - \frac{3}{2} \ln\frac{M_Z^2}{\mu^2}
             + \frac{3}{2} B(m_Z^2,m_Z;m_H^2) \right)
\end{eqnarray}
\begin{eqnarray}
X^{(1)}_W & = & \frac{73}{9} - 3 \ln\frac{M_W^2}{\mu^2}
                           + 2 \ln\frac{M_Z^2}{\mu^2}
                           - \frac{17}{3} B(m_Z^2,m_W^2;m_W^2) + B(m_H^2,m_W^2;m_W^2)
\nonumber \\ 
& + & \frac{m_H^4}{m_W^4} 
       \left( \frac{1}{12} - \frac{1}{12}\ln\frac{M_H^2}{\mu^2}
                           + \frac{1}{12}B(m_H^2,m_W^2;m_W^2) \right)
\nonumber \\ 
& + & \frac{m_H^2}{m_W^2} \left( \frac{7}{12} + \frac{1}{12} \ln\frac{M_W^2}{\mu^2}
        - \frac{1}{2} \ln \left( \frac{M_H^2}{\mu^2} \right) 
        - \frac{1}{3} B(m_H^2,m_W^2;m_W^2) \right)
\nonumber \\ 
& + & \frac{m_Z^4}{m_W^4} 
     \left( \frac{1}{12} - \frac{1}{12} \ln\frac{M_Z^2}{\mu^2}
                         + \frac{1}{12} B(m_Z^2,m_W^2;m_W^2) \right)
\nonumber \\ 
& + & \frac{m_Z^2}{m_W^2} \left(
          \frac{3}{4} + \frac{1}{12} \ln\frac{M_W^2}{\mu^2}
                      - \frac{2}{3} \ln\frac{M_Z^2}{\mu^2}
         + \frac{4}{3} B(m_Z^2,m_W^2;m_W^2)  \right)
\nonumber \\ 
& + & \frac{m_W^2}{m_Z^2} 
\left( - 8 + 4 \ln\frac{M_W^2}{\mu^2} - 4 B(m_Z^2,m_W^2,m_W^2) \right)
\nonumber \\ 
& + & \frac{m_Z^4}{m_W^2 m_H^2} \left( \frac{1}{2} 
         - \frac{3}{2} \ln\frac{M_Z^2}{\mu^2} \right)
         + \frac{m_W^2}{m_H^2} \left( 1 - 3 \ln\frac{M_W^2}{\mu^2} \right)
\end{eqnarray}
\begin{eqnarray}
X^{(1)}_Z & = &  \frac{13}{18} - \frac{1}{6} \ln\frac{M_W^2}{\mu^2}
           + \frac{4}{3} B(m_W^2,m_W;m_Z^2)  
\nonumber \\  
& + & \frac{m_H^4}{m_W^2 m_Z^2} 
   \left( \frac{1}{12} - \frac{1}{12} \ln\frac{M_H^2}{\mu^2}
                       + \frac{1}{12} B(m_H^2,m_Z^2;m_Z^2) \right)
\nonumber \\ 
& + & \frac{m_H^2}{m_W^2} \left( \frac{7}{12} + \frac{1}{12} \ln\frac{M_Z^2}{\mu^2}
          - \frac{1}{2} \ln\frac{M_H^2}{\mu^2} - \frac{1}{3} B(m_H^2,m_Z^2;m_Z^2) \right)
\nonumber \\ 
& + & \frac{m_Z^2}{m_W^2} \left( \frac{2}{9} - \frac{1}{6} \ln\frac{M_Z^2}{\mu^2}
         + \frac{1}{12} B(m_W^2,m_W^2;m_Z^2) + B(m_H^2,m_Z^2;m_Z^2) \right)
\nonumber \\ 
& + & \frac{m_W^2}{m_Z^2} 
\left( - \frac{4}{3} \ln\frac{M_W^2}{\mu^2} - \frac{17}{3} B(m_W^2,m_W^2;m_Z^2)\right)
+ \frac{m_W^4}{m_Z^4} \left( 4 \ln\frac{M_W^2}{\mu^2} - 4 B(m_W^2,m_W^2;m_Z^2) \right)
\nonumber \\ 
& + & \frac{m_W^2}{m_H^2} \left( 1 - 3 \ln\frac{M_W^2}{\mu^2} \right)
    + \frac{m_Z^4}{m_W^2 m_H^2} \left( \frac{1}{2} 
    - \frac{3}{2} \ln\frac{M_Z^2}{\mu^2} \right).
\end{eqnarray}
where we have used the following function:

$$
B(m_1^2,m_2^2;p^2) 
= \int\limits_0^1 dx \ln \Biggl( \frac{m_1^2}{\mu^2} x 
+ \frac{m_2^2}{\mu^2} (1-x) - \frac{p^2}{\mu^2} x(1-x) - \mbox{i} 0 \Biggr)\;\;.
$$

%%%%%%%%%%%%%%%%%%%%%%%%%%%%%%%%%%%%%%%%%%%%%%%%%%%%%%%%%%%%%%%%%%%%%%%%%%%%%%%%%%%%%
\section{Unrenormalized one-loop expressions in $d$ dimension}
\setcounter{equation}{0}

The computation of higher loop corrections requires a deeper expansion
in $\ep$ of lower order terms.
In this Appendix we present for completeness the results for unrenormalized
one-loop corrections to the  relation between pole and $\overline{\rm MS}$
masses of the gauge bosons in arbitrary dimension $d$ without expanding
it in $\ep$.
Using the following notation
\begin{equation}
\frac{M_V^2}{m_{0,V}^2} =
   1 + \left( \frac{g_0^2}{16\pi^2} \right) 
\left[ X^{(1),\mbox{\small boson}}_{0,V} + X^{(1),\mbox{\small fermion}}_{0,V} \right]\;,
\label{X0:def}
\end{equation}
we have
\begin{eqnarray}
X^{(1),\mbox{\small  boson}}_{0,W} & = & 
\frac{1}{4(d-1)} \Biggl[
\frac{m_H^4}{m_W^4}  \left( - B_0(m_H^2,m_W^2,m_W^2) - A_0(m_H^2) \right)
\nonumber \\ && \hspace{1.8cm}
+  \frac{m_H^2}{m_W^2} \left( A_0(m_W^2) + 4 B_0(m_H^2,m_W^2,m_W^2) \right)
\nonumber \\ && \hspace{1.8cm}
+  \frac{m_Z^4}{m_W^4}   \left( - B_0(m_Z^2,m_W^2,m_W^2) - A_0(m_Z^2) \right)
- 16 B_0(m_Z^2,m_W^2,m_W^2)
\nonumber \\ && \hspace{1.8cm}
+  \frac{m_Z^2}{m_W^2} \left( A_0(m_W^2) + 4 A_0(m_Z^2) + 8 B_0(m_Z^2,m_W^2,m_W^2) \right)
                  \Biggr]
\nonumber \\ &&
-  \frac{m_Z^2}{m_W^2} \left(  2 B_0(m_Z^2,m_W^2,m_W^2) +  A_0(m_Z^2) \right)
+  7 B_0(m_Z^2,m_W^2,m_W^2) 
\nonumber \\ &&
+ 2 \frac{m_W^2}{m_Z^2} \left(  2 B_0(m_Z^2,m_W^2,m_W^2) +  A_0(m_W^2) \right)
- B_0(m_H^2,m_W^2,m_W^2) 
\nonumber \\ &&
+  (d-5) A_0(m_W^2) + (d-2) A_0(m_Z^2)
- \frac{d-1}{2} \frac{m_Z^4}{m_H^2 m_W^2} A_0(m_Z^2)
\nonumber \\ &&
- \frac{1}{2} \frac{m_H^2}{m_W^2}  A_0(m_H^2)  - \frac{m_W^2}{m_H^2} (d-1) A_0(m_W^2)
- 2 \frac{1}{d-3} A_0(m_W^2) \left(1-\frac{m_W^2}{m_Z^2} \right) \; ,
\\ 
X^{(1),\mbox{\small  fermion}}_{0,W} & = & 
- \frac{1}{2} \frac{d-2}{d-1} \sum_{\rm lepton}  
\left(B_0(0,m_l^2,m_W^2) + \frac{m_l^2}{m_W^2} A_0(m_l^2) \right) 
+ \sum_{\rm lepton}  2 \frac{m_l^4}{m_W^2 m_H^2}  A_0(m_l^2)
\nonumber \\ &&
+ \frac{1}{2} \frac{d-3}{d-1} \sum_{\rm lepton} 
\frac{m_l^2}{m_W^2}  B_0(0,m_l^2,m_W^2) 
\nonumber \\ &&
+ \frac{1}{2} \frac{1}{d-1} \sum_{\rm lepton} \frac{m_l^4}{m_W^4} 
\left( B_0(0,m_l^2,m_W^2) + A_0(m_l^2) \right)
\nonumber \\ &&
- \frac{N_c}{2(d-1)} \sum_{i,j=1}^3 
\Biggl[ 
\left( 2 \frac{m_{u_i}^2 m_{d_j}^2 }{m_W^4} 
       - \frac{m_{u_i}^4 }{m_W^4} 
       - \frac{m_{d_j}^4 }{m_W^4} 
      \right) 
-  (d-3) \left(    \frac{m_{u_i}^2 }{m_W^2}
                 + \frac{m_{d_j}^2 }{m_W^2}
        \right) 
\nonumber \\ && \hspace{3cm}
+  (d-2) 
\Biggr] \times
K_{ij} K^{*}_{ij} B_0(m^2_{u_i},m^2_{d_j}, m_W^2)
\nonumber \\ &&
- \frac{N_c}{2(d-1)} \sum_{i=1}^3 \frac{ m_{u_i}^2 }{m_W^2} A_0(m_{u_i}^2 )
\Biggl( \sum_{j=1}^3 \frac{m_{d_j}^2}{m_W^2} K_{ij} K^{*}_{ij} 
- \frac{m_{u_i}^2}{m_W^2} 
+ (d-2) 
- 4 (d-1) \frac{ m_{u_i}^2}{m_H^2} 
\Biggr)
\nonumber \\ &&
- \frac{N_c}{2(d-1)} \sum_{i=1}^3 \frac{ m_{d_i}^2 }{m_W^2} A_0(m_{d_i}^2 )
\Biggl( \sum_{j=1}^3 \frac{m_{u_j}^2}{m_W^2} K_{ji} K^{*}_{ji} 
- \frac{m_{d_i}^2}{m_W^2} 
+ (d-2) 
- 4 (d-1) \frac{ m_{d_i}^2}{m_H^2} 
\Biggr)
\nonumber \\ &&
\end{eqnarray}
\begin{eqnarray}
X^{(1),\mbox{\small boson}}_{0,Z} & = &
\frac{1}{4(d-1)} \Biggl[
\frac{m_H^4}{m_W^2 m_Z^2} \left( - B_0(m_H^2,m_Z^2,m_Z^2) - A_0(m_H^2) \right )
\nonumber \\ && \hspace{2.0cm}
+  \frac{m_H^2}{m_W^2} \left( A_0(m_Z^2) + 4 B_0(m_H^2,m_Z^2,m_Z^2) \right)
\nonumber \\ && \hspace{2.0cm}
-  \frac{m_Z^2}{m_W^2}   \left( B_0(m_W^2,m_W^2,m_Z^2) + 2 A_0(m_Z^2) \right)
\nonumber \\ && \hspace{2.0cm}
+ 8 \frac{m_W^2}{m_Z^2}  \left( - 2 B_0(m_W^2,m_W^2,m_Z^2) + A_0(m_W^2)\right)
\nonumber \\ && \hspace{2.0cm}
- 2  \left( A_0(m_W^2) - 4 B_0(m_W^2,m_W^2,m_Z^2) \right)
\Biggr]
\nonumber \\ &&
- \frac{m_W^2}{m_Z^2} \left( 2 A_0(m_W^2) - 7 B_0(m_W^2,m_W^2,m_Z^2) \right)
- \frac{1}{2} \frac{m_H^2}{m_W^2}  A_0(m_H^2)
\nonumber \\ &&
- \frac{m_Z^2}{m_W^2} B_0(m_H^2,m_Z^2,m_Z^2)
- 2 B_0(m_W^2,m_W^2,m_Z^2)
- \frac{d-1}{2} \frac{m_Z^4}{m_W^2 m_H^2} A_0(m_Z^2)
\nonumber \\ &&
+ 2 \frac{m_W^4}{m_Z^4} \left( (d-2)A_0(m_W^2) + 2 B_0(m_W^2,m_W^2,m_Z^2) \right)
- (d-1)\frac{m_W^2}{m_H^2} A_0(m_W^2)
\\ 
X^{(1),\mbox{\small fermion}}_{0,Z} & = &
- \frac{1}{4}\frac{d-2}{d-1} \sum_{\rm lepton} \Biggl [
 \frac{m_Z^2}{m_W^2} B_0(0,0,m_Z^2) 
- \left( 12  - 8 \frac{m_W^2}{m_Z^2} -5 \frac{m_Z^2}{m_W^2} \right) B_0(m^2_l, m^2_l, m_Z^2) 
\nonumber \\ && \hspace{1.0cm}
+ \left (
          10 \frac{m^2_l}{m_W^2} 
        + 16 \frac{m^2_l m_W^2}{m_Z^4} 
        - 24 \frac{m^2_l}{m_Z^2}  
  \right )
\left(  A_0( m^2_l) + \frac{2}{d-2} B_0(m^2_l, m^2_l, m_Z^2) \right)
\Biggr ]
\nonumber \\ &&
- \frac{N_c}{36}\frac{d-2}{d-1} \sum_{\rm up}  \Biggl [
- \left(  40 
      - 17 \frac{m_Z^2}{m_W^2} 
      - 32 \frac{m_W^2}{m_Z^2} \right) B_0(m^2_u, m^2_u, m_Z^2) 
\nonumber \\ && \hspace{1.0cm}
+ \left(  34 \frac{m^2_u}{m_W^2} 
         + 64 \frac{m^2_u m_W^2}{m_Z^4}  
         - 80 \frac{m^2_u}{m_Z^2} \right)
\left(  A_0( m^2_u) + \frac{2}{d-2} B_0(m^2_u, m^2_u, m_Z^2) \right)
\Biggr ] 
\nonumber \\ &&
- \frac{N_c}{36}\frac{d-2}{d-1} \sum_{\rm down} \Biggl [
- \left( 4     
       - 5 \frac{m_Z^2}{m_W^2}  
       - 8 \frac{m_W^2}{m_Z^2} \right) B_0(m^2_d, m^2_d, m_Z^2) 
\nonumber \\ && \hspace{1.0cm}
+ \left( 10 \frac{m^2_d}{m_W^2} 
       + 16 \frac{m^2_d m_W^2}{m_Z^4} 
       - 8 \frac{m^2_d}{m_Z^2}
   \right)
\left( A_0( m^2_d)  + \frac{2}{d-2} B_0(m^2_d, m^2_d, m_Z^2) \right)
\Biggr ]
\nonumber \\ && 
+ \sum_{\rm lepton} \Biggl \{
  2 \frac{m_l^4}{m_W^2 m_H^2} A_0(m_l^2)
+ \frac{1}{2} \frac{m^2_l}{m_W^2}  B_0(m^2_l, m^2_l, m_Z^2) 
\Biggr \}
\nonumber \\ &&
+ N_c\sum_{\rm quark}  \Biggl \{
  2 \frac{m_q^4}{m_W^2 m_H^2} A_0(m^2_q)
+ \frac{1}{2} \frac{m^2_q}{m_W^2} B_0(m^2_q, m^2_q, m_Z^2) 
\Biggr \}
\end{eqnarray}
In above formulae we use the following functions
\begin{eqnarray}
A_0(m_1^2) & = &
 \frac{1}{m_1^2} \left(\frac{\mu^2 e^\gamma}{4\pi}\right)^\ep
 \int \frac{\mbox{d}^d q}{i\pi^{d/2}}
 \frac{1}{ \bigl( q^2-m_1^2 \bigr)} \; ,
\nonumber \\
B_0(m_1^2, m_2^2; p^2) & = &
  \left(\frac{\mu^2 e^\gamma}{4\pi}\right)^\ep
 \int \frac{\mbox{d}^d q}{i\pi^{d/2}}
 \frac{1}{ \bigl( q^2-m_1^2 \bigr) \bigl( (p-q)^2-m_2^2 \bigr)} \; .
\end{eqnarray}
$m_{u_i}$ and $m_{d_i}$ denote the masses of corresponding up- and down-quarks, 
$N_c$ is a number of color ($N_c=3$)
and $K_{ij}$ is the element of the Kobayashi-Maskawa matrix.

%%%%%%%%%%%%%%%%%%%%%%%%%%%%%%%%%%%%%%%%%%%%%%%%%%%%%%%%%%%%%%%%%%%%%%%%%%%%%%%

\section{$\overline{{\rm MS}}$ vs. pole masses at two-loop}

After expansion of the diagrams with respect to $\sin^2 \theta_W$ we
get rid of one of the boson masses and write the functions $X^{(2)}_V$
introduced in (\ref{result}) in the form
\begin{equation}
  X^{(2)}_V = \frac{m_H^4}{m_V^4}\sum_{k=0}^{5} \sin^{2k}\theta_W\, A^V_k.
\label{X2V}
\end{equation}
In particular for the $Z$ boson
propagator we eliminate $m_W$ and vice versa.  Consequently, the
coefficients $A_i^V$ in the above formula are functions of the Higgs mass
and one of the boson masses.  We expand this function with respect to
$m_V^2/m_H^2$ 
$$ 
A_i^V = \sum_{j=0}^5 A_{i,j}^V \left(\frac{m_V^2}{m_H^2}\right)^j  
$$ 
and calculate analytically the first
six coefficients.  This is not a naive Taylor expansion. The general
rules for asymptotic expansions \cite{asymptotic} allow us to extract
also logarithmic dependences, or in other words, to preserve all
analytical properties of the original diagrams.  In the result of the
asymptotic expansion all propagator diagrams are reduced to single
scale massive diagrams (including the two-loop bubbles).  As a
consequence, the finite as well as the $\ep$-part of the corresponding
diagrams, are characterized by a restricted set of transcendental numbers
\cite{numbers} which  may appear in the coefficients $ A_{i,j}$. We find the
following constants:
\begin{eqnarray}
S_0 & = & \frac{\pi}{\sqrt{3}} \sim 1.813799365...,
\hspace{3cm}
S_1 = \frac{\pi}{\sqrt{3}} \ln 3 \sim 1.992662272...,
\nonumber \\
S_2 & = &  \frac{4}{9} \frac{\Cl{2}{\tfrac{\pi}{3}}}{\sqrt{3}} 
\sim 0.260434137632162...,
~~~~~
S_3 = \pi \Cl{2}{\tfrac{\pi}{3}} \sim 3.188533097...
\end{eqnarray}
Furthermore, $\ln(m_H^2)$ denotes
$\ln\left( m_H^2/\mu^2 \right)$ where $\mu$ is the 't Hooft
scale. We also introduce the notation  $w_H = m_W^2/m_H^2$ and $z_H =
m_Z^2 /m_H^2$.

\subsection{Analytical results for the two-loop finite part of W-boson}
\setcounter{equation}{0}

\begin{eqnarray*}
%part1
A_0^W & = &
~~~~\Biggl[
-{\tfrac {359}{128}}
+{\tfrac {243}{32}}\,{ S_2} 
- \tfrac{1}{24}\,{\pi}^{2}
+{\tfrac {33}{16}}\,\ln ({ m_H^2})
-{\tfrac {9}{32}}\, \left( \ln ({ m_H^2}) \right)^2
\Biggr ]
\nonumber \\ & + & 
w_H \Biggl[
{\tfrac {2483}{192}}
-{\tfrac {231}{32}}\,{ S_0}
+{\tfrac  {243}{32}}\,{ S_2}
+{\tfrac {433}{864}}\,{\pi }^{2}
-{\tfrac {161}{32}}\,\ln ({ w_H})
+{\tfrac {69}{16}}\,\ln ({ w_H})\ln ({ m_H^2})
\nonumber \\ && 
-{\tfrac {519}{32}}\,\ln ({ m_H^2})
+{\tfrac {99}{16}}\,\ln ({ m_H^2}){ S_0}
+{\tfrac {43}{8}}\,\left (\ln ({ m_H^2})\right )^{2} \Biggr]
\nonumber \\ & + &
w_H^2 \Biggl [
{\tfrac {8311573}{20736}}
-{\tfrac {153}{8}}\,{ S_3}
-{\tfrac {243}{4}}\,{ S_0}
-{\tfrac {87651}{128}}\,{ S_2}
-{\tfrac {47893}{3456}}\,{\pi }^{2}
+{\tfrac {459}{16}}\,\zeta (3)
-{\tfrac {399943}{1728}}\,\ln ({ w_H})
\nonumber \\ && 
+{\tfrac {1705}{16}}\,\ln ({ w_H}){ S_0}
+{\tfrac {1805}{48}}\,\left (\ln ({ w_H})\right )^{2}
+{\tfrac {1951}{24}}\,\ln ({ w_H})\ln ({ m_H^2})
-{\tfrac {356585}{1728}}\,\ln ({ m_H^2})
\nonumber \\ && 
+{\tfrac {1595}{16}}\,\ln ({ m_H^2}){ S_0}
+{\tfrac {10013}{288}}\,\left (\ln ({ m_H^2})\right )^{2}
\Biggr ]
\nonumber \\ & +& 
w_H^3\,\Biggl [
{\tfrac {10730119}{86400}}
-{\tfrac {3451}{96}}\,{ S_0}
-{\tfrac {35739}{80}}\,{ S_2}
+{\tfrac {3167}{1080}}\,{\pi }^{2}
-{\tfrac {1173881}{4320}}\,\ln ({ w_H})
\nonumber \\ && 
+{\tfrac {517}{8}}\,\ln ({ w_H}){ S_0}
+{\tfrac {6767}{96}}\,\left (\ln ({ w_H})\right )^{2}
+{\tfrac {1743}{16}}\,\ln ({ w_H})\ln ({ m_H^2})
-{\tfrac {30383}{192}}\,\ln ({ m_H^2})
\nonumber \\ && 
+{\tfrac {297}{8}}\,\ln ({ m_H^2}){ S_0}
+{\tfrac {177}{4}}\,\left (\ln ({ m_H^2})\right )^{2}\Biggr ]
\nonumber \\ & + & w_H^{4}\Biggl [
-{\tfrac {276774409}{1296000}}
+{\tfrac {86473}{3200}}\,{ S_0}
-{\tfrac {3807}{20}}\,{ S_2}
+{\tfrac {859}{80}}\,{\pi }^{2}
-{\tfrac {168691}{1152}}\,\ln ({ w_H})
+{\tfrac {10323}{160}}\,\ln ({ w_H}){ S_0}
\nonumber \\ && 
+{\tfrac {90121}{960}}\,\left (\ln ({ w_H})\right )^{2}
+{\tfrac {187}{2}}\,\ln ({ w_H})\ln ({ m_H^2})
+{\tfrac {15733}{240}}\,\ln ({ m_H^2})
+{\tfrac {81}{4}}\,\left (\ln ({ m_H^2})\right )^{2} \Biggr ]
\nonumber \\ 
& + & w_H^{5}\Biggl [
-{\tfrac {13424129921}{14112000}}
+{\tfrac {416057}{2400}}\,{ S_0}
-{\tfrac {452727}{1120}}\,{ S_2}
+{\tfrac {1395311}{30240}}\,{\pi }^{2}
-{\tfrac {3016477}{10080}}\,\ln ({ w_H})
\nonumber \\ && 
+{\tfrac {7739}{40}}\,\ln ({ w_H}){ S_0}
+{\tfrac {16007}{60}}\,\left (\ln ({ w_H})\right )^{2}
+{\tfrac {12347}{72}}\,\ln ({ w_H})\ln ({ m_H^2})
+{\tfrac {6333137}{30240}}\,\ln ({ m_H^2}) \Biggr ]
\end{eqnarray*}

\begin{eqnarray*}
%part2
A_1^W & = & 
{ w_H}
\left [{\tfrac {79}{48}}+{\tfrac {77}{32}}\,{ S_0}+\tfrac{3}{16}\,{\pi }^{2}
+{\tfrac {21}{32}}\,\ln ({ w_H})-{\tfrac {9}{16}}\,\ln ({ w_H})\ln({ m_H^2})
-{\tfrac {25}{32}}\,\ln ({ m_H^2})
-{\tfrac {33}{16}}\,\ln ({ m_H^2}){ S_0}\right ]
\nonumber \\ 
& + & 
w_H^{2}\Biggl [
-{\tfrac {15491}{576}}
+{\tfrac {65}{6}}\,{ S_3}
-{\tfrac {11183}{48}}\,{ S_0}
+66\,{ S_1} 
- {\tfrac {4239}{16}}\,{ S_2}
+ {\tfrac {8837}{216}}\,{\pi }^{2} 
- {\tfrac {22}{3}}\,{\pi }^{2}\ln (3)
+{\tfrac {67}{4}}\,\zeta_3
\nonumber \\ && 
-{\tfrac {149}{216}}\,\ln ({ w_H})
-{\tfrac {209}{12}}\,\ln ({ w_H}){ S_0}
-{\tfrac {277}{72}}\,\left (\ln ({ w_H})\right )^{2}
- \tfrac{5}{4}\,\ln ({ w_H})\ln ({ m_H^2})
+{\tfrac {559}{72}}\,\ln ({  m_H^2})
\nonumber \\ && 
-{\tfrac {473}{24}}\,\ln ({ m_H^2}){ S_0}
-{\tfrac {5}{16}}\,\left (\ln ({ m_H^2})\right )^{2}\Biggr ]
\nonumber \\
& + & w_H^{3}\Biggl [
-{\tfrac {14221}{216}}+{\tfrac {977}{72}}\,{ S_0}+{\tfrac{8181}{64}}\,{ S_2}
+{\tfrac {5705}{1728}}\,{\pi }^{2}-{\tfrac {5923}{216}}\,\ln ({ w_H})
+{\tfrac {77}{3}}\,\ln ({ w_H}){ S_0}
\nonumber \\ && 
+{\tfrac {2801}{144}}\,\left (\ln ({ w_H})\right )^{2}
+{\tfrac {397}{16}}\,\ln ({ w_H})\ln ({ m_H^2})
-{\tfrac {1361}{192}}\,\ln ({ m_H^2})
+{\tfrac {99}{8}}\,\ln ({ m_H^2}){ S_0}
+12\,\left (\ln ({ m_H^2})\right )^{2}\Biggr ]
\nonumber \\ 
& + & w_H^{4} \Biggl [
-{\tfrac {43842631}{288000}}+{\tfrac {259993}{4800}}\,{ S_0}
+{\tfrac {4329}{160}}\,{ S_2}+{\tfrac {21149}{1440}}\,{\pi }^{2}
+{\tfrac {26903}{1200}}\,\ln ({ w_H})+{\tfrac {4431}{80}}\,\ln ({ w_H}){ S_0}
\nonumber \\ &&
+{\tfrac {40579}{720}}\,\left (\ln ({ w_H})\right )^{2}
+{\tfrac {423}{8}}\,\ln ({ w_H})\ln ({ m_H^2})
+{\tfrac {635}{32}}\,\ln ({ m_H^2})+27\,\left (\ln ({ m_H^2})\right )^{2}\Biggr]
\nonumber \\ 
& + & w_H^{5}\Biggl [
-{\tfrac {367144853}{518400}}+{\tfrac{2094431}{7200}}\,{ S_0}
+{\tfrac {29151}{320}}\,{ S_2}+{\tfrac {210541}{2880}}\,{\pi }^{2}
+{\tfrac {8105093}{43200}}\,\ln ({ w_H})
\nonumber \\ &&
+{\tfrac {28277}{120}}\,\ln ({ w_H}){ S_0}
+{\tfrac {120943}{720}}\,\left (\ln ({ w_H})\right )^{2}
-{\tfrac {53}{8}}\,\ln ({ w_H})\ln ({ m_H^2})-{\tfrac {4829}{480}}\,\ln ({ m_H^2})
\Biggr ]
\end{eqnarray*}

\begin{eqnarray*}
%part3
A_2^W & = & 
w_H
\Biggl [ -{\tfrac {29}{32}}+{\tfrac {35}{16}}\,{S_0}
+\tfrac{3}{16}\,{\pi }^{2}
+{\tfrac {21}{32}}\,\ln ({w_H})-{\tfrac {9}{16}}\,\ln ({w_H})\ln ({m_H^2})
+{\tfrac {45}{32}}\,\ln ({m_H^2})-{\tfrac {15}{8}}\,\ln ({m_H^2}){S_0} \Biggr]
\nonumber \\ 
& + & w_H^2 \Biggl[
{\tfrac {439321}{3456}}+{\tfrac {7}{36}}\,{S_3}+{\tfrac{18119}{144}}\,{S_0}
-44\,{S_1}+{\tfrac {14469}{64}}\,{S_2}-{\tfrac {19417}{576}}\,{\pi }^{2}
-16\,{\pi }^{2}\ln (2)+{\tfrac {110}{9}}\,{\pi }^{2}\ln (3)
\nonumber \\ &&
-{\tfrac {751}{24}}\,\zeta_3-{\tfrac {973}{432}}\,\ln ({w_H})
-{\tfrac {233}{12}}\,\ln ({w_H}){S_0}
-{\tfrac {581}{144}}\,\left (\ln ({w_H})\right )^{2}
\nonumber \\ &&
+{\tfrac {45}{16}}\,\ln ({w_H})\ln ({m_H^2})+{\tfrac {281}{96}}\,\ln({m_H^2})
-{\tfrac {71}{4}}\,\ln ({m_H^2}){S_0}
+{\tfrac {55}{32}}\,\left (\ln ({m_H^2})\right )^{2}\Biggr ]
\nonumber \\
& + & w_H^3
\Biggl [ 
-{\tfrac {304691}{3456}}+{\tfrac {15601}{576}}\,{S_0}+{\tfrac {10659}{64}}\,{S_2}
+{\tfrac {9167}{1728}}\,{\pi }^{2}+{\tfrac {4267}{864}}\,\ln ({w_H})
+{\tfrac {1073}{48}}\,\ln ({w_H}){S_0}
\nonumber \\ &&
+{\tfrac {3509}{144}}\,\left (\ln ({w_H})\right )^{2}
+{\tfrac {393}{16}}\,\ln ({w_H})\ln ({m_H^2})+{\tfrac {945}{64}}\,\ln({m_H^2})
+{\tfrac {141}{8}}\,\ln ({m_H^2}){S_0}
+{\tfrac {45}{4}}\,\left (\ln ({m_H^2})\right )^{2} \Biggr ]
\nonumber \\
& + & w_H^{4}
\Biggl[-{\tfrac {107428459}{432000}}+{\tfrac {990193}{9600}}\,{S_0}
+{\tfrac {5103}{40}}\,{S_2}+{\tfrac {111}{5}}\,{\pi }^{2}
+{\tfrac {10042841}{86400}}\,\ln ({w_H})
+{\tfrac {10771}{160}}\,\ln ({w_H}){S_0}
\nonumber \\ &&
+{\tfrac {32969}{320}}\,\left (\ln ({w_H})\right )^{2}
+{\tfrac {801}{8}}\,\ln ({w_H})\ln ({m_H^2})
+{\tfrac {1881}{32}}\,\ln ({m_H^2})
+{\tfrac {99}{2}}\,\left (\ln ({m_H^2})\right )^{2}\Biggr ]
\nonumber \\
& + & w_H^{5}
\Biggl[
-{\tfrac {3726216829}{2592000}}+{\tfrac {2622653}{3600}}\,{S_0}
+{\tfrac {98229}{320}}\,{S_2}+{\tfrac {1160293}{8640}}\,{\pi }^{2}
+{\tfrac {21966593}{43200}}\,\ln ({w_H})
\nonumber \\ &&
+{\tfrac {27221}{60}}\,\ln ({w_H}){S_0}
+{\tfrac {253819}{720}}\,\left (\ln ({w_H})\right )^{2}
+ \tfrac{1}{8}\,\ln ({w_H})\ln ({m_H^2})-{\tfrac {199}{96}}\,\ln ({m_H^2})\Biggr ]
\end{eqnarray*}

\begin{eqnarray*}
%part4w
A_3^W & = & 
{w_H}\,
\Biggl [-2+{\tfrac {49}{18}}\,{S_0}+  \tfrac{3}{16}\,{\pi }^{2}
+{\tfrac {21}{32}}\,\ln ({w_H})-{\tfrac {9}{16}}\,\ln({w_H})\ln({m_H^2})
+{\tfrac {75}{32}}\,\ln ({m_H^2})-\tfrac{7}{3}\,\ln ({m_H^2}){S_0}\Biggr ]
\nonumber \\
& + & w_H^{2}\Biggl [-{\tfrac {799}{12}}-{\tfrac {199}{54}}\,{S_3}
-{\tfrac {26975}{144}}\,{S_0}-\tfrac{5}{3}\,{S_1}+{\tfrac {4745}{24}}\,{S_2}
+{\tfrac {1669}{54}}\,{\pi }^{2}+{\tfrac {104}{27}}\,{\pi }^{2}\ln (3)
-{\tfrac {425}{36}}\,\zeta (3)
\nonumber \\ && 
+{\tfrac {284}{27}}\,\ln ({w_H})
-{\tfrac {1709}{108}}\,\ln ({w_H}){S_0}-{\tfrac {38}{9}}\,\left (\ln({w_H})\right )^{2}
+{\tfrac {55}{8}}\,\ln ({w_H})\ln ({m_H^2})+{\tfrac {47}{4}}\,\ln({m_H^2})
\nonumber \\ && 
-{\tfrac {859}{108}}\,\ln ({m_H^2}){S_0}
+{\tfrac {15}{4}}\,\left (\ln ({m_H^2})\right )^{2}\Biggr ]
\nonumber \\
& + & w_H^{3}\Biggl [
-{\tfrac {857575}{6912}}+{\tfrac {196669}{5184}}\,{S_0}+{\tfrac {49193}{192}}\,{S_2}
+{\tfrac {13715}{1728}}\,{\pi }^{2}+{\tfrac {26065}{576}}\,\ln ({w_H})
+{\tfrac {2429}{432}}\,\ln ({w_H}){S_0}
\nonumber \\ && 
+{\tfrac {7795}{288}}\,\left (\ln ({w_H})\right )^{2}
+{\tfrac {309}{16}}\,\ln ({w_H})\ln ({m_H^2})+{\tfrac {2169}{64}}\,\ln({m_H^2})
+{\tfrac {125}{8}}\,\ln ({m_H^2}){S_0}+8\,\left (\ln ({m_H^2})\right)^{2}\Biggr ]
\nonumber \\
& + & w_H^{4}\Biggl [-{\tfrac {3314383}{9000}}+{\tfrac{1531019}{10800}}\,{S_0}
+{\tfrac {42717}{160}}\,{S_2}+{\tfrac {53773}{1440}}\,{\pi }^{2}
+{\tfrac {6187879}{21600}}\,\ln ({w_H})
+{\tfrac {7507}{120}}\,\ln ({w_H}){S_0}
\nonumber \\ && 
+{\tfrac {19279}{120}}\,\left (\ln ({w_H})\right )^{2}
+{\tfrac {1323}{8}}\,\ln ({w_H})\ln ({m_H^2})
+{\tfrac {3943}{32}}\,\ln ({m_H^2})+81\,\left (\ln ({m_H^2})\right)^{2}\Biggr ]
\nonumber \\
& + & w_H^{5}\Biggl [-{\tfrac {3404318779}{1296000}}+{\tfrac  {1761697}{1296}}\,{S_0}
+{\tfrac {288607}{480}}\,{S_2}
+{\tfrac {1098061}{4320}}\,{\pi }^{2}+{\tfrac {4020149}{3600}}\,\ln({w_H})
\nonumber \\ && 
+{\tfrac {38929}{54}}\,\ln ({w_H}){S_0}
+{\tfrac {43325}{72}}\,\left (\ln ({w_H})\right )^{2}
+{\tfrac {55}{8}}\,\ln ({w_H})\ln ({m_H^2})
+{\tfrac {2839}{480}}\,\ln ({m_H^2})\Biggr ]
\end{eqnarray*}

\begin{eqnarray*}
%part5w
A_4^W & = & 
{w_H}\,\Biggl[
-{\tfrac {3095}{1152}}+{\tfrac {2723}{864}}\,{S_0}
+\tfrac{3}{16}\,{\pi }^{2}
+{\tfrac {21}{32}}\,\ln ({w_H})-{\tfrac {9}{16}}\,\ln ({w_H})\ln({m_H^2})
+{\tfrac {563}{192}}\,\ln ({m_H^2})
\nonumber \\ && 
-{\tfrac{389}{144}}\,\ln({m_H^2}){S_0}\Biggr ]
\nonumber \\
& + & w_H^{2} \Biggl[
-{\tfrac {40045}{432}}
-{\tfrac {841}{216}}\,{S_3}
-{\tfrac{747209}{5184}}\,{S_0}
+\tfrac{1}{9} \,{S_1}
+{\tfrac {125399}{384}}\,{S_2}
+{\tfrac {660755}{31104}}\,{\pi }^{2}
+{\tfrac {68}{27}}\,{\pi }^{2}\ln (3)
-{\tfrac {791}{144}}\,\zeta_3
\nonumber \\ &&
+{\tfrac {4151}{216}}\,\ln ({w_H})
-{\tfrac {26723}{1296}}\,\ln ({w_H}){S_0}
-{\tfrac {635}{144}}\,\left (\ln ({w_H})\right )^{2}
+{\tfrac {175}{16}}\,\ln ({w_H})\ln ({m_H^2})
+{\tfrac {27841}{1728}}\,\ln ({m_H^2})
\nonumber \\ &&
-{\tfrac {7153}{1296}}\,\ln ({m_H^2}){S_0}
+{\tfrac {185}{32}}\,\left (\ln ({m_H^2})\right )^{2}\Biggr ]
\nonumber \\
& + & w_H^{3}\Biggl[
-{\tfrac {803113}{5184}}+{\tfrac {323699}{7776}}\,{S_0}
+{\tfrac{33121}{96}}\,{S_2}+{\tfrac {28757}{2592}}\,{\pi }^{2}
+{\tfrac {85597}{864}}\,\ln ({w_H})-{\tfrac {30247}{1296}}\,\ln({w_H}){S_0}
\nonumber \\ &&
+{\tfrac {7933}{288}}\,\left (\ln ({w_H})\right )^{2}
+{\tfrac {145}{16}}\,\ln ({w_H})\ln ({m_H^2})
+{\tfrac {10643}{192}}\,\ln ({m_H^2})
+{\tfrac {103}{12}}\,\ln ({m_H^2}){S_0}
+\tfrac{9}{4}\,\left (\ln ({m_H^2})\right )^{2}\Biggr]
\nonumber \\
& + & w_H^{4}\Biggl [
-{\tfrac {1919013}{4000}}+{\tfrac {40790531}{259200}}\,{S_0}
+{\tfrac {869239}{1920}}\,{S_2}+{\tfrac {1015361}{17280}}\,{\pi }^{2}
+{\tfrac {16268087}{28800}}\,\ln ({w_H})
+{\tfrac {7991}{360}}\,\ln ({w_H}){S_0}
\nonumber \\ &&
+{\tfrac {660151}{2880}}\,\left (\ln ({w_H})\right )^{2}
+{\tfrac {2025}{8}}\,\ln ({w_H})\ln ({m_H^2})
+{\tfrac {7121}{32}}\,\ln ({m_H^2})
+{\tfrac {495}{4}}\,\left (\ln ({m_H^2})\right )^{2}\Biggr ]
\nonumber \\
& + & w_H^{5}\Biggl [
-{\tfrac {305951057}{72000}}
+{\tfrac {414264923}{194400}}\,{S_0}
+{\tfrac {265819}{240}}\,{S_2}+{\tfrac {955459}{2160}}\,{\pi }^{2}
+{\tfrac {11862391}{5400}}\,\ln ({w_H})
\nonumber \\ &&
+{\tfrac {3022481}{3240}}\,\ln ({w_H}){S_0}
+{\tfrac {40741}{45}}\,\left (\ln ({w_H})\right )^{2}
+{\tfrac {109}{8}}\,\ln ({w_H})\ln ({m_H^2})
+{\tfrac {6673}{480}}\,\ln ({m_H^2}) \Biggr ]
\end{eqnarray*}

\begin{eqnarray*}
%part6w
A_5^W & = & 
{w_H} \Biggl[ 
-{\tfrac {5947}{1920}}+{\tfrac {2975}{864}}\,{S_0}+\tfrac{3}{16}\,{\pi }^{2}
+{\tfrac {21}{32}}\,\ln ({w_H})-{\tfrac {9}{16}}\,\ln ({w_H})\ln({m_H^2})
\nonumber \\ && 
+{\tfrac {1051}{320}}\,\ln ({m_H^2})-{\tfrac {425}{144}}\,\ln({m_H^2}){S_0}\Biggr ]
\nonumber \\
& + & w_H^{2}\Biggl [
-{\tfrac {942115}{5184}}-{\tfrac {37}{9}}\,{S_3}
-{\tfrac {3403793}{19440}}\,{S_0}-{\tfrac {1}{54}}\,{S_1}
+{\tfrac {643159}{1440}}\,{S_2}+{\tfrac {3758107}{116640}}\,{\pi }^{2}
+{\tfrac {10}{9}}\,{\pi }^{2}\ln (3)
+\tfrac{7}{6}\,\zeta_3
\nonumber \\ &&
+{\tfrac {61481}{2160}}\,\ln ({w_H})
-{\tfrac {4435}{162}}\,\ln ({w_H}){S_0}
-{\tfrac {331}{72}}\,\left (\ln ({w_H})\right )^{2}
+15\,\ln ({w_H})\ln ({m_H^2})+{\tfrac {3829}{180}}\,\ln ({m_H^2})
\nonumber \\ &&
-{\tfrac {995}{216}}\,\ln ({m_H^2}){S_0}
+{\tfrac {125}{16}}\,\left (\ln ({m_H^2})\right )^{2}\Biggr ]
\nonumber \\
& + & 
w_H^{3}\Biggl [
-{\tfrac {4522879}{25920}}+{\tfrac {1329751}{38880}}\,{S_0}
+{\tfrac {120049}{288}}\,{S_2}+{\tfrac {39089}{2592}}\,{\pi }^{2}
+{\tfrac {16039}{96}}\,\ln ({w_H})
-{\tfrac {84613}{1296}}\,\ln ({w_H}){S_0}
\nonumber \\ &&
+{\tfrac {929}{36}}\,\left (\ln ({w_H})\right )^{2}
-{\tfrac {99}{16}}\,\ln ({w_H})\ln ({m_H^2})
+{\tfrac {77011}{960}}\,\ln ({m_H^2})-{\tfrac {35}{9}}\,\ln({m_H^2}){S_0}
-6\,\left (\ln ({m_H^2})\right )^{2}\Biggr ]
\nonumber \\
& + & w_H^{4}\Biggl [
-{\tfrac {239678009}{432000}}+{\tfrac {15484933}{129600}}\,{S_0}
+{\tfrac {635063}{960}}\,{S_2}+{\tfrac {771181}{8640}}\,{\pi }^{2}
+{\tfrac {42518293}{43200}}\,\ln ({w_H})-{\tfrac {9928}{135}}\,\ln({w_H}){S_0}
\nonumber \\ &&
+{\tfrac {74083}{240}}\,\left (\ln ({w_H})\right )^{2}
+{\tfrac {2943}{8}}\,\ln ({w_H})\ln ({m_H^2})
+{\tfrac {58767}{160}}\,\ln ({m_H^2})+180\,\left (\ln ({m_H^2})\right)^{2}\Biggr ]
\nonumber \\
& + & w_H^{5}\Biggl [
-{\tfrac {16217189681}{2592000}}+{\tfrac {138228439}{48600}}\,{S_0}
+{\tfrac {1072891}{576}}\,{S_2}+{\tfrac {6308171}{8640}}\,{\pi }^{2}
+{\tfrac {171948913}{43200}}\,\ln ({w_H})
\nonumber \\ &&
+{\tfrac {604271}{648}}\,\ln ({w_H}){S_0}
+{\tfrac {895867}{720}}\,\left (\ln ({w_H})\right )^{2}
+{\tfrac {163}{8}}\,\ln ({w_H})\ln ({m_H^2})+{\tfrac {10507}{480}
}\,\ln ({m_H^2})\Biggr ]
\end{eqnarray*}

%%%%%%%%%%%%%%%%%%%%%%%%%%%%%%%%%%%%%%%%%%%%%%%%%%%%%%%%%%%%%%%%%%%%%%%%%
\subsection{Analytical results for the two-loop finite part of Z-boson}

\begin{eqnarray*}
%part1z
A_0^Z & = & 
 \hspace{0.5cm} \Biggl[
-{\tfrac {359}{128}}+{\tfrac {243}{32}}\,{S_2}-\tfrac{1}{24}\,{\pi }^{2}
+{\tfrac {33}{16}}\,\ln ({m_H^2})
-{\tfrac {9}{32}}\,\left (\ln ({m_H^2})\right )^{2}
\Biggr] \frac{m_Z^4}{m_W^4}
\nonumber \\ & + & 
{z_H}\,\Biggl[{\tfrac {2483}{192}}-{\tfrac {231}{32}}\,{S_0}
+{\tfrac {243}{32}}\,{S_2}+{\tfrac {433}{864}}\,{\pi }^{2}
-{\tfrac {161}{32}}\,\ln ({z_H})+{\tfrac {69}{16}}\,
\ln ({z_H})\ln ({m_H^2})
\nonumber \\ &&
-{\tfrac {519}{32}}\,\ln ({m_H^2})
+{\tfrac {99}{16}}\,\ln ({m_H^2}){S_0}
+{\tfrac {43}{8}}\,\left (\ln ({m_H^2})\right )^{2}\Biggr]
\nonumber \\ & + & 
z_H^{2}\Biggl[
{\tfrac {8311573}{20736}}-{\tfrac {153}{8}}\,{S_3}-{\tfrac  {243}{4}}\,{S_0}
-{\tfrac {87651}{128}}\,{S_2}-{\tfrac {47893}{3456}}\,{\pi }^{2}
+{\tfrac {459}{16}}\,\zeta (3)-{\tfrac {399943}{1728}}\,\ln ({z_H})
\nonumber \\ &&
+{\tfrac {1705}{16}}\,\ln ({z_H}){S_0}
+{\tfrac {1805}{48}}\,\left (\ln ({z_H})\right )^{2}+{\tfrac {1951}{24}}\,\ln ({z_H})
\ln ({m_H^2})
-{\tfrac {356585}{1728}}\,\ln ({m_H^2})
\nonumber \\ &&
+{\tfrac {1595}{16}}\,\ln({m_H^2}){S_0}
+{\tfrac {10013}{288}}\,\left (\ln ({m_H^2})\right )^{2}\Biggr]
\nonumber \\ & + & 
z_H^{3}\Biggl[
{\tfrac {10730119}{86400}}-{\tfrac {3451}{96}}\,{S_0}
-{\tfrac {35739}{80}}\,{S_2}+{\tfrac {3167}{1080}}\,{\pi }^{2}
-{\tfrac {1173881}{4320}}\,\ln ({z_H})+{\tfrac {517}{8}}\,\ln({z_H}){S_0}
\nonumber \\ &&
+{\tfrac {6767}{96}}\,\left (\ln ({z_H})\right )^{2}
+{\tfrac {1743}{16}}\,\ln ({z_H})\ln ({m_H^2})-{\tfrac {30383}{192}}\,\ln ({m_H^2})
+{\tfrac {297}{8}}\,\ln ({m_H^2}){S_0}+{\tfrac {177}{4}}\,\left (\ln ({m_H^2})\right )^{2}
\Bigg] \nonumber \\ & + & 
z_H^{4}\Biggl[
-{\tfrac {276774409}{1296000}}+{\tfrac {86473}{3200}}\,{S_0}
-{\tfrac {3807}{20}}\,{S_2}+{\tfrac {859}{80}}\,{\pi }^{2}
-{\tfrac {168691}{1152}}\,\ln ({z_H})+{\tfrac {10323}{160}}\,\ln({z_H}){S_0}
\nonumber \\ &&
+{\tfrac {90121}{960}}\,\left (\ln ({z_H})\right )^{2}
+{\tfrac {187}{2}}\,\ln ({z_H})\ln ({m_H^2})
+{\tfrac {15733}{240}}\,\ln ({m_H^2})+{\tfrac {81}{4}}\,\left (\ln ({m_H^2})\right)^{2}\Biggr ]
\nonumber \\ & + & z_H^{5}\Biggl[
-{\tfrac {13424129921}{14112000}}
+{\tfrac {416057}{2400}}\,{ S_0}
-{\tfrac {452727}{1120}}\,{ S_2}
+{\tfrac {1395311}{30240}}\,{\pi }^{2}
-{\tfrac {3016477}{10080}}\,\ln ({ z_H})
\nonumber \\ &&
+{\tfrac {7739}{40}}\,\ln ({ z_H}){ S_0}
+{\tfrac {16007}{60}}\,\left (\ln ({ z_H})\right )^{2}
+{\tfrac {12347}{72}}\,\ln ({ z_H})\ln ({ m_H^2})
+{\tfrac {6333137}{30240}}\,\ln ({ m_H^2})
\Biggr ]
\end{eqnarray*}

\begin{eqnarray*}
%part2z
A_1^Z & = & 
{z_H}\,\Biggl[-{\tfrac {2243}{288}}+{\tfrac {77}{8}}\,{S_0}+{\tfrac  {243}{16}}\,{S_2}
+{\tfrac {271}{432}}\,{\pi }^{2}
+\tfrac{7}{6}\,\ln ({z_H})-\ln ({z_H})\ln({m_H^2})
\nonumber \\ &&
+{\tfrac {91}{12}}\,\ln ({m_H^2})-{\tfrac {33}{4}}\,\ln({m_H^2}){S_0}\Biggr] 
\nonumber \\ & + & 
z_H^{2}\Biggl[
-{\tfrac {4637683}{10368}}+{\tfrac {589}{12}}\,{S_3}-{\tfrac {12295}{24}}\,{S_0}
+66\,{S_1}+{\tfrac {157725}{64}}\,{S_2}+{\tfrac {99227}{1728}}\,{\pi  }^{2}
+\tfrac{11}{3}\,{\pi }^{2}\ln (3)
\nonumber \\ &&
-{\tfrac {853}{8}}\,\zeta (3)
+{\tfrac {225601}{864}}\,\ln ({z_H})-{\tfrac {4279}{24}}\,\ln ({z_H}){S_0}
-{\tfrac {733}{18}}\,\left (\ln ({z_H})\right )^{2}
-{\tfrac {632}{9}}\,\ln ({z_H})\ln ({m_H^2})
\nonumber \\ &&
+{\tfrac {203531}{864}}\,\ln ({m_H^2})-{\tfrac {3839}{24}}\,\ln ({m_H^2}){S_0}
-{\tfrac {5119}{144}}\,\left (\ln ({m_H^2})\right )^{2}\Biggr] 
\nonumber \\ & + & 
z_H^{3}\Biggl[
-{\tfrac {4480681}{43200}}+{\tfrac {6445}{144}}\,{S_0}
+{\tfrac {126999}{160}}\,{S_2}-{\tfrac {10643}{1440}}\,{\pi }^{2}
+{\tfrac {694519}{2160}}\,\ln ({z_H})-{\tfrac {2167}{12}}\,\ln({z_H}){S_0}
\nonumber \\ &&
-{\tfrac {11395}{144}}\,\left (\ln ({z_H})\right )^{2}
-122\,\ln ({z_H})\ln ({m_H^2})+{\tfrac {5381}{24}}\,\ln ({m_H^2})
-99\,\ln ({m_H^2}){S_0}-{\tfrac {129}{2}}\,\left (\ln ({m_H^2})\right)^{2}\Biggr] 
\nonumber \\ & + & 
z_H^{4}\Biggl[
{\tfrac {284392801}{1296000}}-{\tfrac {155881}{960}}\,{S_0}
+{\tfrac {37971}{40}}\,{S_2}-{\tfrac {661}{18}}\,{\pi }^{2}
+{\tfrac {6123413}{43200}}\,\ln ({z_H})-{\tfrac {3837}{16}}\,\ln({z_H}){S_0}
\nonumber \\ &&
-{\tfrac {11129}{96}}\,\left (\ln ({z_H})\right )^{2}
-{\tfrac {19}{4}}\,\ln ({z_H})\ln ({m_H^2})-{\tfrac {763}{240}}\,\ln({m_H^2})
-{\tfrac {27}{2}}\,\left (\ln ({m_H^2})\right )^{2}\Biggr ]
\nonumber \\ & + & 
z_H^{5}\Biggl[
 {\tfrac {6259218463}{3969000}}
-{\tfrac {1671301}{1800}}\,{ S_0}
+{\tfrac {685701}{224}}\,{ S_2}
-{\tfrac {5834099}{30240}}\,{\pi }^{2}
-{\tfrac {270677}{6048}}\,\ln ({ z_H})
-{\tfrac {25747}{30}}\,\ln ({ z_H}){ S_0}
\nonumber \\ &&
-{\tfrac {204503}{360}}\,\left (\ln ({ z_H})\right )^{2}
+{\tfrac {163}{2}}\,\ln ({ z_H})\ln ({ m_H^2})
+{\tfrac {89983}{840}}\,\ln ({ m_H^2})
\Biggr ]
\end{eqnarray*}

\begin{eqnarray*}
%part3z
A_2^Z & = & 
{z_H}\,\Biggl[
{\tfrac {3307}{576}}-{\tfrac {7}{32}}\,{S_0}+{\tfrac {729}{32}}\,{S_2}
+{\tfrac {217}{288}}\,{\pi }^{2}+{\tfrac {119}{96}}\,\ln ({z_H})
-{\tfrac {17}{16}}\,\ln ({z_H})\ln ({m_H^2})
-{\tfrac {383}{96}}\,\ln ({m_H^2})
\nonumber \\ &&
+\tfrac{3}{16}\,\ln ({m_H^2}){S_0}
-\tfrac{1}{8}\,\left (\ln ({m_H^2})\right )^{2}\Biggr] 
\nonumber \\ & + & 
z_H^{2}\Biggl[-{\tfrac {14576777}{20736}}-{\tfrac  {2923}{72}}\,{S_3}
+{\tfrac {110455}{72}}\,{S_0}-176\,{S_1}-{\tfrac {349773}{128}}\,{S_2}
-{\tfrac {111577}{1152}}\,{\pi }^{2}-{\tfrac {121}{9}}\,{\pi }^{2}\ln(3)
\nonumber \\ &&
+{\tfrac {8731}{48}}\,\zeta (3)+{\tfrac {25009}{576}}\,\ln ({z_H})
-{\tfrac {87}{16}}\,\ln ({z_H}){S_0}
+ \tfrac{1}{3}\,\left (\ln ({z_H})\right )^{2}
+{\tfrac {1195}{72}}\,\ln ({z_H})\ln ({m_H^2})
\nonumber \\ &&
+{\tfrac {112861}{1728}}\,\ln ({m_H^2})
-{\tfrac {731}{48}}\,\ln ({m_H^2}){S_0}
+{\tfrac {239}{32}}\,\left (\ln ({m_H^2})\right )^{2}\Biggr] 
\nonumber \\ & + & 
z_H^{3}\Biggl[
-{\tfrac {452509}{3200}}+{\tfrac {6145}{144}}\,{S_0}
+{\tfrac {2913}{80}}\,{S_2}+{\tfrac {2279}{270}}\,{\pi }^{2}
-{\tfrac {251521}{1440}}\,\ln ({z_H})+{\tfrac {1943}{12}}\,\ln({z_H}){S_0}
\nonumber \\ &&
+{\tfrac {5623}{288}}\,\left (\ln ({z_H})\right )^{2}
+{\tfrac {485}{16}}\,\ln ({z_H})\ln ({m_H^2})-{\tfrac {2951}{64}}\,\ln({m_H^2})
+{\tfrac {735}{8}}\,\ln ({m_H^2}){S_0}
+{\tfrac {37}{4}}\,\left (\ln ({m_H^2})\right )^{2}\Biggr] 
\nonumber \\ & + & 
z_H^{4}\Biggl[-{\tfrac {106557211}{144000}}
+{\tfrac {3265133}{9600}}\,{S_0}+{\tfrac {41949}{160}}\,{S_2}
+{\tfrac {65291}{1440}}\,{\pi }^{2}-{\tfrac {16156451}{86400}}\,\ln({z_H})
+{\tfrac {51121}{160}}\,\ln ({z_H}){S_0}
\nonumber \\ &&
+{\tfrac {109207}{960}}\,\left (\ln ({z_H})\right )^{2}
+{\tfrac {139}{2}}\,\ln ({z_H})\ln ({m_H^2})
+{\tfrac {6093}{80}}\,\ln ({m_H^2})
+{\tfrac {63}{4}}\,\left (\ln ({m_H^2})\right )^{2}\Biggr ]
\nonumber \\ & + & 
z_H^{5}\Biggl[
-{\tfrac {109475704943}{25401600}}
+{\tfrac {15528551}{7200}}\,{ S_0}
+{\tfrac {1335741}{1120}}\,{ S_2}
+{\tfrac {9380087}{30240}}\,{\pi }^{2}
+{\tfrac {34762379}{151200}}\,\ln ({ z_H})
\nonumber \\ &&
+{\tfrac {188237}{120}}\,\ln ({ z_H}){ S_0}
+{\tfrac {36539}{45}}\,\left (\ln ({ z_H})\right )^{2}
+{\tfrac {10021}{72}}\,\ln ({ z_H})\ln ({ m_H^2})
+{\tfrac {5529703}{30240}}\,\ln ({ m_H^2})
\Biggr ]
\end{eqnarray*}

\begin{eqnarray*}
%part4z
A_3^Z & = & 
{z_H}\,\Biggl[
{\tfrac {53}{24}}+{\tfrac {77}{144}}\,{S_0}+{\tfrac {243}{8}}\,{S_2}
+{\tfrac {95}{108}}\,{\pi }^{2}+{\tfrac {21}{16}}\,\ln ({z_H})
-{\tfrac {9}{8}}\,\ln ({z_H})\ln ({m_H^2})-{\tfrac {15}{16}}\,\ln({m_H^2})
\nonumber \\ &&
-{\tfrac {11}{24}}\,\ln ({m_H^2}){S_0}- \frac{1}{4}\,\left (\ln ({m_H^2})\right)^{2}\Biggr] 
\nonumber \\ & + & 
z_H^{2}\Biggl[{\tfrac {6479761}{5184}}+{\tfrac {365}{54}}\,{S_3}
-{\tfrac {11525}{8}}\,{S_0}+{\tfrac {457}{3}}\,{S_1}+{\tfrac  {102293}{96}}\,{S_2}
+{\tfrac {21419}{288}}\,{\pi }^{2}+{\tfrac {377}{27}}\,{\pi }^{2}\ln(3)
\nonumber \\ &&
-{\tfrac {4889}{36}}\,\zeta (3)-{\tfrac {65147}{432}}\,\ln ({z_H})
+{\tfrac {9091}{108}}\,\ln ({z_H}){S_0}+\tfrac{5}{9}\,\left (\ln ({z_H})\right)^{2}
+{\tfrac {87}{4}}\,\ln ({z_H})\ln ({m_H^2})
\nonumber \\ &&
-{\tfrac {65201}{432}}\,\ln ({m_H^2})
+{\tfrac {2339}{27}}\,\ln ({m_H^2}){S_0}
+{\tfrac {695}{72}}\,\left (\ln ({m_H^2})\right )^{2}\Biggr] 
\nonumber \\ & + & 
z_H^{3}\Biggl[
{\tfrac {315263}{43200}}-{\tfrac {160771}{2592}}\,{S_0}
+{\tfrac {92279}{480}}\,{S_2}-{\tfrac {3983}{4320}}\,{\pi }^{2}
+{\tfrac {187433}{2160}}\,\ln ({z_H})-{\tfrac {12461}{216}}\,\ln({z_H}){S_0}
\nonumber \\ &&
-{\tfrac {437}{72}}\,\left (\ln ({z_H})\right )^{2}
-{\tfrac {99}{8}}\,\ln ({z_H})\ln ({m_H^2})+{\tfrac {9839}{96}}\,\ln({m_H^2})
-{\tfrac {149}{4}}\,\ln ({m_H^2}){S_0}-{\tfrac {29}{2}}\,\left (\ln ({m_H^2})
\right )^{2}\Biggr] 
\nonumber \\ & + & z_H^{4}\Biggl[
{\tfrac {275866331}{1296000}}-{\tfrac  {3508117}{10800}}\,{S_0}
+{\tfrac {13353}{20}}\,{S_2}-{\tfrac {1403}{90}}\,{\pi }^{2}
+{\tfrac {1202389}{10800}}\,\ln ({z_H})-{\tfrac {2932}{15}}\,\ln({z_H}){S_0}
\nonumber \\ &&
-{\tfrac {179}{40}}\,\left (\ln ({z_H})\right )^{2}
+{\tfrac {287}{4}}\,\ln ({z_H})\ln ({m_H^2})+{\tfrac  {15001}{240}}\,\ln ({m_H^2})
+9\,\left (\ln ({m_H^2})\right )^{2}\Biggr] 
\nonumber \\ & + & 
z_H^{5}\Biggl[
 {\tfrac {178942516547}{63504000}}
-{\tfrac {871139}{324}}\,{ S_0}
+{\tfrac {8825401}{3360}}\,{ S_2}
-{\tfrac {6422629}{30240}}\,{\pi }^{2}
-{\tfrac {5656397}{50400}}\,\ln ({ z_H})
\nonumber \\ &&
-{\tfrac {162991}{108}}\,\ln ({ z_H}){ S_0}
-{\tfrac {11701}{24}}\,\left (\ln ({ z_H})\right )^{2}
+{\tfrac {7087}{36}}\,\ln ({ z_H})\ln ({ m_H^2})
+{\tfrac {3910009}{15120}}\,\ln ({ m_H^2})
\Biggr]
\end{eqnarray*}

\begin{eqnarray*}
%part5z
A_4^Z & = & 
{z_H}\,\Biggl[
{\tfrac {7}{9}}+{\tfrac {371}{864}}\,{S_0}+{\tfrac {1215}{32}}\,{S_2}
+{\tfrac {869}{864}}\,{\pi }^{2}+{\tfrac {133}{96}}\,\ln ({z_H})
-{\tfrac {19}{16}}\,\ln ({z_H})\ln ({m_H^2})
+{\tfrac {5}{16}}\,\ln ({m_H^2})
\nonumber \\ &&
-{\tfrac {53}{144}}\,\ln({m_H^2}){S_0}
-\tfrac{3}{8}\,\left (\ln ({m_H^2})\right )^{2}\Biggr] 
\nonumber \\ & + & 
z_H^{2}\Biggl[-{\tfrac {12699875}{20736}}+{\tfrac  {793}{216}}\,{S_3}
+{\tfrac {1537817}{2592}}\,{S_0}-{\tfrac {365}{9}}\,{S_1}
-{\tfrac {22477}{128}}\,{S_2}-{\tfrac {1229275}{31104}}\,{\pi }^{2}
-{\tfrac {95}{27}}\,{\pi }^{2}\ln (3)
\nonumber \\ &&
+{\tfrac {3767}{144}}\,\zeta (3)
+{\tfrac {28561}{1728}}\,\ln ({z_H})-{\tfrac {18155}{1296}}\,\ln({z_H}){S_0}
+{\tfrac {7}{9}}\,\left (\ln ({z_H})\right )^{2}
+{\tfrac {1937}{72}}\,\ln ({z_H})\ln ({m_H^2})
\nonumber \\ &&
+{\tfrac {34757}{1728}}\,\ln ({m_H^2})
-{\tfrac {17485}{1296}}\,\ln ({m_H^2}){S_0}
+{\tfrac {3409}{288}}\,\left (\ln ({m_H^2})\right )^{2}\Biggr] 
\nonumber \\ & + & 
z_H^{3}\Biggl[-{\tfrac {3647159}{25920}}+{\tfrac  {142763}{7776}}\,{S_0}
+{\tfrac {23303}{96}}\,{S_2}+{\tfrac {4529}{2592}}\,{\pi }^{2}
-{\tfrac {1921}{144}}\,\ln ({z_H})+{\tfrac {743}{324}}\,\ln({z_H}){S_0}
\nonumber \\ &&
-{\tfrac {1775}{288}}\,\left (\ln ({z_H})\right )^{2}
-{\tfrac {209}{16}}\,\ln ({z_H})\ln ({m_H^2})
+{\tfrac {9841}{192}}\,\ln ({m_H^2})+{\tfrac {53}{24}}\,\ln({m_H^2}){S_0}
-{\tfrac {69}{4}}\,\left (\ln ({m_H^2})\right )^{2}\Biggr] 
\nonumber \\ & + & 
z_H^{4}\Biggl[-{\tfrac {696762899}{1296000}}
+{\tfrac {36167453}{259200}}\,{S_0}+{\tfrac {832139}{960}}\,{S_2}
+{\tfrac {38173}{8640}}\,{\pi }^{2}-{\tfrac {1862941}{17280}}\,\ln({z_H})
+{\tfrac {56477}{1440}}\,\ln ({z_H}){S_0}
\nonumber \\ &&
+{\tfrac {86023}{2880}}\,\left (\ln ({z_H})\right )^{2}
+92\,\ln ({z_H})\ln ({m_H^2})+{\tfrac {21263}{240}}\,\ln ({m_H^2})
+{\tfrac {45}{4}}\,\left (\ln ({m_H^2})\right )^{2}\Biggr] 
\nonumber \\ & + & 
z_H^{5}\Biggl[
-{\tfrac {430433439517}{127008000}}
+{\tfrac {343121543}{194400}}\,{ S_0}
+{\tfrac {3744143}{1120}}\,{ S_2}
+{\tfrac {49949}{672}}\,{\pi }^{2}
-{\tfrac {54474521}{151200}}\,\ln ({ z_H})
\nonumber \\ &&
+{\tfrac {2240561}{3240}}\,\ln ({ z_H}){ S_0}
+{\tfrac {11753}{60}}\,\left (\ln ({ z_H})\right )^{2}
+{\tfrac {6109}{24}}\,\ln ({ z_H})\ln ({ m_H^2})
+{\tfrac {3370111}{10080}}\,\ln ({ m_H^2})
\Biggr]
\end{eqnarray*}

\begin{eqnarray*}
%part6z
A_5^Z & = & 
{z_H}\,\Biggl[
-{\tfrac {281}{720}}+{\tfrac {7}{24}}\,{S_0}+{\tfrac {729}{16}}\,{S_2}
+{\tfrac {163}{144}}\,{\pi }^{2}+{\tfrac {35}{24}}\,\ln ({z_H})
-\tfrac{5}{4}\,\ln ({z_H})\ln ({m_H^2})
\nonumber \\ &&
+{\tfrac {107}{80}}\,\ln ({m_H^2})
-\tfrac{1}{4}\,\ln ({m_H^2}){S_0}
-\tfrac{1}{2}\,\left (\ln ({m_H^2})\right )^{2}\Biggr] 
\nonumber \\ & + & 
z_H^{2}\Biggl[{\tfrac {33773}{384}}-{\tfrac {1}{108}}\,{S_3}
-{\tfrac {2421059}{19440}}\,{S_0}-{\tfrac {103}{54}}\,{S_1}
+{\tfrac {219449}{2880}}\,{S_2}+{\tfrac {1585397}{233280}}\,{\pi }^{2}
+\tfrac{1}{27}\,{\pi }^{2}\ln (3)
\nonumber \\ &&
-{\tfrac {23}{72}}\,\zeta (3)
-{\tfrac {20023}{1440}}\,\ln ({z_H})-{\tfrac {19}{24}}\,\ln({z_H}){S_0}
+\left (\ln ({z_H})\right )^{2}+{\tfrac {577}{18}}\,\ln ({z_H})\ln ({m_H^2})
\nonumber \\ &&
-{\tfrac {13429}{1440}}\,\ln ({m_H^2})-{\tfrac {493}{648}}\,\ln({m_H^2}){S_0}
+{\tfrac {673}{48}}\,\left (\ln ({m_H^2})\right )^{2}\Biggr] 
\nonumber \\ & + & 
z_H^{3}\Biggl[
-{\tfrac {32327203}{259200}}+{\tfrac {90853}{38880}}\,{S_0}
+{\tfrac {202459}{720}}\,{S_2}+{\tfrac {17237}{6480}}\,{\pi }^{2}
-{\tfrac {33671}{4320}}\,\ln ({z_H})-{\tfrac {311}{648}}\,\ln({z_H}){S_0}
\nonumber \\ &&
-{\tfrac {901}{144}}\,\left (\ln ({z_H})\right )^{2}
-{\tfrac {55}{4}}\,\ln ({z_H})\ln ({m_H^2})
+{\tfrac {16597}{240}}\,\ln ({m_H^2})-{\tfrac {7}{18}}\,\ln({m_H^2}){S_0}
-20\,\left (\ln ({m_H^2})\right )^{2}\Biggr] 
\nonumber \\ & + & 
z_H^{4} \Biggl[-{\tfrac {81023989}{216000}}-{\tfrac
  {727891}{43200}}\,{S_0}+{\tfrac {164217}{160}}\,{S_2}
+{\tfrac {10243}{1440}}\,{\pi }^{2}-{\tfrac {78513}{1600}}\,\ln({z_H})
-{\tfrac {3853}{2160}}\,\ln ({z_H}){S_0}
\nonumber \\ &&
+{\tfrac {53587}{1440}}\,\left (\ln ({z_H})\right )^{2}
+{\tfrac {449}{4}}\,\ln ({z_H})\ln ({m_H^2})+{\tfrac {8863}{80}}\,\ln
({m_H^2})+{\tfrac {27}{2}}\,\left (\ln ({m_H^2})\right )^{2}\Biggr] 
\nonumber \\ & + & 
z_H^{5} \Biggl[
-{\tfrac {23064502157}{31752000}}
-{\tfrac {107194699}{194400}}\,{ S_0}
+{\tfrac {19739897}{5040}}\,{ S_2}
+{\tfrac {361861}{15120}}\,{\pi }^{2}
+{\tfrac {4365419}{151200}}\,\ln ({ z_H})
\nonumber \\ &&
-{\tfrac {206327}{1620}}\,\ln ({ z_H}){ S_0}
+{\tfrac {25421}{360}}\,\left (\ln ({ z_H})\right )^{2}
+{\tfrac {2810}{9}}\,\ln ({ z_H})\ln ({ m_H^2})
+{\tfrac {1550081}{3780}}\,\ln ({ m_H^2})
\Biggr]
\end{eqnarray*}

%%%%%%%%%%%%%%%%%%%%%%%%%%%%%%%%%%%%%%%%%%%%%%%%%%%%%%%%%%%%%%%%%%%%%%%%%%%%%%

%%%%%%%%%%%%%%%%%%%%%%%%%%%%%%%%%%%%%%%%%%%%%%%%%%%%%%%%%%%%%%%%%%%%%%%%%%%%
\newpage

\begin{figure}[t]
\vskip -20mm
\vspace*{-10mm}
\centerline{\vbox{\epsfysize=80mm \epsfbox{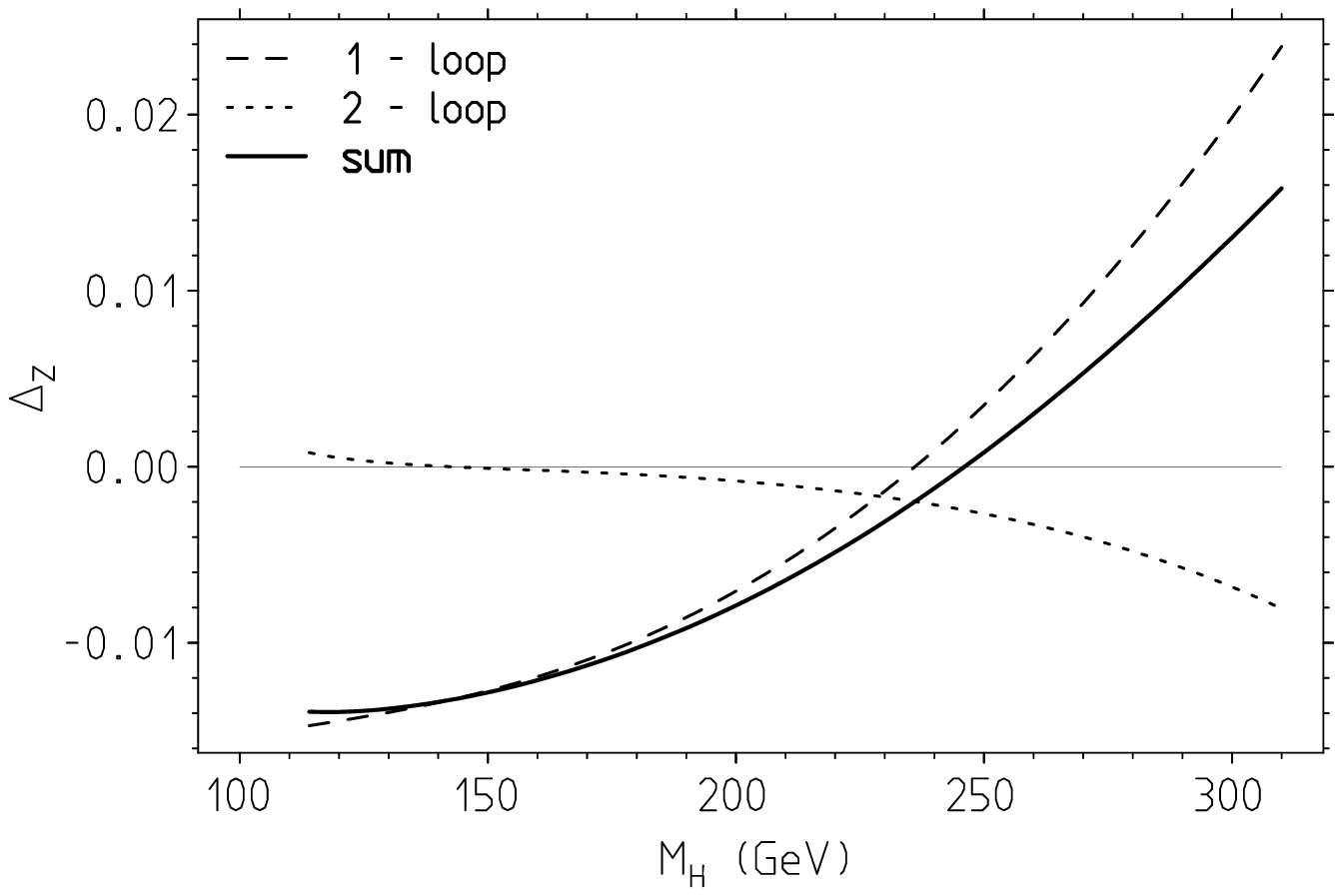}}}
\caption{\label{zlight}
One- and two-loop corrections to the relation
$\Delta_Z \equiv m^2_Z(M_Z)/M^2_Z-1 $
as a function of the Higgs mass $M_H$ for intermediate Higgs masses.}
\end{figure}

\begin{figure}[b]
%\vskip -20mm
\vspace*{-10mm}
\centerline{\vbox{\epsfysize=80mm \epsfbox{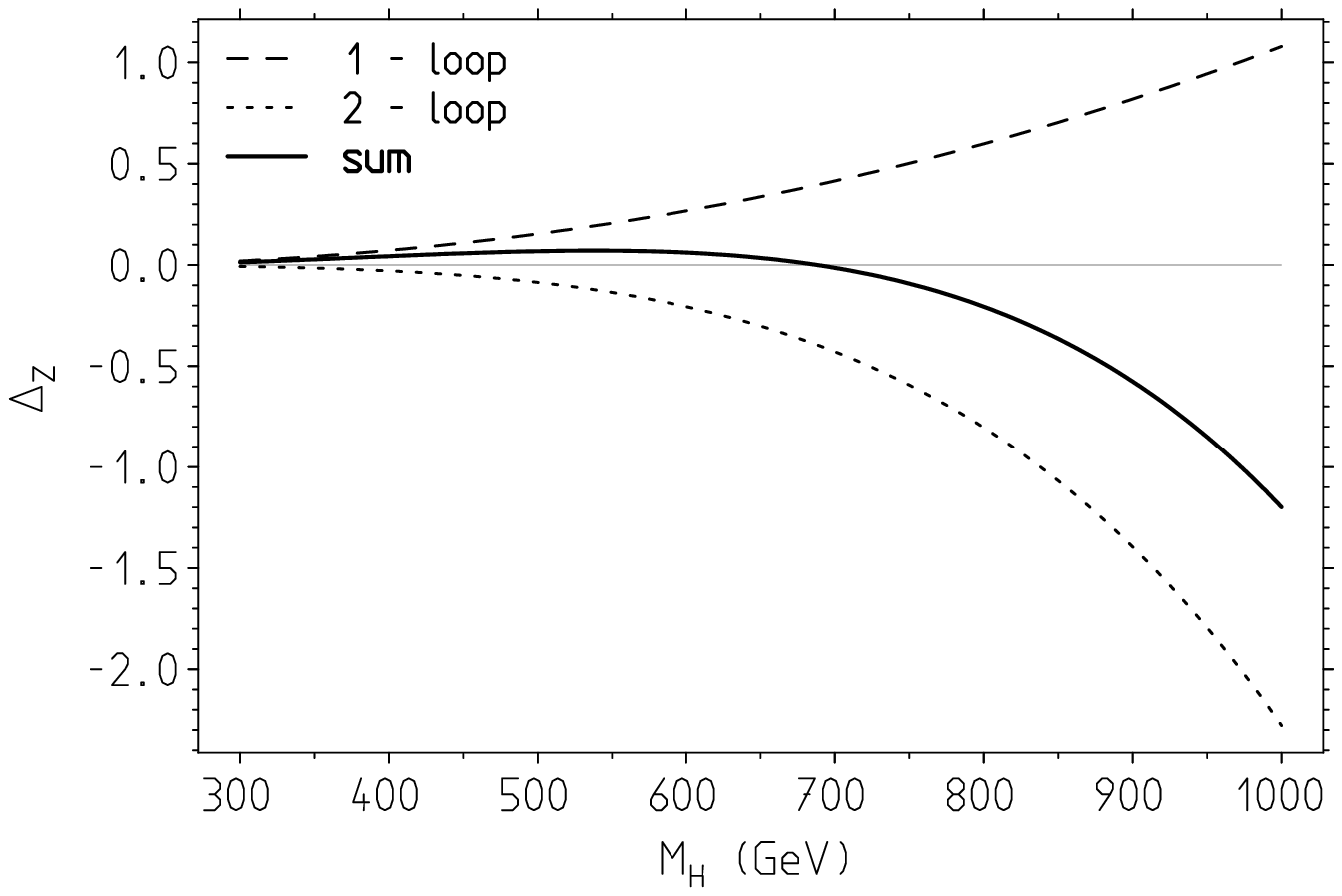}}}
\caption{\label{zheavy}
One- and two-loop corrections to the relation
$\Delta_Z \equiv m^2_Z(M_Z)/M^2_Z-1 $
as a function of the Higgs mass $M_H$ for heavy Higgs masses.}
\end{figure}

%%%%%%%%%%%%%%%%%%%%%%%%%%%%%%%%%%%%%%%%%%%%%%%%%%%%%%%%%%%%%%%%%%%%%%%%%%%%
\begin{figure}[t]
\vspace*{-10mm}
\centerline{\vbox{\epsfysize=80mm \epsfbox{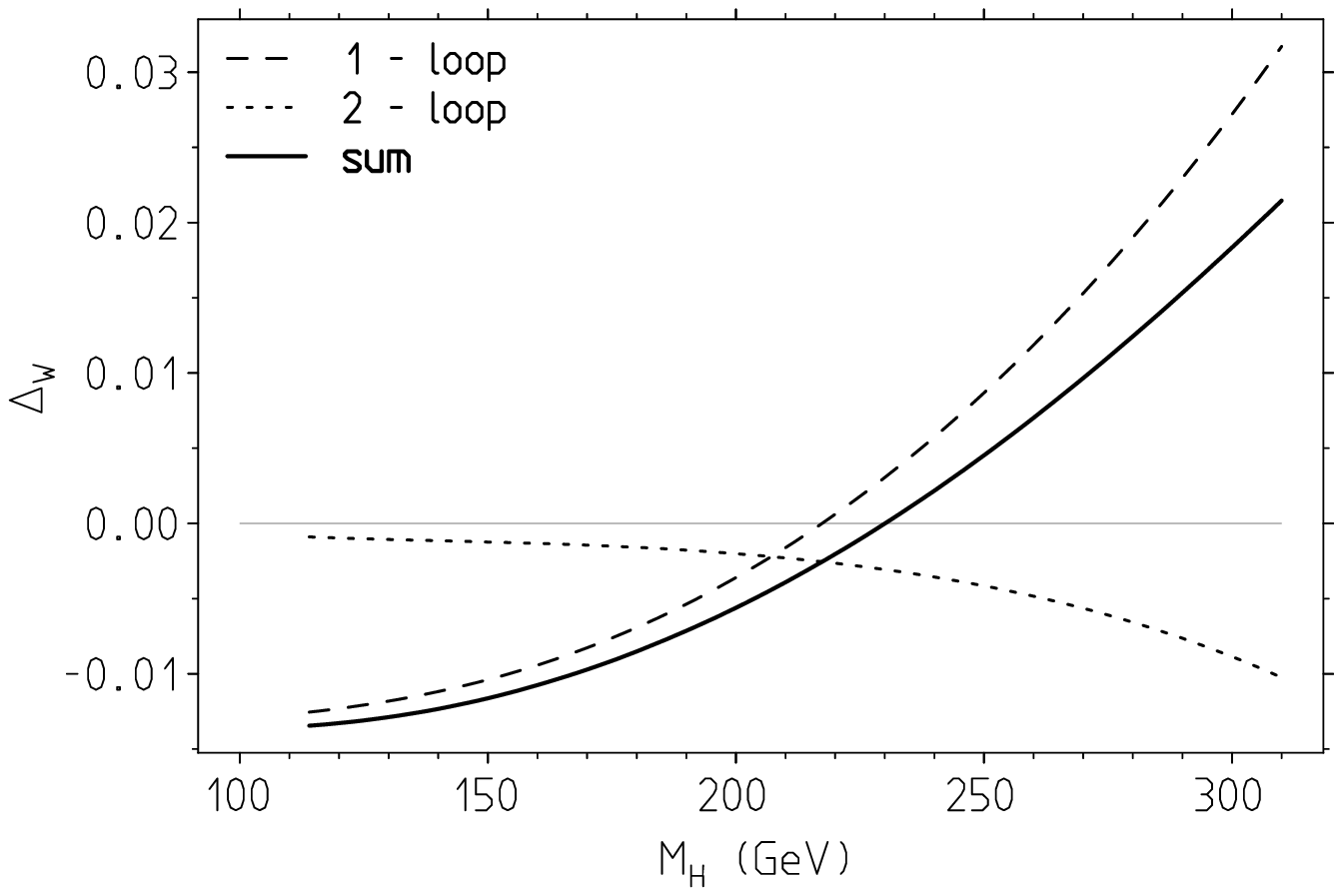}}}
\caption{\label{wlight} 
One- and two-loop corrections to the relation
$\Delta_W \equiv m^2_W(M_W)/M^2_W-1$ 
as a function  of the Higgs mass $m_H$ for intermediate Higgs masses.
%Dashed and dot lines correspond to the one- and two-loop
%corrections, respectively. The thick line is their sum.
}
\end{figure}

\begin{figure}[b]
\vspace*{-10mm}
\centerline{\vbox{\epsfysize=80mm \epsfbox{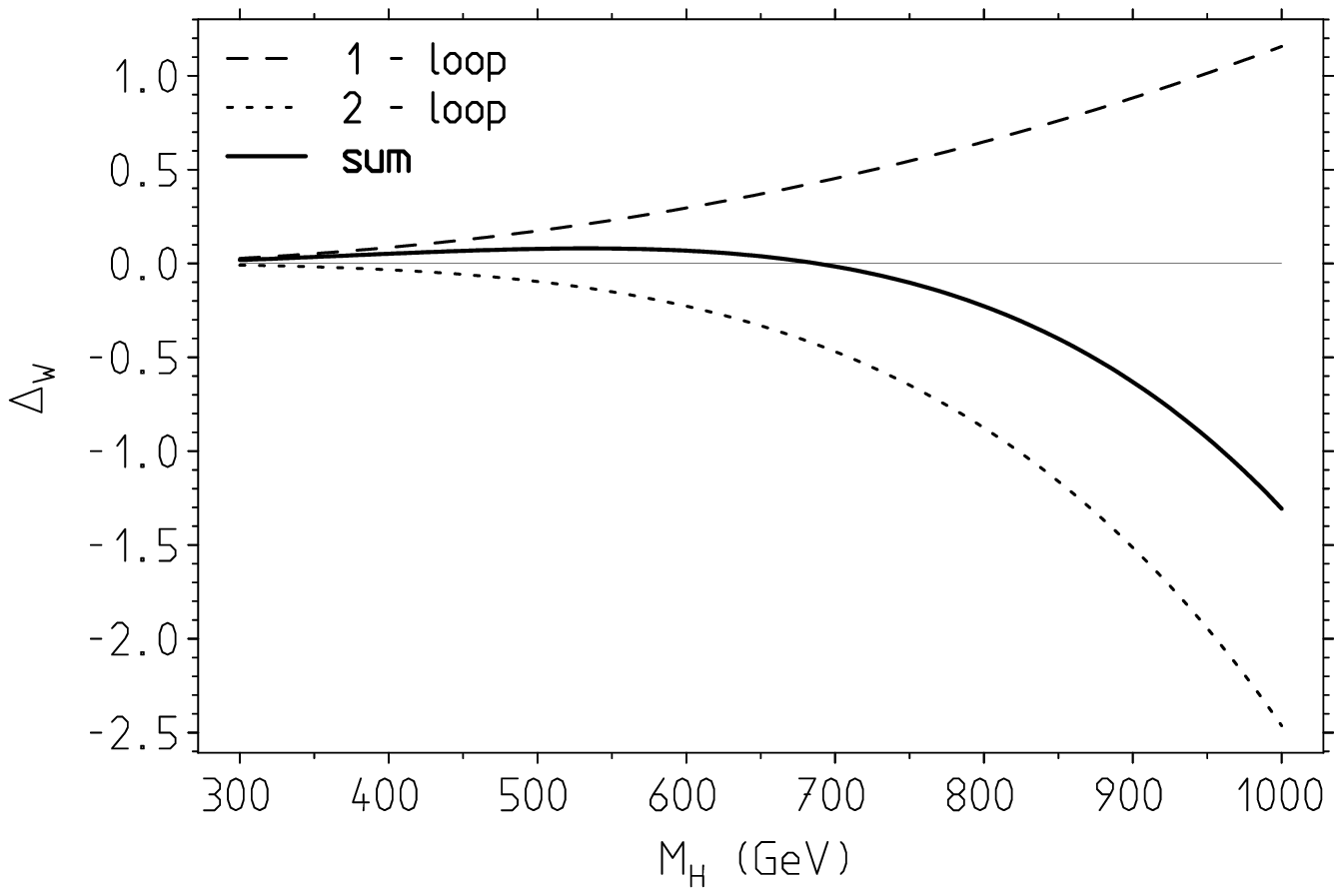}}}
\caption{\label{wheavy}
One- and two-loop corrections to the relation
%$\Biggl[ m^2_W(M_W)/M^2_W-1 \Biggr]$ 
$\Delta_W \equiv m^2_W(M_W)/M^2_W-1$ 
as a function of the Higgs mass $m_H$ for heavy Higgs masses.
%Dashed and dot lines correspond to the one- and two-loop
%corrections, respectively. The thick line is their sum.
}
\end{figure}

\begin{figure}[t]
\vspace*{-10mm}
\centerline{\vbox{\epsfysize=80mm \epsfbox{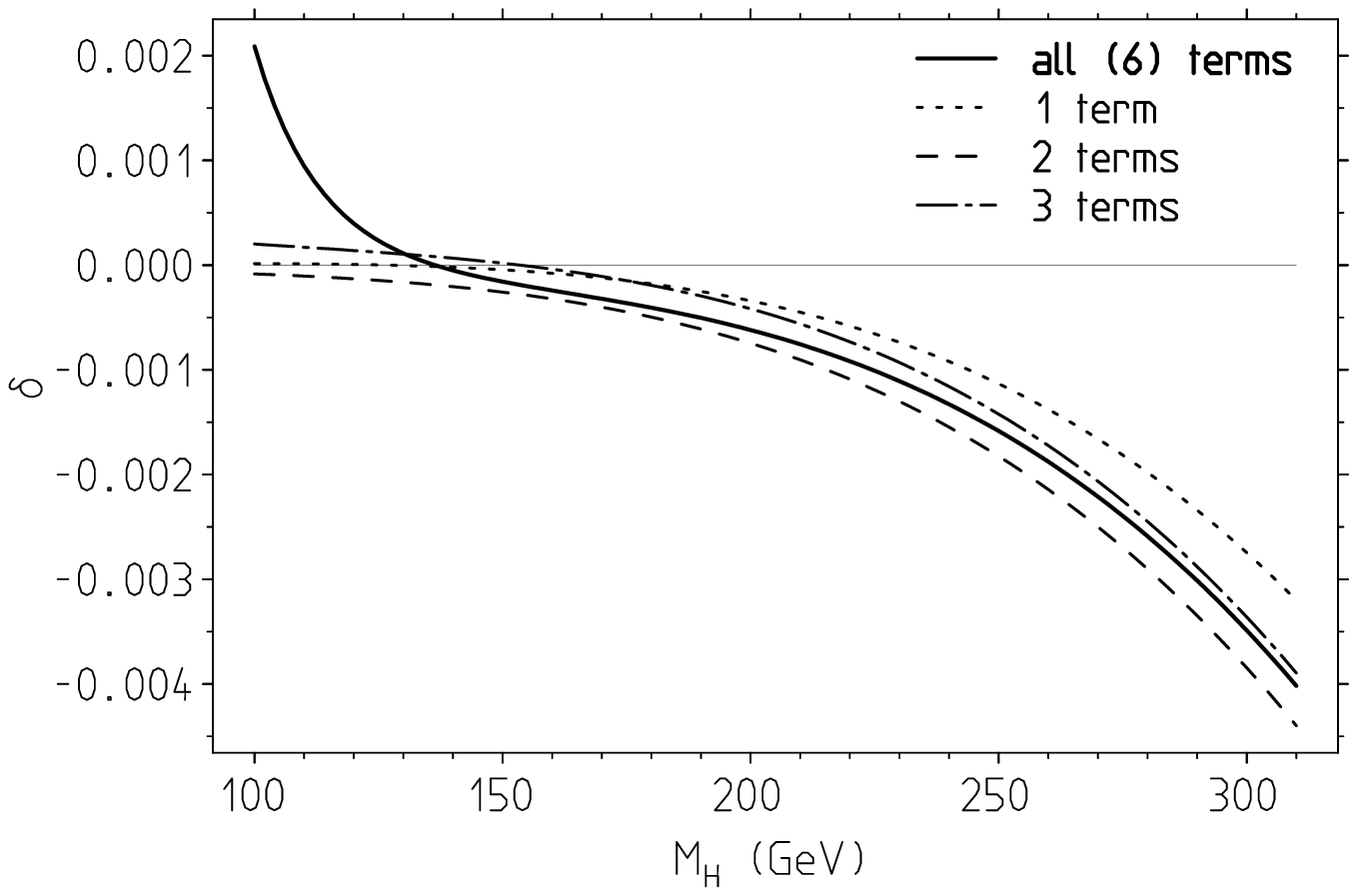}}}
\caption{\label{z-higgs-light}
The dependence on the number of coefficients of the expansion
(\ref{result}) used for the evaluation of the two-loop corrections.
We show 
$\delta \equiv 
-\:\left(\hat{\Pi}_Z^{(2)} + \hat{\Pi}_Z^{(1)}\hat{\Pi}_Z^{(1)}{}'\right)/M_Z^2$ 
(see \ref{MS2:subtracted}) as a 
function of the Higgs mass. The dotted, dashed, dot-dashed and full lines show
results obtained with the first one, two, three and all calculated (six)
coefficients, respectively.}
\end{figure}

\begin{figure}[b]
\vspace*{-10mm}
\centerline{\vbox{\epsfysize=80mm \epsfbox{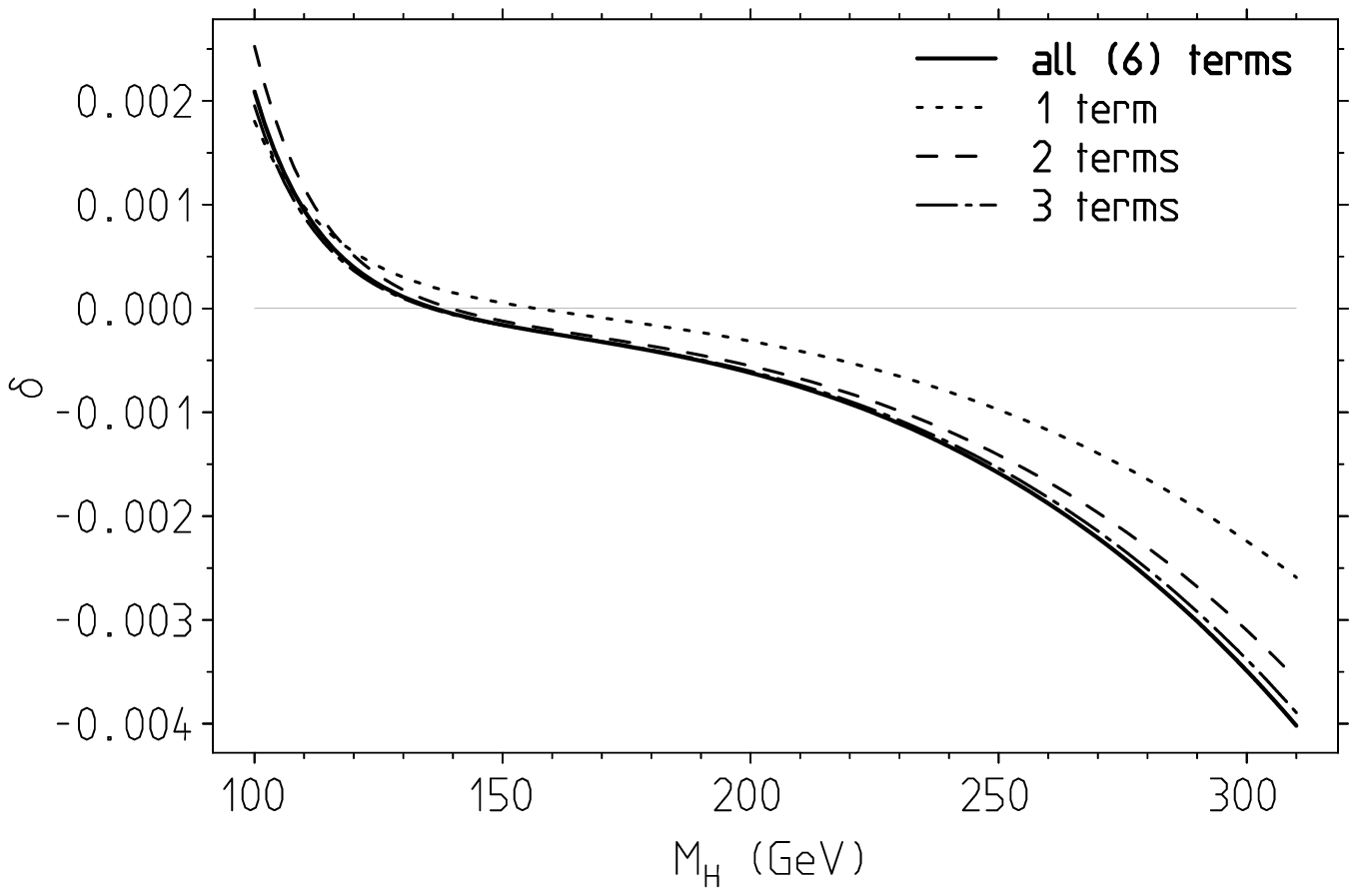}}}
\caption{\label{z-sin-light}
The dependence on the number of coefficients of the expansion with
respect to $\sin^2\theta_W$ of the two-loop corrections.
We show $\delta \equiv -\:\left(\hat{\Pi}_Z^{(2)} + \hat{\Pi}_Z^{(1)}
\hat{\Pi}_Z^{(1)}{}'\right)/M_Z^2$ (see \ref{MS2:subtracted})
as a function of the Higgs mass. The dotted, dashed, dot-dashed and full lines show
results obtained with the first one, two, three and all calculated (six)
coefficients, respectively.}
\end{figure}

\begin{figure}[t]
\vspace*{-10mm}
\centerline{\vbox{\epsfysize=80mm \epsfbox{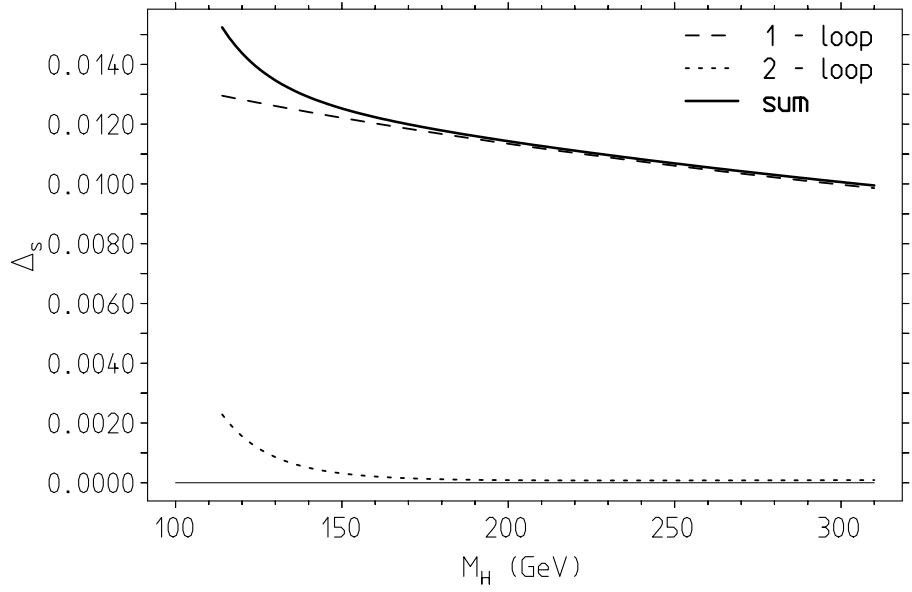}}}
\caption{\label{sin-light} 
One- and two-loop corrections to $\Delta_s=
\sin^2\theta^{\msb}_W/\sin^2\theta^{\rm OS}_W-1$ (see (\ref{sinus}))
 as a function of the Higgs mass $m_H$ for intermediate Higgs masses ($\mu=M_Z$).
%Dashed and dot lines correspond to the one- and two-loop
%corrections, respectively. The thick line is their sum.
}
\end{figure}

\begin{figure}[b]
\vspace*{-10mm}
\vspace*{-10mm}
\centerline{\vbox{\epsfysize=80mm \epsfbox{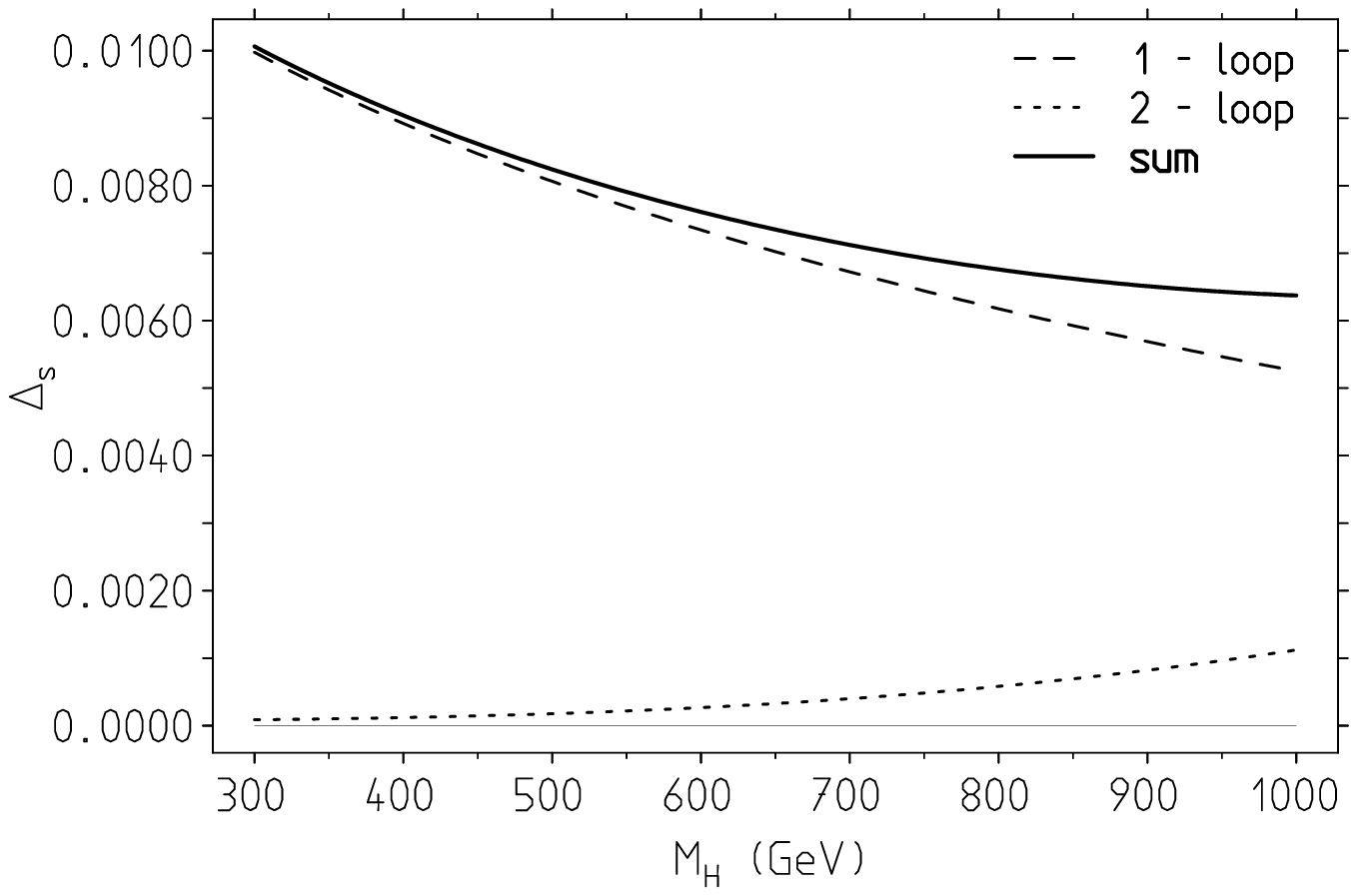}}}
\caption{\label{sin-heavy} 
One- and two-loop corrections to $\Delta_s=
\sin^2\theta^{\msb}_W/\sin^2\theta^{\rm OS}_W-1$
(see (\ref{sinus}))
as a function  of the Higgs mass $m_H$ for heavy Higgs masses ($\mu=M_Z$).
%Dashed and dot lines correspond to the one- and two-loop
%corrections, respectively. The thick line is their sum.
}
\end{figure}
%%%%%%%%%%%%%%%%%%%%%%%%%%%%%%%%%%%%%%%%%%%%%%%%%%%%%%%%%%%%%%%%%%%%%%%%%%%


\begin{thebibliography}{99}

\bibitem{Abbaneo:2001ix}
D.~Abbaneo et al.  [ALEPH, DELPHI, L3 and OPAL Collaborations,
LEP Electroweak Working Group, SLD Heavy Flavor and Electroweak Groups],
%``A combination of preliminary electroweak measurements and constraints  on the standard model,''
hep-ex/0112021.
%%CITATION = HEP-EX 0112021;%%

\bibitem{YR89}
M. Consoli, W. Hollik, F. Jegerlehner, in {\em Z Physics at LEP1}, \\
eds. G. Altarelli et al., CERN 89-08 (1989) 1. \\
For a more recent review see \\
D.~Bardin and G.~Passarino, 
{\it The Standard Model in the Making}, \\ Oxford, UK: Clarendon, 1999.

\bibitem{Aguilar-Saavedra:2001rg}
J.~A.~Aguilar-Saavedra et al.  [ECFA/DESY LC Physics Working Group Collaboration],
%``TESLA Technical Design Report Part III: Physics at an e+e- Linear Collider,''
hep-ph/0106315.
%%CITATION = HEP-PH 0106315;%%

\bibitem{QED-running}
S.~Eidelman and F.~Jegerlehner, 
Z. Phys. {\bf C67} (1995) 585; \\
%%CITATION = HEP-PH 9502298;%% 
M.~Steinhauser, Phys.Lett. {\bf B429} (1998) 158.
%%CITATION = HEP-PH 9803313;%%

\bibitem{LTL}
J.J.~van der Bij and M.~Veltman,  Nucl. Phys. {\bf B231} (1984) 205; \\
%%CITATION = NUPHA,B231,205;%%
J.J.~van der Bij,  Nucl. Phys. {\bf B248} (1984) 141; \\
%%CITATION = NUPHA,B248,141;%%
J.J.~van der Bij and F.~Hoogeven,  Nucl. Phys. {\bf B283} (1987) 477; \\
%%CITATION = NUPHA,B283,477;%%
A.~Djouadi, Nuovo Cim. {\bf A100} (1988) 357; \\
%%CITATION = NUCIA,A100,357;%%
M.~Consoli, W.~Hollik and  F.~Jegerlehner,
Phys. Lett. {\bf B227} (1989) 167; \\
%%CITATION = PHLTA,B227,167;%%
B.~A.~Kniehl, Nucl. Phys. {\bf B347} (1990) 86;\\
%%CITATION = NUPHA,B347,86;%%
F.~Halzen and  B.A.~Kniehl, Nucl. Phys. {\bf B353} (1991) 567; \\
%%CITATION = NUPHA,B353,567;%%
%J.~Fleischer, O.~V.~Tarasov, F.~Jegerlehner and P.~Raczka, 
J.~Fleischer et. al.,
Phys. Lett. {\bf B293} (1992) 437; \\
%%CITATION = PHLTA,B293,437;%%
K.~G.~Chetyrkin, A.~Kwiatkowski and M.~Steinhauser, 
Mod. Phys. Lett. {\bf A8} (1993) 2785; \\
%%CITATION = MPLAE,A8,2785;%%
G.~Degrassi, Nucl. Phys. {\bf B407} (1993) 271; \\
%%CITATION = HEP-PH 9302288;%%
G.~Buchalla and  A.~J.~Buras, Nucl. Phys. {\bf B398} (1993) 285; \\
%%CITATION = NUPHA,B398,285;%%
%R.~Barbieri, M.~Beccaria, P.~Ciafalani, G.~Curci and A.~Vicere,
R.~Barbieri et al.,
Phys. Lett. {\bf B288} (1992) 95; {\bf B312} (1993) 511(E); \\
%CITATION = PHLTA,B288,95;%%
Nucl. Phys. {\bf B409} (1993) 105; \\
%%CITATION = NUPHA,B409,105;%%
J.~Fleischer, O.V.~Tarasov and F.~Jegerlehner, 
Phys. Lett. {\bf B319} (1993) 249; \\
%%CITATION = PHLTA,B319,249;%%
%L.~Avdeev, J.~Fleischer, S.~Mikhailov and O.~Tarasov,
L.~Avdeev et al.,
Phys.Lett. {\bf B336} (1994) 560; {\bf B349} (1995) 597(E); \\
%%CITATION = HEP-PH 9406363;%%
K.~G.~Chetyrkin,  J.H.~K\"uhn and  M.~Steinhauser, 
Phys. Lett. {\bf B351} (1995) 331; \\
%%CITATION = HEP-PH 9502291;%%
K.G.~Chetyrkin, J.H.~K\"uhn and A.~Kwiatkowski,
Phys. Lett. {\bf B282} (1992) 221;\\
%%CITATION = PHLTA,B282,221;%%
Phys. Rept. {\bf 277} (1996) 189;\\
%%CITATION = PRPLC,277,189;%%
A.~Czarnecki and  J.~H.~K\"uhn, 
Phys. Rev. Lett. {\bf 77} (1996) 395; \\
%%CITATION = HEP-PH 9608366;%%
R.~Harlander, T.~Seidensticker and M.~Steinhauser,
Phys. Lett. {\bf B426} (1998) 125;\\
%%CITATION = HEP-PH 9712228;%%
%J.~Fleischer, F.~Jegerlehner, M.~Tentyukov and  O.~Veretin,
J.~Fleischer et al.,
Phys. Lett. {\bf B459} (1999) 625; \\
%%CITATION = PHLTA,B459,625;%%
G.~Degrassi, S.~Fanchiotti and P.~Gambino,
%``Current algebra approach to heavy top effects in delta (rho),''
Int.\ J.\ Mod.\ Phys.{\bf A10} (1995) 1377 \\
%%CITATION = HEP-PH 9403250;%%
%G.~Degrassi, S.~Fanchiotti, F.~Feruglio, B.P.~Gambino and A.~Vicini, 
G.~Degrassi et al.,
Phys.\ Lett. {\bf B350} (1995) 75; \\
%%CITATION = PHLTA,B350,75;%%
T. van Ritbergen and  R.G.~Stuart,
Phys. Rev. Lett. {\bf 82} (1999) 488; \\
%%CITATION = HEP-PH 9808283;%%
D.~Bardin et al., 
Comput. Phys. Commun. {\bf 133} (2001) 229.
%%CITATION = HEP-PH 9908433;%%


\bibitem{2loop-rho}
%S.~Fanchiotti, F.~Feruglio, B.P.~Gambino and A.~Vicini, 
G.~Degrassi et al., 
Phys.\ Lett. {\bf B350} (1995) 75.
%%CITATION = PHLTA,B350,75;%%

\bibitem{2loop-muon:SM}
G.~Degrassi, P.~Gambino and A.~Vicini, 
Phys. Lett. {\bf B383} (1996) 219; \\
%%CITATION = PHLTA,B383,219;%%
G.~Degrassi, P.~Gambino and A.~Sirlin, 
Phys. Lett. {\bf B394} (1997) 188; \\
%%CITATION = PHLTA,B394,188;%%
%G.~Degrassi, P.~Gambino, M.~Passera  and A.~Sirlin, 
G.~Degrassi et al.,
Phys. Lett. {\bf B418} (1997) 209; \\
%%CITATION = PHLTA,B418,209;%%
S.~Bauberger and G.~Weiglein, 
Phys. Lett. {\bf B419} (1998) 333; \\
%%CITATION = PHLTA,B419,333;%%
%A.~Freitas, W.~Hollik, W.~Walter and  G.~Weiglein, 
A.~Freitas et al.,
Phys.\ Lett.\  {\bf B495} (2000) 338.
%%CITATION = PHLTA,B495,338;%%

\bibitem{2loop:Z}
G.~Degrassi and  P.~ Gambino, Nucl.\ Phys.\ {\bf B567} (2000) 3.
%%CITATION = NUPHA,B567,3;%%

\bibitem{2loop-muon:QED}
T. van Ritbergen and  R.G.~Stuart, 
Phys. Rev. Lett. {\bf 82} (1999) 488;\\ 
%%CITATION = PRLTA,82,488;%%
Nucl. Phys. {\bf B564} (2000) 343; \\
%%CITATION = NUPHA,B564,343;%%
M.~Steinhauser and T.~Seidensticker, 
Phys. Lett. {\bf B467} (1999) 271.
%%CITATION = PHLTA,B467,271;%%

\bibitem{FJLL94}
F.~Jegerlehner,
Nucl.\ Phys.\ Proc.\ Suppl.\  {\bf 37B} (1994) 129.
%%CITATION = NUPHZ,37B,129;%%

\bibitem{ASLP99}
A.~Sirlin,
%``Ten years of precision electroweak physics,''
in {\it Proc. of the 19th Intl. Symp. on Photon and Lepton 
Interactions at High Energy LP99 } 
ed. J.A. Jaros and M.E. Peskin, Int.\ J.\ Mod.\ Phys.\ A {\bf 15S1} (2000) 398.
%%CITATION = HEP-PH 9912227;%%

\bibitem{asymptotic}
V.A.~Smirnov, 
Commun. Math. Phys. 134 (1990), 109;
%%CITATION = CMPHA,134,109;%%
Mod. Phys. Lett. {\bf A10} (1995) 1485;
%%CITATION = MPLAE,A10,1485;%%
{\it Renormalization and asymptotic expansions}
(Bikrh\"auser, Basel, 1991).

\bibitem{2loop-muon:numeric}
P.~Gambino, A.~Sirlin and G.~Weiglein, J.High Energy Phys. {\bf 04} (1999) 025.
%%CITATION = JHEPA,9904,025;%%

\bibitem{onshell-scheme}
G.~'t Hooft and M.~Veltman, Nucl.\ Phys.\ {\bf B50} (1972) 318;\\
%%CITATION = NUPHA,B50,318;%%
D.A.~Ross and  J.C.~Taylor, 
Nucl.\ Phys.\ {\bf B51} (1973) 125; \\
%%CITATION = NUPHA,B51,125;%%
D.Yu.~Bardin, P.~Khristova and  O.M.~Fedorenko, 
Nucl. Phys. {\bf B175} (1980) 435; \\
%%CITATION = NUPHA,B175,435;%%
Nucl. Phys. {\bf B197} (1982) 1; \\
%%CITATION = NUPHA,B197,1;%%
S.~Sakakibara, 
Phys.\ Rev.\ {\bf D24} (1981) 1149; \\
%%CITATION = PHRVA,D24,1149;%%
%K.~Aoki, Z.~Hioki, R.~Kawabe, M.~Konuma and T.~Muta, 
K.~Aoki et al.,
Prog. Theor. Phys. Suppl. {\bf 73} (1982) 1; \\
%%CITATION = PTPSA,73,1;%%
M.~B\"ohm, H.~Spiesberger and W.~Hollik, 
Fortsch.\ Phys.\ {\bf 34} (1986) 687; \\
%%CITATION = FPYKA,34,687;%%
A. Denner, Fortsch.\ Phys.\ {\bf 41} (1993) 307.
%%CITATION = FPYKA,41,307;%%

\bibitem{Sirlin}
A. Sirlin, Phys. Rev. {\bf D22} (1980) 971.
%%CITATION = PHRVA,D22,971;%%

\bibitem{FJ}
J.~Fleischer and F.~Jegerlehner, Phys. Rev. {\bf D23} (1981) 2001;\\
%%CITATION = PHRVA,D23,2001;%%
F.~Jegerlehner, in ``Radiative Corrections in $SU(2)_L \times U(1)$'',\\
eds. B.W.~Lynn, J.F.~Wheater, World Scientific Publ., Singapore, 1984.
%in: "Testing of the Standard Model", Eds. M. Zra\l ek, R. Manka, World Scientific, Singapore,  1988;
%in: "Testing the Standard Model", Eds. M. Cvetic, P. Langacker, World Scientific, Singapore, 1991.

\bibitem{ftt}
J.~Fleischer,  O.~V.~Tarasov and  M.~Tentyukov, 
Nucl. Phys. Proc. Suppl. {\bf 89} (2000) 112.
%%CITATION = NUPHZ,89,112;%%

\bibitem{bubble}
C.~Ford, I.~Jack and D.R.T.~Jones, Nucl.\ Phys.\ {\bf B387} (1992) 373;
{\bf B504} (1997) 551(E); \\ 
%%CITATION = NUPHA,B387,373;%%
A.I.~Davydychev and J.B.~Tausk, Nucl.\ Phys.\ {\bf B397} (1993) 123.
%%CITATION = NUPHA,B397,123;%%

\bibitem{2loop-fermion}
G.~Weiglein, R.~Scharf and M. B\"ohm, Nucl. Phys. {\bf B416} (1994) 606.
%%CITATION = NUPHA,B416,606;%%

\bibitem{2loop-analytic-a}
D.J.~Broadhurst, J.~Fleischer and O.V.~Tarasov,
Z.\ Phys.\ {\bf C60} (1993) 287; \\
%%CITATION = ZEPYA,C60,287;%%
%F.A.~Berends, M.~B\"ohm, M.~Buza and R.~Scharf, 
F.A.~Berends et. al., Z.\ Phys.\ {\bf C63} (1994) 227; \\
%%CITATION = ZEPYA,C63,227;%%
S.~Bauberger, F.A.~Berends, M.~B\"ohm and M.~Buza, 
Nucl. Phys. {\bf B434}  (1995) 383. 
%%CITATION = NUPHA,B434,383;%%

\bibitem{2loop-analytic-b}
R.~Scharf and J.B.~Tausk, Nucl. Phys. {\bf B412} (1994) 523.
%%CITATION = NUPHA,B412,523;%%

\bibitem{2loop-numeric}
F.A.~Berends and J.B.~Tausk, Nucl. Phys. {\bf B421}  (1994) 456; \\
%%CITATION = NUPHA,B421,456;%% 
A.~Ghinculov and J.J.~van der Bij, Nucl. Phys. {\bf B436} (1995) 30; \\
%%CITATION = NUPHA,B436,30;%%
S.~Bauberger and M.~B\"ohm, Nucl. Phys. {\bf B445 } (1995) 25.
%%CITATION = NUPHA,B445,25;%%

\bibitem{gammaZ}
L.~Baulieu and R.~Coquereaux,  Ann. Phys. {\bf 140} (1982) 163.
%%CITATION = APNYA,140,163;%%

\bibitem{qgraf}
P. Nogueira, J. Comput. Phys. {\bf 105} (1993) 279.
%%CITATION = JCTPA,105,279;%%

\bibitem{diana}
J. Fleischer and M.N.~Tentyukov,
Comp. Phys. Commun. {\bf 132} (2000) 124.
%%CITATION = CPHCB,132,124;%%

\bibitem{FORM}
J. A. M. Vermaseren, {\it Symbolic manipulation with FORM}, \\ Amsterdam,
Computer Algebra Nederland, 1991.

\bibitem{tarasov-propagator}
O.V.~Tarasov, Nucl.\ Phys.\ {\bf B502} (1997) 455;
%%CITATION = NUPHA,B502,455;%%
Phys. Rev. {\bf D54} (1996) 6479.
%%CITATION = PHRVA,D54,6479;%%

\bibitem{2region}
J. Fleischer, M.Yu. Kalmykov and O.L. Veretin, 
Phys.\ Lett.\ {\bf B427} (1998) 141.
%%CITATION = PHLTA,B427,141;%%

\bibitem{tlamm}
%L.V.~Avdeev,  J.~Fleischer, M.Yu.~Kalmykov and M.N.~Tentyukov,
L.V.~Avdeev et al.,
Nucl.\ Ins.\ Meth. {\bf A389} (1997) 343;\\ 
%%CITATION = NUIMA,A389,343;%%
Comp.\ Phys.\ Commun.\ {\bf 107} (1997) 155.
%%CITATION = CPHCB,107,155;%%

\bibitem{top}
%J.~Fleischer, F.~Jegerlehner, M.~Tentyukov and  O.~Veretin, 
J.~Fleischer et al.,
Phys. Lett. {\bf B459} (1999) 625.
%%CITATION = PHLTA,B459,625;%%

\bibitem{diagramatic}
M.~Veltman, {\it Diagramatica}, 
Cambridge  University Press, 1994.

\bibitem{onshell2}
J.~Fleischer and O.V. Tarasov,
Comp. Phys. Commun. {\bf 71} (1992) 193; \\
%%CITATION = CPHCB,71,193;%%
J.~Fleischer and  M.~Yu.~Kalmykov,
Comp. Phys. Commun. {\bf 128} (2000) 531.
%%CITATION = CPHCB,128,531;%%

\bibitem{oleg-onshell}
%J.~Fleischer, F.~Jegerlehner, O.V.~Tarasov and  O.L.~Veretin,
J.~Fleischer et al.,
Nucl. Phys. {\bf B539} (1999) 671; {\bf B571} (2000) 511(E).
%%CITATION = NUPHA,B539,671;%%

\bibitem{onshell-master}
J.~Fleischer, A.V.~Kotikov and O.L.~Veretin,
Nucl. Phys. {\bf B547} (1999) 343; \\
%%CITATION = NUPHA,B547,343;%%
J.~Fleischer, M.Yu.~Kalmykov and  A.V.~Kotikov,
Phys.\ Lett.\ {\bf B462} (1999) 169; \\ {\bf B467} (1999) 310(E); \\
%%CITATION = PHLTA,B462,169;%%
A.I.~Davydychev and  M.Yu.~Kalmykov, 
Nucl. Phys. {\bf B605} (2001) 266 (hep-th/0012189). 
%%CITATION = HEP-TH 0012189;%%

\bibitem{pole:QCD}
R.~Tarrach, Nucl. Phys.~{\bf B183} (1981) 384; \\
%%CITATION = NUPHA,B183,384;%%
A.S.~Kronfeld, Phys.~Rev.~{\bf D58} (1998) 051501.
%%CITATION = PHRVA,D58,051501;%%

\bibitem{pole:SM}
P.~Gambino and P.A.~Grassi, Phys.~Rev.~{\bf D62} (2000) 076002.
%%CITATION = PHRVA,D62,076002;%%

\bibitem{dimreg}
G.~'t Hooft and M.~Veltman,
Nucl.\ Phys.\ {\bf B44} (1972) 189;\\
%%CITATION = NUPHA,B44,189;%%
C.G.~Bollini and J.J.~Giambiagi,
Nuovo~Cim. {\bf B12} (1972) 20; \\
%%CITATION = NUCIA,B12,20;%%
J.F.~Ashmore,
Lett.\ Nuovo Cim.\ {\bf 4} (1972) 289;\\
%%CITATION = NCLTA,4,289;%%
G.M.~Cicuta and E.~Montaldi,
Lett.\ Nuovo Cim.\ {\bf 4} (1972) 329.
%%CITATION = NCLTA,4,329;%%

\bibitem{RG_1loop}
D.J.~Gross and F.~Wilczek,  
Phys. Rev. {\bf D8} (1973) 3633. 
%%CITATION = PHRVA,D8,3633;%%

\bibitem{tHooft:RG}
G.~'t Hooft, Nucl. Phys. {\bf B61} (1973) 455.
%%CITATION = NUPHA,B61,455;%%

\bibitem{RG_2loop}
D.R.T.~Jones, Nucl. Phys. {\bf B87} (1975) 127; 
%%CITATION = NUPHA,B87,127;%%
Phys. Rev. {\bf D25} (1982) 581; \\
%%CITATION = PHRVA,D25,581;
M.E.~Machacek and M.T.~Vaughn, Nucl.Phys. {\bf B222} (1983) 83. 
%%CITATION = NUPHA,B249,70;%%

\bibitem{modify:MS}
D.J.~Broadhurst,
Z. Phys. {\bf C54} (1992) 599.
%%CITATION = ZEPYA,C54,599;%%

\bibitem{mass-ratio}
%N.~Gray, D.J.~Broadhurst, W.~Grafe and K.~Schilcher, 
N.~Gray et al.,
Z. Phys. {\bf C48} (1990) 673; \\
%%CITATION = ZEPYA,C48,673;%%
L.V.~Avdeev and M.Yu.~Kalmykov, 
Nucl. Phys. {\bf B502} (1997) 419.
%%CITATION = NUPHA,B502,419;%%

\bibitem{Veltman:screening}
M.~Veltman, Acta Phys. Polon. {\bf B8} (1977) 475.
%%CITATION = APPOA,B8,475;%%

\bibitem{finite}
G.~'t Hooft and M.~Veltman, 
Nucl. Phys. {\bf B153} (1979) 365.
%%CITATION = NUPHA,B153,365;%%

\bibitem{one-loop-propagator}
U.~Nierste, D.~M\"uller and M.~B\"ohm, 
Z. Phys. {\bf C57} (1993) 605.
%%CITATION = ZEPYA,C57,605;%%

\bibitem{D-ep}
A.I.~Davydychev,
Proc.\ Workshop ``AIHENP-99'', Heraklion, Greece, April 1999
(Parisianou S.A., Athens, 2000), p.~219 (hep-th/9908032); 
%%CITATION = HEP-TH 9908032;%%
Phys.\ Rev.\ {\bf D61} (2000) 087701.
%%CITATION = PHRVA,D61,087701;%%

\bibitem{1loop-analytic}
A.I.~Davydychev and M.Yu.~Kalmykov,
Nucl.\ Phys.\ B (Proc.\ Suppl.) {\bf 89} (2000) 283 (hep-th/0005287);
%%CITATION = HEP-TH 0005287;%%
Nucl. Phys. {\bf B605} (2001) 266 (hep-th/0012189). 
%%CITATION = HEP-TH 0012189;%%

\bibitem{Lewin}
L.~Lewin, {\it Polylogarithms and associated functions}
(North-Holland, Amsterdam, 1981).

\bibitem{BDS}
F.~A.~Berends, A.~I.~Davydychev and V.~A.~Smirnov, 
Nucl. Phys. {\bf B478} (1996) 59.
%%CITATION = HEP-PH 9602396;%%

\bibitem{BD-TMF}
E.E.~Boos and A.I.~Davydychev,
Theor.\ Math.\ Phys.\ {\bf 89} (1991) 1052.
%%CITATION = TMPHA,89,1052;%%

\bibitem{numbers}
D.J.~Broadhurst, hep-th/9604128;
%%CITATION = HEP-TH 9604128;%%
Eur.\ Phys.\ J.\ {\bf C8} (1999) 311; \\
%%CITATION = EPHJA,C8,311;%%
J.~Fleischer and M.~Yu.~Kalmykov,
Phys.\ Lett.\ {\bf B470} (1999) 168; \\
%%CITATION = PHLTA,B470,168;%%
M.Yu.~Kalmykov and O.~Veretin, 
Phys.\ Lett.\ {\bf B483} (2000) 315.
%%CITATION = PHLTA,B483,315;%%

\bibitem{FHWW02}
A.~Freitas, W.~Hollik, W.~Walter, G.~Weiglein,
``Electroweak two-loop corrections to the $M_W-M_Z$ mass correlation in the
Standard Model'', hep-ph/0202131.

\end{thebibliography}
\end{document}